\documentclass[hyper]{JHEP3}

\usepackage{epsfig}
\usepackage{latexsym}
\usepackage{amsfonts}
\usepackage{amsmath}
\usepackage{amsthm}
\usepackage{amssymb}
\usepackage{amsbsy}
\usepackage{multirow}

\newcommand{\Dp}[1]{\mathrm{D {#1}}}
\newcommand{\antiDp}[1]{\overline{\Dp{#1}}}

\newcommand{\GmSix}{\Gamma_6}
\newcommand{\GmSixb}{\Gamma_{\overline{6}}}
\newcommand{\GmA}{\Gamma_a}

\def\cala         {{\cal A}}

\def\calh         {{\cal H}}

\def\calk         {{\cal K}}
\def\call         {{\cal L}}

\def\caln         {{\cal N}}

\newsavebox{\uuunit}
\sbox{\uuunit}
    {\setlength{\unitlength}{0.825em}
     \begin{picture}(0.6,0.7)
        \thinlines
        \put(0,0){\line(1,0){0.5}}
        \put(0.15,0){\line(0,1){0.7}}
        \put(0.35,0){\line(0,1){0.8}}
       \multiput(0.3,0.8)(-0.04,-0.02){12}{\rule{0.5pt}{0.5pt}}
     \end {picture}}

\def\be{\begin{equation}}
\def\ee{\end{equation}}
\def\bea{\begin{eqnarray}}
\def\eea{\end{eqnarray}}

\def\a{\alpha}
\def\b{\beta}

\def\G{\Gamma}

\def\n{\nu}
\def\o{\omega}
\def\O{\Omega}

\def\S{\Sigma}

\title{Quantizing $\caln=2$ Multicenter Solutions} \author{Jan de Boer$^1$,
Sheer El-Showk$^1$, Ilies Messamah$^1$ and Dieter Van den Bleeken$^2$

\\

$^1$ Instituut voor Theoretische Fysica, Universiteit Amsterdam, \\
Valckenierstraat 65, 1018XE Amsterdam, The Netherlands\\
\\

$^2$ Instituut voor Theoretische Fysica, KU Leuven, \\
Celestijnenlaan 200D, B-3001 Leuven, Belgium \\
}

\abstract{${\cal N}=2$ supergravity
in four dimensions, or equivalently ${\cal N}=1$ supergravity in five
dimensions, has an interesting set of BPS solutions that each correspond to a number of charged centers.  This set
contains black holes, black rings and their bound states, as well as many smooth
solutions.  Moduli spaces of such solutions carry a natural symplectic form
which we determine, and which allows us to study their quantization. By counting
the resulting wavefunctions we come to an independent derivation of some of the
wall-crossing formulae. Knowledge of the explicit form of these wavefunctions
allows us to find quantum resolutions to some apparent classical paradoxes
such as  solutions with barely bound centers and those with an infinitely deep
throat. We show that quantum effects seem to cap off the throat at a finite
depth and we give an estimate for the corresponding mass gap in the dual
CFT.  This is an interesting example of a system where quantum effects cannot be neglected
at macroscopic scales even though the curvature is everywhere small.}
\preprint{ITFA-2008-28\\KUL-TF-08/18}

\begin{document}

\section{Introduction}

One way to try to understand black holes in string theory is to follow them as
we slowly decrease the value of the string coupling constant. Eventually
gravitational interactions will become so weak that the black hole will turn
into weakly interacting strings and branes. The number of states in the latter
system with given quantum numbers can often be determined via a direct
computation in weakly coupled string or field theory. In suitable supersymmetric
situations, where the number of states can be argued to be independent of $g_s$,
this provides a microscopic computation of the entropy of the black hole as was
first accomplished in \cite{Strominger:1996sh}.

In some cases, notably for half-BPS states in $\mathcal{N}=4$ SYM
\cite{Lin:2004nb}\cite{Mandal:2005wv} and for half-BPS states in the D1-D5
system \cite{Lunin:2001fv} one can do much better and show that the phase space
of classical smooth supergravity solutions with the same quantum numbers as the
black hole agrees with (the supersymmetric sector of) the phase space of
the weakly coupled dual field theory. In particular, the symplectic form
obtained in supergravity by restricting the full symplectic form to the relevant
family of solutions is identical to that obtained in the dual field theory
\cite{Grant:2005qc}\cite{Maoz:2005nk}\cite{Donos:2005vs}\cite{Rychkov:2005ji}. In these cases the
entire phase space is subject to a non-renormalization theorem and one can see
the individual microstates arising from supergravity directly. Furthermore, the black hole
is to be viewed as a coarse grained, thermodynamical description of the
corresponding family of smooth solutions \cite{Balasubramanian:2005mg}. This latter statement is the essence
of the fuzzball proposal
\cite{Lunin:2001jy,Lunin:2002qf,Mathur:2002ie,Lunin:2002bj}. With such
a detailed matching between supergravity solutions and the dual field theory at
hand we can, in principle, ask very detailed questions about black hole
geometries.  In particular, we can try to analyze how and where (semi)classical
supergravity breaks down. A precise answer to the latter question is bound to
shed considerable light on the nature of quantum gravity and the
information loss paradox.

Unfortunately, so far full agreement has only been achieved in cases where no
large macroscopic black holes exist. Motivated by this we consider, in this
paper, families of solutions of the supergravity equations of motion in four and
five dimensions with fixed asymptotics that preserve four out of eight
supercharges. Our analysis applies both to the case of asymptotically flat and
asymptotically AdS${}_3$ solutions.  Among such solutions there exist large,
supersymmetric and extremal black holes, bound states of black holes and black
rings, as well as many smooth solutions. For a fixed set of given charged
centers there is a multi-parameter space of BPS solutions that we refer to as
the {\it solution space} of that configuration.  At small values of $g_s$ such a
configuration of charged centers can be related to a 0+1 dimensional quiver
gauge theory \cite{Denef:2002ru}, whose BPS moduli space coincides with the
corresponding solution space in supergravity (i.e. they are parameterized by the
same variables satisfying the same constraints).  This gauge theory describes
branes wrapped on internal cycles of the compactification manifold and a
non-renormalization theorem relates the symplectic form derived in this context
to the symplectic form on the supergravity solution space (i.e.  which makes the
latter a phase space).  Thus, rather than directly evaluating the symplectic
form in supergravity, we will evaluate it in a weakly coupled dual description
and rely on the non-renormalization theorem to extend our results to strong
coupling.  We should note that extending the symplectic form in this way would
not make any sense if the moduli space of solutions was itself not protected by
the same renormalization theorem.  The fact that this space can actually be
explicitly computed in the small and large $g_s$ regime and matches perfectly is
an encouraging confirmation of the validity of our approach.

Obviously, the BPS phase space so obtained is a restricted subset of the full
phase space of the supergravity theory so great care must be taken in using the
associated wave-functions.  One immediate subtlety that emerges, for instance,
is that quantization does not completely commute with reduction.  Specifically,
first imposing a BPS condition and then quantizing should give rise to slight
deformations of states obtained by first quantizing and then imposing a BPS
condition.  In the latter case the wave functions will also have some small
support away from supersymmetric solutions \cite{Denef:2002ru}, whereas in the
former they do not. Thus the detailed form of the wave-functions will be
different.  The wave-functions with support on the BPS phase space will not be
appropriate for many dynamical questions, but only for those for which the
path-integral localizes on supersymmetric field configurations.  A more detailed
discussion of the subtleties emerging in such a quantization can be found in
\cite{Maoz:2005nk}.

%
%

These subtleties notwithstanding, we will find that explicit knowledge of the
phase space and corresponding wave functions in supergravity is very useful in
understanding and clarifying the nature of the various supergravity solutions.
In particular, for a specific set of configurations we will be able to give an
alternative derivation of the wall-crossing formula which expresses how many
states appear or disappear from the spectrum whenever we cross a wall of
marginal stability \cite{Denef:2007vg}, including non-primitive crossings.  What
is more, our derivation relies on the explicit phase space structure rather than
the ability to decompose configurations by tuning moduli at infinity (as in
\cite{Denef:2007vg}) so we are able to compute the number of states for scaling solutions
\cite{Bena:2004tk} \cite{Denef:2007vg}\cite{Bena:2007qc} which exist for all
values of the moduli and hence have no known split attractor flow description.
Extending our results to scaling solutions with an arbitrary number of centers
may shed some light on the idea that these solutions are
candidate “microstate” geometries for macroscopic $\caln = 2$ black holes.  This
is something we hope to come back to in the near future.

Our results also point to what might well be more general lessons regarding the
quantum structure of black holes.  Of particular importance is understanding
when quantum effects can no longer be neglected.  A general lesson that could be
drawn from our analysis is that the combination of gravity and quantum mechanics
can lead to large scale quantum structure \cite{Mathur:2007sc}
\cite{Mathur:2005zp}.  We find, for instance, that a small phase space volume
(one plank unit) can sometimes be spread over a macroscopically large physical
volume.  There exist a class of nominally ``classical'' solutions with large
actions that differ from each other on macroscopic scales but that nonetheless
occupy the same plank volume of phase space and thus do not represent ``good''
classical geometries because Heisenberg's uncertainty principle implies that it
is impossible to localize a wavefunction on one of these solutions.  This
applies, in particular, to the aforementioned scaling solutions and implies
that, although classically they seem to correspond to smooth geometries with
infinitely deep throats, quantum effects actually cap the throats off at some
finite distance.

By analysing the explicit wavefunctions we discover that the probability density
vanishes at the point in phase space that corresponds, classically, to the
infinitely deep throat. So it seems that, taking into account quantum effects,
the infinitely deep throat solution is not a good classical solution and, at
some depth, the naive classical geometry breaks down.  We study this using
an ``effective'' capped-throat whose depth is computed in the state that
localizes as close to the scaling point as possible. In this state we compute
the expectation value of the size of the ``cap'' in terms of the charges and
relate it to the mass gap of the dual CFT. Indeed, without this quantum effect
that cuts off the throat, we would be forced to understand the strange
appearance of a continuous spectrum in the CFT because the energy of excitations
localized deep down the throat are red-shifted by an arbitrary amount when
measured from infinity.

%
%

One point to stress is that we will be completely explicit in describing the
phase space and the corresponding wave functions in cases with two or three
centers and also in a restricted class of solutions with an arbitrarily large
number of centers. The main technical tool that allows us full analytic control
over these wavefunctions is the observation that, in many cases, the solution
spaces of multicentered solutions are toric manifolds. Using some technology
from toric K\"ahler geometry allows us to perform the geometric quantization of
these solution spaces explicitly.

Using some results of \cite{deBoer:2008fk} our analysis carries over
straightforwardly to solutions that are asymptotically
AdS${}_3\times$S${}^2\times$CY. Quantization of such families of solutions
yields states dual to those in $N=(0,4)$ MSW conformal field theory, and we hope
that this will help shed some light on the mysterious nature of this SCFT
\cite{Maldacena:1997de}\cite{Minasian:1999qn}\cite{deBoer:2008fk}.

The outline of this paper is as follows. In section~2, we review the multicenter
solutions in asymptotically flat and AdS${}_3$ space-times that we use in the
remainder of the paper and we summarize the split attractor flow conjecture
which allows us to distinguish bona fide solutions from ill-defined ones. In
section~3 we discuss various aspects of solutions obtained in this way with
particular emphasis on two-centered and three-centered ones. Section~4 contains
the main result of the paper, which is an explicit expression for the symplectic
form which is then used to quantize the space of two- and three-centered
solutions using methods of toric K\"ahler geometry.  We also discuss how to
properly incorporate the fermionic degrees of freedom in the phase space
description and show that these are crucial in order to obtain agreement with
the wall-crossing formula. In sections~5 and 6 we generalize our procedure to
certain classes of solutions with an arbitrary number of centers but only up to
three different types of charges. We again compare to wall-crossing formulae,
this time for non-primitive charges, and again find perfect agreement.
In section~7 we consider what happens when centers can move off to infinity in
order to clarify which solutions correspond to bound and which to unbound
states. Finally, in section 8 we discuss the quantum structure of solution spaces with a
scaling point and compute the way in which quantum effects cap off the throat of
these solutions that classically can become infinitely deep. We use these
results to give some estimate of the mass gap in the dual CFT. We conclude in
section~9 with some comments about the status and meaning of the fuzzball
proposal in this context and some indications for future work.

There are also three appendices where we have collected some technical results
and background material. In appendix A we describe three center solution spaces
in full generality and detail, proving that the size of the angular momentum is
a smooth coordinate on that space. In the second appendix we review some results
in symplectic toric geometry that will be of use in the main text. Appendix C
contains a toy model computation of the mass gap using AdS/CFT by considering a
free scalar field on an approximate geometry.

\section{Review of Multicentered Black Holes}

Let us begin with a brief review of multicentered black hole solutions of
$\mathcal{N}=2$ supergravity in 4 dimensions.  From the 10 dimensional point of
view we will be considering Type IIA on a Calabi-Yau threefold $X$ (though
most of our results carry over to IIB).  Furthermore
we shall work in the limit where the volume of $X$ is very large so we can restrict
ourselves to just the cubic part of the prepotential.

The multicenter solutions are determined by specifying a number of charges,
$\Gamma_a$, and distributing their location in the spatial $\mathbb{R}^3$. These
charged centers correspond in the 10 dimensional picture to branes wrapping even
cycles in the CY$_3$. There are $2b_2+2$ independent such cycles in homology,
with $b_2$ the second Betti-number of $X$, each giving rise to one of the
$2b_2+2$ abelian vector fields of the $\caln=2$ supergravity.  We will often
denote brane configurations by their coefficients in a basis of homology, i.e.
$\G=(p^0,p^A,q_A,q_0)=p^0+p^AD_A+q_A\tilde D^A+q_0 V$, where the $D_A$ form a
basis of $\mathrm{H}^2(X,\mathbb{Z})$, the $\tilde D_A$ make up a dual basis and
$V$ is the unit volume element of X, dual to 1.  There is a natural intersection
product on $\mathrm{H}^*(X, \mathbb{Z})$ given by

\begin{equation}
\langle \Gamma_1,\Gamma_2 \rangle := - p_1^0 q^2_0 + p_1^A q^2_A - q^1_A p_2^A
+  q^1_0 p_2^0.
\end{equation}

Furthermore the moduli of the Calabi-Yau appear as scalar fields in the
four-dimensional effective theory. In the solutions we will be considering the
hyper-multiplet moduli will be kept constant while the ones in the
vector-multiplet will be allowed to vary dynamically. It is thus necessary to
choose boundary values for the scalar moduli at infinity.  We will briefly
discuss the {\em split attractor flow conjecture} which relates the existence of
solutions at particular values of the moduli at infinity to the existence of
certain flow trees in moduli space.

We will be very brief in our review as these solutions are discussed in great
detail in the references \cite{Denef:2000nb}, \cite{Bates:2003vx} and
\cite{Denef:2007vg}.  Although below we will discuss primarily the four dimensional
version of these solutions it is well known that these solutions can be uplifted
to five dimensions \cite{Gaiotto:2005gf} yielding multicentered solutions with a
$U(1)$ isometry and a Gibbons-Hawking base that can be made smooth by required
all the four-dimensional charges to come from fluxed $D6$-branes
\cite{Bena:2005va} \cite{Berglund:2005vb}.  These are solutions of 11
dimensional supergravity compactified on the same Calabi-Yau, $X$.

\subsection{Four Dimensional Solutions}

Our starting point is the class of multicentered black holes solutions first
discussed in \cite{Behrndt:1997ny} \cite{Denef:2000nb} and later explicitly constructed
in \cite{Bates:2003vx}.  They are given by the following four dimensional metric,
gauge fields and moduli:
\begin{eqnarray}
 ds^2 & = & -\frac{1}{\S}(dt+\o)^2+\S\, dx^idx^i\,, \nonumber\\
 \cala^0 & = & \frac{\partial \log \S}{\partial H_0}\left(dt+\o\right)+\omega_0\,,\label{multicenter}\\
 \cala^A & = & \frac{\partial \log \S}{\partial H_A}\left(dt+\o\right)+\cala_d\,,\nonumber\\
 t^A&=&B^A+i\,J^A=\frac{H^A-i\frac{\partial \S}{\partial H_A}}{H^0+i\frac{\partial \S}{\partial H_0}},\nonumber
\end{eqnarray}
All functions appearing in this solution can be expressed in terms of $2b_2+2$
harmonic functions:
\begin{equation} \label{harmonics}
H=(H^0,H^A,H_A,H_0)=\sum_a\frac{\G_a}{|x-x_a|}-2\mathrm{Im}(e^{-i\alpha}\O)|_\infty\,,
\end{equation}
where the components of the $\G_a$ take values in $H_{\textrm{ev}}^*(X, \mathbb{Z})$, the
integral even cohomology of the Calabi-Yau $X$, $e^{i\a}$ is the phase of the
total central charge at infinity (i.e.
$Z(\G)|_\infty=\langle\sum_s\G_s,\O|_\infty\rangle$ and
$e^{i\a}=\frac{Z}{|Z|}|_\infty$) and
$\Omega=-\frac{e^{B+iJ}}{\sqrt{\frac{4J^3}{3}}}$.

%
%
The function $\Sigma$ appearing in the metric in (\ref{multicenter}) is known as
the entropy function.  When evaluated at $\vec{x}_a$ it is proportional to the
entropy of the black hole whose horizon lies at $\vec{x}_a$.  This follows from the
Bekenstein-Hawking relation and the fact that $\Sigma$ determines the area of
the horizon at $\vec{x}_a$.
%
%

The rest of the solution is defined in terms of the harmonic functions as
follows
\begin{eqnarray}\label{metric-functions}
 d\omega_0 & = & \star dH^0 \,,\nonumber\\
 d\cala_d^A & = & \star dH^A \,,\nonumber\\
 \star d\o & = & \langle dH,H\rangle \,\nonumber
 \end{eqnarray}
If we take the prepotential to be cubic, which is tantamount to taking the
large volume limit in IIA, we can find an explicit form for the entropy
function \cite{Shmakova:1996nz}
\begin{eqnarray}
 \S&=&\sqrt{\frac{Q^3-L^2}{(H^0)^2}}\,,\label{conditions}\\
 L&=&H_0(H^0)^2+\frac{1}{3}D_{ABC}H^AH^BH^C-H^AH_AH^0\,,\nonumber\\
 Q^3&=&(\frac{1}{3}D_{ABC}y^Ay^By^C)^2\,,\nonumber\label{defQ} \\
 D_{ABC}y^Ay^B&=&-2H_CH^0+D_{ABC}H^AH^B\,,\nonumber
 \end{eqnarray}
where one should note that the Hodge star $\star$ is the one of flat $\mathbb{R}^3$. Finally there are $N-1$ consistency conditions on the relative
positions of the $N$ centers. These conditions come from
integrability of the equations of motion for $\o$. They take the simple form
\begin{equation}
  \langle H,\G_s\rangle|_{x=x_s}=0\,,
 \label{consistency}
\end{equation}
or written out more explicitly\footnote{In this sum we imply summation over $a$ different from some fixed $b$. In case we don't give a separate mention of the summation index we mean a double sum. E.g. $\sum_{a<b}$ is sum over $a$ {\bf and} $b$ with $a$ smaller than $b$, while $\sum_{a,\,a<b}$ would mean a sum over only $a$ such that $a$ is smaller than some fixed $b$.}
\begin{equation}
        \sum_{a,\,a\neq  b}
	\frac{\langle
	\G_a,\G_b\rangle}{r_{ab}}=\langle h,\G_a\rangle\,, \label{consistency2}
\end{equation}
where $r_{ab}=|x_{ab}|=|x_a-x_b|$.

An important property of a configuration with a sufficient number of centers is
that although the centers bind to each other there is some freedom left to
change their relative positions. These possible movements can be thought of as
flat directions in the interaction potential. Equation (\ref{consistency2})
constrains the locations of the centers to the points where this potential is
zero. As, for a system with $N$ centers, there are $N-1$ such equations for $3N-3$
coordinate variables (neglecting the overall center of mass coordinate) there
is, in general, a $2N-2$ dimensional moduli space of solutions for fixed
charges and asymptotics. This space may or may not be connected and it may even
have interesting topology. We will refer to this as the moduli space of
solutions or {\em solution space}; the latter terminology will be preferred as it is
less likely to be confused with the moduli space of the Calabi-Yau in which the
scalar fields $t^A$ take value. The shape of this solution space does, in fact,
depend quite sensitively on where the moduli at infinity, $t^A|_\infty$, lie in
the Calabi-Yau moduli space (as the latter determine $h$ on the RHS of eqn.
(\ref{consistency2})).  Quantization of some low-dimensional instances of this
space will be the primary goal of the rest of the paper.

Let us briefly recall, once more, that these
four-dimensional solutions can be uplifted to five-dimensions
\cite{Gaiotto:2005gf}.  Moreover, for certain choices of charges, namely pure
D6's with only abelian flux turned on\footnote{Note that in the duality frame we
are working in it is purely fluxed D6-centers that give rise to smooth five
dimensional solutions. In other duality frames this can be purely fluxed D4's,
etc.  It is probably more precise to consider zero entropy centers instead of
only smooth centers, as the former is a duality invariant statement.}, the
five-dimensional uplift of the solutions is smooth
\cite{Bena:2007kg}\cite{Berglund:2005vb}\cite{Cheng:2006yq}\cite{Balasubramanian:2006gi};
there are, at worst, conical singularities at the location of each center.  The
five dimensional solutions are, in general, topologically non-trivial
multi-Taub-NUT spaces with fluxes supported on the two cycles between the
Taub-NUT centers.  The fact that such smooth solutions, with neither horizons
nor singularities, can be found is important for the fuzzball proposal and some
of the discussion that follows will involve such configurations within the
context of this proposal.  For a more details and discussion of these smooth
solutions the reader is referred to the references given above.

\subsection{Angular Momentum} \label{angular-momentum}

These solutions are stationary and, as shown in \cite{Denef:2000nb}, they carry
an angular momentum equal to
\begin{equation}
J=\frac{1}{2}\sum_{a<  b} \frac{\langle \G_a,\G_b\rangle\,x_{ab}}{r_{ab}}\label{angularmomentum}\,.
\end{equation}
Note that $J$ is quantized in half integer units.

As angular momentum will play an important role in the rest of this paper,
providing a natural coordinate on the solution space, we will derive some more
useful ways of expressing it.  Multiplying the condition (\ref{consistency2}) by
$x_a$ and then summing over the different centers shows that
\begin{equation}
J=\frac{1}{2}\sum_a\langle h,\G_a\rangle x_a\,. \label{angularmomentum2}
\end{equation}
By using the fact that $\sum_a\langle h,\G_a\rangle=0$ one can rewrite the last
expression as
\begin{equation}
	J=\frac{1}{2}\sum_{a, a\neq b}\langle h,\G_a\rangle x_{ab}\,.\label{formula4}
\end{equation}
Starting from this formula we can show that the size of the angular momentum can
be compactly written in terms of the inter-center distances $r_{ab}$. Squaring
(\ref{formula4}) gives
\begin{equation}
J^2=\frac{1}{4}\sum_{a, a\neq b}\sum_{c,\, c\neq b}\langle h,\G_a\rangle\langle h,\G_c\rangle\,x_{ab}\cdot x_{cb}\,.
\end{equation}
Now we can use that for any three points labeled by $a,b,c$ there is the
relation $x_{ab}\cdot x_{cb}=\frac{1}{2}(r_{ab}^2+r_{cb}^2-r_{ac}^2)$ and a
little more algebra reveals that
\begin{equation}
|J|=\frac{1}{2}\sqrt{-\sum_{a<b}\langle h,\G_a\rangle\langle h,\G_b\rangle\,r_{ab}^2}\,.\label{angularnorm}
\end{equation}

\subsection{The Split Attractor Flow Conjecture and Wall Crossing}\label{afconjecture}

%
%

So far we have reviewed a class of four dimensional solutions but these
solutions are relatively complicated and it is non-trivial to determine if they
are well-defined everywhere.  In particular the entropy function, $\Sigma$, that
appears in the solution involves a square root and may take imaginary values in
some regions (when uplifted to five dimensions this can lead to closed timelike
curves  \cite{Cheng:2006yq} \cite{Berglund:2005vb} \cite{Bena:2005va}).  In
\cite{Denef:2000nb} and \cite{Denef:2007vg} a simplified criterion was proposed
for the well behavedness of such solutions which we will now relate (very
briefly).

In \cite{Denef:2000nb} a conjecture was proposed whereby the existence of
multicentered solutions is equivalent to the existence of an {\em attractor flow
tree}.  The latter is a graph in the moduli space beginning at the moduli at
infinity and ending at the attractor points for each center.  The edges
correspond to single center flows towards the attractor point for the sum of
charges further down the tree.  Vertices can only occur where single center
flows (for a charge $\Gamma = \Gamma_1 + \Gamma_2$) cross walls of marginal
stabilities so that the central charges are all aligned ($|Z(\Gamma)| =
|Z(\Gamma_1)| + |Z(\Gamma_2)|$).  The actual (spatially dependent) flow of the
moduli, $t^A$, for a multicentered solution will then be a thickening of this
graph.  A sample tree is displayed in figure \ref{fig_flow}.

As mentioned, the main purpose of the attractor flow tree is to allow us to
determine if a solution is well defined.  For a single centered black hole the
entropy function, $\Sigma$, undergoes a monotonic flow from infinity to the
horizon.  At infinity the value of $\Sigma$ depends on the choice of moduli
(boundary conditions) while at the horizon it flows to a fixed value depending
only on the charges, as the moduli are fixed by an attractor mechanism.
Spherical symmetry dictates that the moduli depend only on a radial variable so
the flow through moduli space is indeed just a single line from the moduli at
infinity to the attractor value.  If $\Sigma$ should become imaginary
somewhere along this flow the solution would suffer from pathologies.  However,
since the flow is monotonic it need only be checked at its initial (the moduli
at infinity) and final points (the attractor point).

For a multicentered system the moduli depend on three variables and the flow is
no longer monotonic in a straightforward way (it is not even a one dimensional
tree but rather a ``fat graph'').  By assuming that solutions could be built
constructively by bringing in centers from infinity \cite{Denef:2000nb} was able
to conjecture that even for multicentered configurations we can study a flow
{\em tree} in the moduli space (recall the actual flow will be a ``fat'' version
of this) and study each leg of the flow to check for pathologies.  The
conjecture is then that if the tree exists (each leg is pathology free) then the
full solution is actually well behaved  (see \cite{Denef:2000nb}
\cite{Denef:2000ar} for more details).  There is considerable evidence for this
conjecture
\cite{Denef:2000nb}\cite{Denef:2002ru}\cite{Denef:2000ar}\cite{Denef:2007vg} and
our computation in Section \ref{comparison} will provide even further support.

Although our treatment of attractor trees here is very brief we will discuss one
more important result from \cite{Denef:2007vg}, the so called wall crossing formulae. These are formulas for the index of BPS states of given total charge. They are derived using that the index of the BPS Hilbert space corresponding to an attractor flow tree can be determined in terms of contributions from each vertex and node.  A given vertex for a
split $\Gamma \rightarrow \Gamma_a + \Gamma_b$ contributes a factor $|\langle
\Gamma_a, \Gamma_b \rangle|$, associated to the number of states in the angular
momentum multiplet coming from the crossed electric and magnetic fields of the
centers $\Gamma_a$ and $\Gamma_b$.  Each node, meanwhile, contributes a factor
$\Omega(\Gamma_a)$ where $\Omega(\Gamma_a)$ represents the index associated to
the charge $\Gamma_a$ itself (i.e. coming from ``internal'' degrees of freedom and determined by the entropy formula (\ref{conditions}) with
$H$ being replaced by $\G_a$).
Thus, for instance, the index corresponding to figure \ref{fig_flow}
is
\be\label{afc_entropy}
|\langle \Gamma_3, \Gamma_1 + \Gamma_2\rangle| \, |\langle \Gamma_1, \Gamma_2
\rangle | \, \Omega(\Gamma_1) \, \Omega(\Gamma_2) \, \Omega(\Gamma_3)
\ee
Note that the spacetime contribution (from the vertices) and the ``internal''
contributions (nodes) are easily and clearly separated in this computation.  For
more details, including a derivation (assuming the split attractor conjecture)
the reader is referred to \cite{Denef:2000nb} and \cite{Denef:2007vg}.

%
%

\FIGURE{
\includegraphics[page=3,scale=1]{pictures_quantizing.pdf}
\caption{Three centered attractor flow tree.  The system is composed of three
centers of charge $\Gamma_1$, $\Gamma_2$ and $\Gamma_3$ and the moduli at
infinity are at the value labelled by the yellow circle.  Each leg of the tree
above represents a single center flow towards the attractor value associated
with the {\em total} charge below that point.  Thus the first flow is towards
the attractor point for $\Gamma = \Gamma_1 + \Gamma_2 + \Gamma_3$.  After the
first split the flows are towards the attractor points for charges $\Gamma_3$
and $\Gamma_4 = \Gamma_1 + \Gamma_2$. In each case the split occurs along walls
of marginal stability (thick blue lines).  The first, horizon, line of MS
corresponds to $|Z(\Gamma)| = |Z(\Gamma_3)| + |Z(\Gamma_4)|$ while the second is
for $|Z(\Gamma_4)| = |Z(\Gamma_1)| + |Z(\Gamma_2)|$.}\label{fig_flow}
}

\section{Simple Solution Spaces}

Let us describe some simple moduli spaces of solutions in order to have some
feeling for the spaces we wish to quantize (in section
\ref{symplecticandquantization}). We begin with the simple case of the two
centers solution and then discuss the three centers case.

\subsection{The Two Center Case} \label{2center-moduli}

The solution space for two centers, when it exists, is two dimensional.  The
constraint equation (\ref{consistency2}),
\begin{equation}
   \frac{\langle\G_1,\G_2\rangle}{x_{12}} = \langle h, \G_1 \rangle
\end{equation}
admits solutions if $\langle\G_1,\G_2\rangle/\langle h, \G_1 \rangle$ is
positive and finite.  These solutions have fixed inter-center distance, $x_{12}$,
and are hence parameterized only by the direction of the inter-center axis
(recall that we are factoring out the center-of-mass coordinate).  Thus the
solution space is topologically an $S^2$.

If $\langle \G_1,\G_2 \rangle = \langle h, \G_1\rangle = 0$ then the centers are
unbound and can be located anywhere.  In this case the solutions carry no
angular momentum and are static.  To quantize static solutions one needs to include velocities, as was
e.g. done in \cite{Michelson:1999dx}. Whether or not static solutions give rise to proper BPS
states is an interesting and subtle question on which we further comment in
our conclusions in section \ref{conclusions}.

\subsection{The Three Center Case} \label{sec_3cntr_short}

The three center solution space is four dimensional.  Placing one
center at the origin (fixing the translational degrees of freedom) leaves six
coordinate degrees of freedom but these are constrained by two equations.  This
leaves four degrees of freedom, of which three correspond to rotations in SO(3)
and one of which is related to the separation of the centers.

The constraint equations take the form
\begin{align}\label{3cntr_constraint}
	\frac{a}{u} - \frac{b}{v} &=	\frac{\Gamma_{12}}{r_{12}} -
	\frac{\Gamma_{31}}{r_{31}} = \langle h, \Gamma_1 \rangle
	=: \alpha \\
	\frac{b}{v} - \frac{c}{w} &= \frac{\Gamma_{31}}{r_{31}} -
	\frac{\Gamma_{23}}{r_{23}} = - \langle h,
	\Gamma_3 \rangle =: \beta
\end{align}
in a self-evident notation.  The nature of the solution space simplifies
considerably if either $\alpha$ or $\beta$ vanish so let us first consider this
case (if both vanish there is an overall scaling degree of freedom and the
centers are unbound; we will, in section \ref{sec_infinity}, see that this
corresponds to a degenerate symplectic form and is thus not amenable to
quantization using the methods described in this paper).  For definiteness we
will take $\alpha = 0$; in this case $\sum_p \langle h, \Gamma_p \rangle =0$
which implies $\langle h, \Gamma_2 \rangle = \beta$.  Thus from
(\ref{angularmomentum2}) we find
\be\label{jvec}
\vec{J} = \frac{\beta}{2} \, r_{23} \hat{z}
\ee
with $\hat{z}$ defined to be parallel to $\vec{x}_2 - \vec{x}_3$.

The solution has an angular momentum vector $J^i$ directed between the centers 2
and 3 and the direction of this vector defines an S$^2$ in the phase space which
we will coordinatize using $\theta$ and $\phi$.  The third center is free to
rotate around the axis defined by this vector (since this does not change any of
the inter-center distances) providing an additional $U(1)$, which we will
coordinatize by an angle $\sigma$, fibred non-trivially over the S$^2$.  Finally
the angular momentum has a length which may be bounded from both below and above
and this provides the final coordinate in the phase space, $j = |\vec{J}|$.
This construction is perhaps not the most obvious one from a spacetime
perspective but, as we will see, in these coordinates the symplectic form takes a simple and
convenient form.  When $\alpha=0$ it is clear from (\ref{jvec}) that $j$ is a
good coordinate on the solution space but this is not immediately obvious for the
more complicated case of $\alpha \neq 0$.  This is nonetheless true and, as shown
in Appendix \ref{3center-moduli}, this is always a good coordinatization of the
three center solution space (though for $\alpha \neq 0$ the relation between $(j,
\sigma, \theta, \phi)$ and the coordinates $\vec{x}_p$ is not as
straightforward).

The quantization of these solutions is particularly interesting in certain
cases where classical reasoning leads to pathologies.  Before
proceeding with quantization we will first briefly review these solutions.

\subsubsection{Scaling Solutions}\label{scaling_sol}

As noted in \cite{Bena:2006kb} and \cite{Denef:2007vg}, for certain choices of
charges it is possible to have points in the solution space where the coordinate
distances between the centers goes to zero.  Moreover, this occurs for any
choice of moduli so it is, in fact, a property of the charges alone.

Such solutions occur as follows.  We take the inter-center distances to be given
by $r_{ab} = \lambda \Gamma_{ab} + \mathcal{O}(\lambda^2)$ (fixing the order of
the $ab$ indices by requiring the leading term to always be positive).  As
$\lambda \rightarrow 0$ we can always solve (\ref{3cntr_constraint}) by
tweaking the $\lambda^2$ and higher terms.  The leading behaviour will be $r_{ab}
\sim \lambda \G_{ab}$ but clearly this is only possible if the $\G_{ab}$ satisfy
the triangle inequality.  Thus any three centers with intersection products $\G_{ab}$ satisfying the
triangle inequalities define a scaling solution.

We will in general refer to such solutions as {\em scaling solutions} meaning,
in particular, supergravity solutions corresponding to $\lambda \sim 0$.
The space of supergravity solutions continuously connected (by varying the
$\vec{x}_p$ continuously) to such solutions will be referred to as {\em scaling
solution spaces}.  We will, however, occasionally lapse and use the term scaling
solution to refer to the entire solution space connected to a scaling solution.
We hope the reader will be able to determine, from the context, whether a specific
supergravity solution or an entire solution space is intended.

These scaling solutions are interesting because (a) they exist for all values of the
moduli; (b) the coordinate distances between the centers go to zero; and (c) an
infinite throat forms as the scale factor in the metric blows up as
$\lambda^{-2}$.
Combining (b) and (c) we see that, although the centers naively collapse on top
of each other, the actual metric distance between them remains finite in the
$\lambda \rightarrow 0$ limit.  In this limit an infinite throat develops
looking much like the throat of a single center black hole with the same charge
as the total charge of all the centers.  Moreover, as this configuration exists
at any value of the moduli, it looks a lot more like a single center black hole
(when the latter exists) than generic non-scaling solutions.  As a consequence
of the moduli independence of these solutions it is not clear how to understand
them in the context of attractor flows; the techniques we develop in this paper
provide an alternative method to quantize these solutions that applies even when
the attractor tree does not.

Unlike the throat of a normal single center black hole the bottom of the scaling
throat has non-trivial structure.  If the charges, $\G_a$, are zero entropy bits
(e.g. D6's with abelian flux) then the 5-dimensional uplifts of these solutions will
yield smooth solutions in some duality frame and the throat will not end in a horizon but will be
everywhere smooth, even at the bottom of the throat.  Outside the throat,
however, such solutions are essentially indistinguishable from single center
black holes.  Thus such solutions have been argued to be ideal candidate
``microstate geometries''\footnote{The unfortunate terminology ``microstate
geometry'' is not intended to suggest that the classical solution necessarily
corresponds to a black hole microstate.  Rather, as we will see later,
quantizing the space of such solutions as a phase space provides a set of
quantum states that may or may not have a classical realization.} corresponding
to single center black holes.  In \cite{Denef:2007vg} it was noticed that some
of these configurations, when studied in the Higgs branch of the associated
quiver gauge theory, enjoy an exponential growth in the number of states unlike
their non-scaling cousins which have only polynomial growth in the charges.

\subsubsection{Barely Bound Centers}

We will be brief here as such configurations are discussed in more detail in
section \ref{sec_infinity}.  For certain values of charges and moduli we will
see that it is possible for some centers to move off to infinity.  Although this
would normal signal the decay of any associated states (as happens, for
instance, for two centers at a wall of marginal stability \cite{Denef:2000nb})
we will see that this is not always the case.  In particular, it is important to
distinguish between cases when centers are forced to infinity (marginal
stability) versus those where there is simply an infinite (flat) direction in
the solution space (threshold stability; see \cite[Appendix B]{deBoer:2008fk}).
Although the first case clearly signals the decay of a state, in the second
case, when centers move off to infinity along one direction of the solution
space but may also stay within a finite distance in other regions of the
solution space, it is still possible to have bound states, as we will
demonstrate.  Quite essential to this argument is the fact that in some cases,
although the solution space may seem naively non-compact (in the standard metric
on $\mathbb{R}^{2N-2}$), its symplectic volume is actually finite and it admits
normalizable wave-functions (whose expectation values, we argue in section
\ref{sec_infinity}, are finite).  There are also cases with unbound centers
where the symplectic form on the solution space is degenerate and, in such
cases, it is not clear if there is a bound state.  Such cases are not amenable
to treatment by the methods developed here, see also the discussion on this point in section \ref{conclusions}.

\section{Symplectic Structure and Solution Space Quantization.}\label{symplecticandquantization}

In this section we study the moduli space of our solutions using a
symplectic form which is derived from the quiver quantum mechanics
action of multiple intersecting branes in the coulomb branch
\cite{Denef:2002ru}. We determine this symplectic form explicitly
for the three center case and show it to be non-degenerate (with
some important exceptions where degenerations have a clear
physical interpretation) and proceed to count the number of states
in this moduli space via geometric quantization.  The estimated
number of states at large charge nicely matches the expected
number of states from the wall-crossing formula given in
\cite{Denef:2007vg}.

Motivated by a non-renormalization theorem \cite{Denef:2002ru} and
the exact agreement of our state counting with Denef and Moore's
wall-crossing formula we propose that the same symplectic from
should be derivable from the super-gravity action following the
logic in \cite{Grant:2005qc} (see also \cite{Lee:1990nz} and
references therein). Actually a similar structure to the quiver
quantum mechanics symplectic form emerges from the gauge field
contribution of the supergravity action. So we can rephrase our
conjecture in the following way: the other putative terms
contributing to the symplectic form from supergravity cancel or
only change the normalization \footnote{The last possibility is
seen for example in \cite{Grant:2005qc}.}.

Note that it is not {\em a priori} clear that our moduli space of
solutions corresponds to a full phase space rather than a
configuration space.  The fact that the solution spaces are always
even, ($2N-2$), dimensional is a positive indication that they are
indeed phase spaces as is the fact that the symplectic form is
non-degenerate in the two and three center cases.  Moreover, when
degenerations do occur they have
well-defined physical meaning further supporting this claim.

\subsection{Symplectic Form from Quiver QM}

In order to be able to quantize the space of solutions described
above, we need to determine its symplectic form which is induced
from the supergravity Lagrangian. The idea is very simple, one
first determines the symplectic structure on the full space of
solutions to the supergravity equations of motion, which takes the
form
\be \label{sym1}
\Omega = \int d\Sigma_l \,\, \delta\left( \frac{\partial
L}{\partial (\partial_l \phi^A)}\right) \wedge \delta \phi^A\, .
\ee
where $L$ is the relevant supergravity Lagrangian.  Here, the integral is over a
Cauchy surface, $\phi^A$ represents a basis of the fields that appear in the
Lagrangian and it is assumed that the Lagrangian does not contain second and
higher order time derivatives.  The induced symplectic form on the space of BPS
solutions is then simply the restriction of (\ref{sym1}) to that space.

In general, there is no guarantee that the symplectic form obtained in this way
is non-degenerate. If it turns out to be degenerate then we should have included
further degrees of freedom to arrive at a non-degenerate symplectic form. This
happens for example when all centers are mutually BPS, i.e. if all inner
products $\langle \Gamma_p,\Gamma_q\rangle$ vanishes. In this case the
symplectic form, as we will see below, is identically zero, and the solution
space is therefore better thought of as being a configuration space. In order to
obtain a non-degenerate symplectic form, we could try to include, for example,
small velocities for the centers \cite{Michelson:1999dx}. It is not clear,
however, whether this can be done while saturating the BPS bound.  We return to
this point briefly in our concluding remarks in section \ref{conclusions}.

It may sound surprising that generically the BPS solutions spaces will carry a
non-degenerate symplectic form, since they are all time-independent. Crucially,
the solutions we consider are stationary but in general not static. Stationary
solutions do carry non-trivial momentum despite being time-independent and this
is what gives rise to the non-trivial symplectic form.  As a consequence of this
we will see that the Hilbert space decomposes into angular momentum multiplets.
The symplectic form (\ref{sym1}), when evaluated on a family of static
solutions, will simply vanish.

The idea to quantize spaces of BPS supergravity solutions using the restriction
of (\ref{sym1}) was successfully applied in
\cite{Grant:2005qc}\cite{Maoz:2005nk}\cite{Rychkov:2005ji} (see also
\cite{Raju:2007uj} for an extensive list of older references and
\cite{Crnkovic:1986ex}\cite{Zuckerman:1989cx}\cite{Lee:1990nz} for more
recent, similar work).  However, when we try to apply similar methods to the
space of multicentered black hole solutions, the expressions become very
lengthy and tedious and we run into serious technical difficulties due to the
complicated nature of the multicentered solutions. Therefore we will proceed
differently and try to derive the required symplectic form from a dual open
string description of the supergravity solutions.

From the open string point of view, which is appropriate for small
values of the string coupling constant, black hole bound states
correspond to supersymmetric vacua of a suitable quiver gauge
theory. The connection between the supergravity solutions and
gauge theory vacua becomes clear once we study the Coulomb branch
of the gauge theory. For simplicity we will assume that all
centers have primitive charges; the extension to non-primitive
charges is straightforward, as the supergravity solutions are not
specifically sensitive to whether charges are primitive or not.

For primitive charges, the quiver gauge theory has as gauge group
a product of $U(1)$'s, one for each center, plus bifundamentals.
The number of bifundamentals minus the number of
anti-bifundamentals charged under a pair of $U(1)$'s is equal to
$\langle \Gamma_p,\Gamma_q\rangle$, the inner product of the
charges. To study the Coulomb branch of this theory, we integrate
out massive bifundamentals, resulting in an effective theory for the
vector multiplets only. The degrees of freedom of the vector
multiplet in $0+1$ dimension consist of three scalars $x^i$, four
fermions, one auxiliary field $D$ and one gauge field $A$. As was
shown in \cite{Denef:2002ru}, the low-energy effective action for
$N$ abelian vector multiplets necessarily takes the form
\be \label{effact}
L=\sum_p (-U_p D_p + A_p \cdot \dot{x}_p) + \,\,{\rm
fermions}\,\,+\,\, \textrm{higher order terms},
\ee
where $p$ labels a brane localized at $\vec{x}_p$ and
\be \label{aux1}
\nabla_p U_q =\nabla_q U_p = \frac{1}{2} (\nabla_p \times A_q +
\nabla_q \times A_p).
\ee
This form of the action is determined by requiring that it is
invariant under supersymmetry. The solutions to (\ref{aux1}) are
given in terms of harmonic functions, in particular we find that
\be \label{aux2}
U_p= \langle \Gamma_p,H(x_p) \rangle
\ee
where $H=\theta+\sum_q \frac{\Gamma_q}{|x-x_q|}$ and the pole at
$x=x_p$ in $H$ does not contribute to (\ref{aux2}). We therefore
see that except for the tree-level constant term $\theta$, there
are only one-loop contributions to $U_p$ and $A_p$. Supersymmetry
prohibits higher loop contributions to $U_p$ and $A_p$. Luckily,
we only need $U_p$ and $A_p$ to match the gauge theory to
supergravity and to extract the symplectic form, and since they do
not receive higher loop corrections we can safely extrapolate the
results from the gauge theory regime at small $g_s$ to the
supergravity regime at large $g_s$.

It would be interesting to rephrase the above non-renormalization theorem more
directly in superspace. The superspace formulation of the $0+1$-dimensional
vector multiplet is in terms of a linear superfield $\Sigma_{\alpha\beta}$ with
$D^2\Sigma=\bar{D}^2\Sigma=0$ \cite{Ivanov:1990jn} \cite{Diaconescu:1997ut}.
It is therefore tempting to try to write (\ref{effact}) in terms of a twisted
superpotential
\be L = \int d\theta^{\alpha} d\bar{\theta}^{\beta}
H_{\alpha\beta}(\Sigma)
\ee
with $H$ some sort of prepotential for $U_p$ and $A_p$. A first
investigation indicates that this is not possible; if it were
possible, then it would be clear that the string coupling, which
can be viewed as the expectation value of the lowest component of
an auxiliary spurious chiral superfield, cannot contribute to the
twisted superpotential. Nevertheless, one can verify explicitly
that it is not possible to write down first order terms like
(\ref{effact}) which involve a non-trivial coupling between a
linear superfield and a chiral superfield, thereby giving an
alternative proof of the non-renormalization theorem.

Given (\ref{effact}), the space of supersymmetric vacua of the gauge theory is
given by the solutions of the D-term equations $U_p=0$. These are identical to
the supergravity constraint equations (\ref{consistency}), (\ref{consistency2}),
as $\theta_p$ can be identified with the constant term in the supergravity harmonic
functions.  The symplectic form then follows immediately by restricting
$\sum_p \delta x_p \wedge \delta A_p $ obtained from
(\ref{effact}) to the solution space $\bigcap_p \{ U_p=0\}$. This symplectic
form is independent of $g_s$ and must therefore agree with the symplectic form
obtained from supergravity.

Phrased in terms of the supergravity solution, the symplectic form
becomes
\be
\tilde{\Omega} = \frac{1}{2} \sum_p \delta x^i_p \wedge \langle \G_p ,\delta {\cal
A}_d^i(x_p)\rangle .
\ee
where ${\cal A}_d$ is the ``spatial'' part of the 4d gauge field
defined in (\ref{multicenter}). Notice that
\be
\delta {\cal A}_d^i(x_p) = (\delta {\cal A}_d^i(x))|_{x=x_p}  +
(\delta x_p^k \partial_k {\cal A}_d^i(x) )|_{x=x_p} .
\ee
To proceed further we denote
\be f_p =
\frac{1}{|{\mathbf x} - {\mathbf x}_p|} .
\ee
With some work, one can show that
\be \label{con1a}
\delta {\cal A}_d = \sum_p \Gamma_p \epsilon_{ijk} \delta x^i_p \,
\partial_j f_p \, dx^k
\ee

Using the above form of $\delta {\cal A}_d$ and
the fact that $\partial^k {\cal A}_d^i(x)$ can be replaced by
$\frac{1}{2} (\partial^k {\cal A}_d^i(x) -\partial^i {\cal
A}_d^k(x)) = \frac{1}{2} {\cal F}_d^{ki}(x)$ we finally obtain for
the symplectic form
\be \label{con3}
\tilde{\Omega} = \frac{1}{4} \sum_{p\neq q} \langle \Gamma_p,\Gamma_q\rangle
\frac{\epsilon_{ijk} (\delta (x_p-x_q)^i \wedge \delta
(x_p-x_q)^j) \, (x_p-x_q)^k }{|{\mathbf x}_p - {\mathbf x}_q|^3} .
\ee
Note that the overall translational mode does not contribute to this
symplectic form.

The symplectic form (\ref{con3}) applies for any number of centers but it must
still be restricted to the solution space defined by the constraint eqns.
(\ref{consistency2}).  Since these spaces are quite complicated we will only be
able to analyse this restriction for the two and three center case and some
simple examples with arbitrary number of centers.

\subsection{The Solution Space as a Phase Space}

The symplectic form of the phase space must be closed and non-degenerate in
order for the latter to be a well-defined symplectic manifold.  We would
like to demonstrate these properties for $\tilde{\Omega}$ in general but the
second property depends on the structure of the solution space (and in fact does
not hold in some degenerate cases as we will see) so we will only be able to
demonstrate it for the two and three center cases.  As previously mentioned it is not
{\em a priori} evident that the solution space is a phase space so in principle
$\tilde{\Omega}$ might have been degenerate.

A direct calculation shows that $d\tilde{\Omega}$ is closed as a two-form on the solution space

\begin{equation}
    d\tilde{\Omega} \sim \sum_{p\neq q} \frac{\langle \Gamma_p, \Gamma_q
	\rangle}{|x_{pq}|} \biggl(\epsilon^{ijk} - 3 \frac{\epsilon_{lij} x_{pq}^k x_{pq}^l}{|x_{pq}|^2}\biggr)
    \delta x_{pq}^i \wedge\delta x_{pq}^k \wedge\delta x_{pq}^j
    = 0
\end{equation}
The last equality is most easily worked out in a coordinate basis.

\subsection{Quantization of the Solution Space}

In sections~\ref{2center-moduli} and Appendix~\ref{3center-moduli} we described the
solution space of two and three-centered solutions. Here, we will
described the quantization of these solution spaces using the
symplectic form (\ref{con3}). It turns out that the structure of
the solution space is determined to a large extent by the symmetries
of the problem. Recall that we already removed the overall
translational degree of freedom from the solution space, which we
can do for example by fixing one center to be at the origin, e.g.
$\mathbf{x}_1=0$, or by fixing the center of mass to be at the
origin. The translational degrees of freedom and their dual
momenta give give rise to a continuum of BPS states, but this
continuum yields a fixed overall contribution to the BPS partition
function. By factoring out this piece we are left with the reduced
BPS partition function, and the quantization of the solution space
we consider here yields contributions to this reduced BPS
partition function.

Besides the overall translational symmetries, the constraint equations
(\ref{consistency2}) are also invariant under global $SO(3)$ rotations of the
${\mathbf x}_p$. These rotations do appear in a non-trivial way in the
symplectic form. Indeed, if we insert $\delta x_p^i = \epsilon^{iab} n^a x_p^b$,
which corresponds to an infinitesimal rotation around the ${\mathbf
n}$-axis\footnote{In other words, we compute $\imath_X \tilde{\Omega}$ with $X$
the vector field $\sum_p \epsilon^{iab} n^a x_p^b \frac{\partial}{\partial
x_p^i}$.}, in $\tilde{\Omega}$ we obtain
\bea
\label{redsymp} \tilde{\Omega} & = & \frac{1}{4} \sum_{p\neq q}
\langle \Gamma_p,\Gamma_q\rangle \frac{\epsilon_{ijk}
\epsilon_{iab} n^a (x_p-x_q)^b\delta (x_p-x_q)^j \, (x_p-x_q)^k
}{|{\mathbf x}_p - {\mathbf x}_q|^3} \nonumber \\
& = & n^i \delta J^i
\eea
where $J^i$ are the components of the angular momentum vector (see section (\ref{angular-momentum}))
\be \label{jan22}
J^i = \frac{1}{4} \sum_{p\neq q} \langle \Gamma_p,\Gamma_q\rangle \frac{
x_p^i-x_q^i }{|{\mathbf x}_p - {\mathbf x}_q|}
\ee
%
The fact that the symplectic form takes the simple form in (\ref{redsymp}) is
not surprising. This is merely a reflection of the fact that angular momentum is
the generator of rotations. Fortunately, this simple form completely determines the
symplectic form in the two and three-center case.

\subsubsection{Two-Center Case}\label{2case}

The two-center case is easy to describe. There is only a regular
boundstate for $\langle \Gamma_1,\Gamma_2 \rangle \neq 0$
 and $\langle
h,\Gamma_p\rangle \neq 0$, and the constraint equations
immediately tell us that $x_{12}$ is fixed and given by
\be
x_{12} = \frac{\langle h,\Gamma_1 \rangle}{\langle \Gamma_1 ,
\Gamma_2 \rangle}.
\ee
In other words, ${\mathbf x}_1-{\mathbf x}_2$ is a vector of fixed
length but its direction is not constrained. The solution space is
simply the two-sphere, and the symplectic form is proportional to
the standard volume form on the two-sphere. In terms of standard
spherical coordinates it is given by
\be
\tilde{\Omega} = \frac{1}{2} \langle \Gamma_1 , \Gamma_2 \rangle \sin
\theta\, d\theta \wedge d\phi = |J|\sin \theta\, d\theta \wedge
d\phi .
\ee
We can now quantize the solution space using the standard rules of geometric
quantization (see e.g. \cite{Ritter:2002zg} \cite{EcheverriaEnriquez:1999jr}).
We introduce a complex variable $z$ by
\be
z^2 = \frac{1+\cos\theta}{1-\cos\theta} e^{2i\phi}
\ee
and find that the K\"ahler potential corresponding to
$\tilde{\Omega}$ is given by
\be
\calk = -2|J| \log(\sin \theta) = - |J| \log\biggl(\frac{z \bar{z}}{(1 + z
\bar{z})^2}\biggr).
\ee
The holomorphic coordinate $z$ represents a section of the line-bundle
$\mathcal{L}$ (over S$^2$, the solution space) whose first Chern class equals
$\tilde{\Omega}/(2\pi)$. The Hilbert space consists of global holomorphic
sections of this line bundle and a basis of these is given by $\psi_m(z)=z^m$.
However, not all of these functions are globally well-behaved. For example,
regularity at $z=0$ requires that $m\geq 0$, and to examine regularity at
$z=\infty$ one could e.g. change coordinates $z\rightarrow 1/z$ and use the
transition functions of $\mathcal{L}$ to find out that $m\leq 2|J|$.
Equivalently, we can directly examine the norm of $\psi_m$ by computing
\be \label{norm1}
|\psi_m|^2 \sim \int d{\rm vol}\, e^{-\calk} |\psi_m(z)|^2
\ee
where $d{\rm vol}$ is the volume form induced by the symplectic
form. In our case we therefore find
\be \label{norm2}
|\psi_m|^2 \sim \int d\cos\theta \, d\phi\,
(1+\cos\theta)^{|J| + m} (1-\cos\theta)^{|J| -m}
\ee
and clearly $\psi_m$ only has a finite norm if $ -|J| \leq m\leq |J|$. The total
number of states equals $2|J|+1$. This is in agreement with the wall-crossing
formula up to a shift by 1. We will see later that the inclusion of fermions
will get rid of this extra constant.

The integrand in (\ref{norm2}) is a useful quantity as it is also the phase
space density associated to the state $\psi_m$.  According to the logic in
\cite{Balasubramanian:2005mg,Alday:2006nd} the right bulk description of one of
the microstates $\psi_m$ should be given by smearing the gravitational solution
against the appropriate phase space density, which here is naturally given by
the integrand in (\ref{norm2}). We will come back to this point later, but
observe, already, that since there are only $2|J|+1$ microstates, we cannot
localize the angular momentum arbitrarily sharply on the $S^2$, but it will be
spread out over an area of approximately $\pi/|J|$ on the unit two-sphere. It is
therefore only in the limit of large angular momentum that we can trust the
description of the two-centered solution (with two centers at fixed positions)
in supergravity.

\subsubsection{Three Center (Non-Scaling) Case} \label{3-centers}

We now return to the symplectic form (\ref{redsymp}). In the three-center case,
we expect four degrees of freedom. As discussed in section
\ref{sec_3cntr_short}, three of those are related to the possibilities to rotate
the system, whilst the fourth one can be taken to be the size of the angular
momentum vector $|\vec{J}|$. The specific form of (\ref{redsymp}) strongly
suggests that these are also the variables in which the symplectic form takes
the nicest form.

We therefore take as our basic variables $J^i$ and $\sigma$, where
$\sigma$ represents an angular coordinate for rotations around the
${\mathbf J}$-axis. Obviously, $\sigma$ does not correspond to a
globally well-defined coordinate, but rather should be viewed as a
local coordinate on an $S^1$-bundle over the space of allowed
angular momenta. Ignoring this fact for now, the rotation
$\delta x_p^i = \epsilon^{iab} n^a x_p^b$ that we used
in (\ref{redsymp}) corresponds to the vector field
\be
X_n = \frac{n^i J^i}{|J|}\frac{\partial}{\partial \sigma} +
\epsilon^{ijk} n^j J^k \frac{\partial}{\partial J^i}.
\ee
The second term is obvious, as ${\mathbf J}$ is rotated in the
same way as the ${\mathbf x}_p$. The first term merely states that
there is also a rotation around the ${\mathbf J}$-axis given by
the component of $n$ in the ${\mathbf J}$-direction. The final
result in (\ref{redsymp}) therefore states that
\be
\tilde{\Omega}(X_n,m^i \frac{\partial}{\partial J^i})= n^i m^i
\qquad \tilde{\Omega}(X_n, \frac{\partial}{\partial \sigma}) = 0.
\ee
It is now easy to determine that
\be \label{jan23}
\tilde{\Omega}(\frac{\partial}{\partial
J^i},\frac{\partial}{\partial J^j} ) = \epsilon_{ijk}
\frac{J^k}{|J|^2},\qquad \tilde{\Omega}(\frac{\partial}{\partial
J^i},\frac{\partial}{\partial \sigma} )= - \frac{J^i}{|J|} .
\ee
Denoting $|\vec{J}|$ as $j$, and parameterize $J^i$ in terms of $j$ and
standard spherical coordinates $\theta,\phi$, the symplectic form
defined by (\ref{jan23}) becomes
\be \label{jan24}
 \tilde{\Omega} = j\sin\theta\, d\theta\wedge d\phi - dj\wedge
d\sigma.
\ee
However, we clearly made a mistake since this two-form is not
closed. The mistake was that $\sigma$ was not a well-defined
global coordinate but rather a coordinate on an $S^1$-bundle. We
can take this into account by including a parallel transport in
$\sigma$ when we change $J^i$. The result at the end of the day is
that the symplectic form is modified to
\be \label{jan25}
\tilde{\Omega} = j\sin\theta\, d\theta\wedge d\phi - dj\wedge
D\sigma
\ee
with $D\sigma = d\sigma - A$, and $dA=\sin\theta\, d\theta\wedge
d\phi$, so that $A$ is a standard monopole one-form on $S^2$. A
convenient choice for $A$ is $A=-\cos\theta\,d\phi$ so that
finally the symplectic form can be written as a manifestly closed
two-form as\footnote{This result is re-derived in a
more straightforward but tedious way in Section \ref{deconstruction}.}
\be \label{fsym}
\tilde{\Omega} = - d (j\cos\theta)\wedge d\phi - dj \wedge
d\sigma.
\ee
This answer looks very simple but in order to quantize the solution space, we have
to understand what the range of the variables is.  Since $\theta,\phi$ are
standard spherical coordinates on $S^2$, $\phi$ is a good coordinate but
degenerates at $\theta=0,\pi$. The magnitude of the angular momentum vector $j$
is bounded as can be seen from (\ref{jan22}). By carefully examining the various
possibilities in the three-center case (see appendix \ref{3center-moduli}), one
finds that generically $j$ takes values in an interval $j\in [j_-,j_+]$, where
$j=j_-$ or $j=j_+$ only if the three-centers lie on a straight line. An
exceptional case is if $j_-=0$ implying that the three-centers can sit
arbitrarily close to each other as seen in appendix \ref{3center-moduli}.  Note
that this latter case corresponds exactly to the scaling solutions described in
section \ref{scaling_sol}.

As we mentioned above, at $j=j_-$ and $j=j_+$ the centers align, and rotations
around the ${\mathbf J}$-axis act trivially. In other words, at $j=j_{\pm}$ the
circle parametrized by $\sigma$ degenerates. Actually, we have to be quite
careful in determining exactly which $U(1)$ degenerates where. Fortunately, what
we have here is a toric K\"ahler manifold, with the two $U(1)$ actions given by
translations in $\phi$ and $\sigma$, and we can use results in theory of toric
K\"ahler manifolds from \cite{Guillemin} (see also \cite{abreu-2000} and
Appendix \ref{app_toric}) to describe the quantization of this space.

We start by defining $x=j$ and $y=j\cos\theta$ to
be two coordinates on the plane. Then the ranges of the variables
$x$ and $y$ are given by
\be x-j_-\geq 0, \quad j_+-x\geq 0, \quad x-y\geq 0,\quad x+y\geq
0. \label{polytope}
\ee
Together these four\footnote{In the case $j_- = 0$ the first equation is
redundant and is not part of the characterization of the polytope.} inequalities
define a Delzant polytope in $\mathbb R^2$ which completely specifies the toric
manifold (see Appendix \ref{app_toric}). At the edges a $U(1)$ degenerates, at
the vertices both $U(1)$'s degenerate. The geometry and quantization of the
solution space can be done purely in terms of the combinatorial data of the
polytope (see figure \ref{fig_polytope}). We have to distinguish two cases.
First, when $j_-=0$, which corresponds to a scaling point inside the solution
space, our toric manifold is topologically $\mathbb{CP}^{2}_{(1,1,2)}$. In the
case $j_->0$, where a scaling point is absent, the solution space becomes the
blow-up of $\mathbb{CP}^{2}_{(1,1,2)}$, which can be identified as the second
Hirzebruch surface $F_2$. These statements can all be verified by e.g.
constructing the normal fan to the relevant polytopes.

\FIGURE{
\includegraphics[page=5,scale=1]{pictures_quantizing.pdf}
\caption{
{\bf (Left)} The polytope for $j_- = 0$ corresponding to $\mathbb{CP}^2_{(1,1,2)}$.
{\bf (Right)} The polytope for $j_- > 0$.  This corresponds to the
second Hirzebruch surface, $F_2$, a blow-up of $\mathbb{CP}^2_{(1,1,2)}$.  }\label{fig_polytope}
}

For the purpose of quantizing the system we will assume that $j_- > 0$; we will
return to the case of $j_- = 0$ in section \ref{deepthroat}.  Thus the results
of the rest of this section only apply for $j_- > 0$.  Furthermore we assume
that all three centers carry different charges; if two centers carry
identical charge one needs to take into account their indistinguishability,
quantum mechanically, and take a quotient of the corresponding solution space. We
won't consider this possibility in this section but come back to it in detail in
section \ref{haldeg}.

The construction of
canonical complex coordinates is done by constructing a function $g$ as
\begin{align} \label{jan51}
g = \frac{1}{2} &\left[ (x-j_-) \log(x-j_-) + (j_+-x)\log(j_+-x)  \right.
\nonumber  \\
&\left.  + (x-y)\log(x-y) + (x+y) \log(x+y) \right]
\end{align}
which is related in an obvious way to the four inequalities in
(\ref{polytope}). Then the complex coordinates can be chosen to be
$\exp(\partial_x g + i \sigma)$ and $\exp(\partial_y g + i \phi)$.
Explicitly, and after removing some irrelevant numerical factors,
the complex coordinates are
\bea
 z^2 & = & j^2 \sin^2 \theta \frac{j-j_-}{j_+-j} e^{2i\sigma}
 \nonumber \\
 w^2 & = & \frac{1+\cos\theta}{1-\cos\theta} e^{2i\phi}
\eea
and the K\"ahler potential ends up being equal to
\be
\calk=j_- \log(j-j_-) - j_+ \log (j_+-j) + 2j.
\ee
Again, a basis for the Hilbert space is given by wave functions $\psi_{m,n}=z^m
w^n$.  Note that, as in the two center case, these wave functions are, by
construction, sections of a line bundle, $\call$, whose curvature is given, once
more, by $\tilde{\Omega}/(2\pi)$.

To find the range of $n,m$ we look at the
norm
\be \label{3norm}
|\psi_{m,n}|^2 \sim \int e^{-2r}
\frac{(j_+-r)^{j_+}}{(r-j_-)^{j_-}} \left( r^2 \sin^2 \theta
\left( \frac{r-j_-}{j_+-r} \right) \right)^n \left(
\frac{1+\cos\theta}{1-\cos\theta} \right)^m \,
r\,dr\,d\cos\theta\,d\phi\,d\sigma.
\ee
This is finite if $j_- \leq n\leq j_+$ and $-n\leq m \leq n$. Not
surprisingly, these equations are identical to the original
inequalities that defined the polytope, and the number of states
is equal to the number of lattice points in the polytope; notice
that this is not quite identical to the area of the polytope. In
our case that number of points is
\be \label{ndof}
{\cal N} = (j_+ - j_- +1)(j_+ + j_- + 1).
\ee
This connection holds quite generally for toric K\"ahler
manifolds.  As in the two center case fermionic contributions will correct both
the state count, (\ref{ndof}), and the phase space measure, (\ref{3norm}); this
will be discussed in the next section.

The integrand in (\ref{3norm}) can once more (as in the two center case) be
viewed as a phase space density against which the supergravity solution has to
be smeared in order to find the gravitational dual of each microstate
\cite{Balasubramanian:2005mg} \cite{Alday:2006nd}.

\subsubsection{Including the Fermionic Degrees of Freedom} \label{fermion-dof}

From the open string point of view \cite{Denef:2002ru} we know
that (\ref{ndof}) is incorrect and that we must include fermionic
degrees of freedom in order to account for all the BPS states
(e.g. in the two center case).  This is because, in the open
string description, the centers are described by  $\mathcal{N}=4$,
d=1 supersymmetric quiver quantum mechanics (QQM) with the
position of each center encoded in the scalars of a vector
multiplet and the latter also includes fermionic components which
must be accounted for in any quantization procedure.

Since we expect to see the same number of BPS states in both the
open and closed description and since the bosonic phase spaces in
both cases match exactly (and the symplectic forms agree in view
the non-renormalization theorem discussed above) we may ask what
the closed string analog of the fermions in the QQM is?

Consider our phase space: the coordinates, $x_p$, subject to the constraint
(\ref{consistency2}), parameterize the space of purely bosonic BPS solutions
but, for each such solution, we may still be able to excite fermions if doing so
is allowed by the equations of motion. If we consider only infinitesimal
fermionic perturbations of the bosonic solutions then the former will always
appear linearly in the equations of motion, acted on by a (twisted) Dirac
operator.  Thus fermions which are zero modes of this operator may be excited
without altering the bosonic parts of the solution (to first order).

Determining the actual structure of these zero modes is quite
non-trivial.  A natural guess is that the bosonic coordinates of
the centers must be augmented by fermionic partners (making the
solution space a superspace) as is argued in
\cite{Aichelburg:1987hy} \cite{Aichelburg:1987ia} where there is
no potential. The fact that the bosonic coordinates are
constrained by a potential complicates the problem in our case so
we will simply posit the simplest and most natural guess and
justify it, {\em a posteriori}, by reproducing the necessary
correction to match the open string picture,  the explicit two
center and halo quantization of \cite{Denef:2002ru}, as well as
the split attractor conjecture \cite{Denef:2007vg}.

Thus we will posit that the full solution space is actually the
total space of the spin bundle over the K\"ahler phase space
described in Section \ref{3-centers}.  The correct phase space
densities are now harmonic spinors on the original phase space.
This is natural from a mathematical point of view
\cite{vaisman:1994} and can be argued physically as follows.

The space of solutions in the open string picture is spanned by letting the
bosonic coordinates take their allowed, constant, values and setting the
fermionic coordinates to zero (except for the center of
mass degrees of freedom which we will ignore in what follows). We
could of course try to restrict the symplectic form to this
space, and then quantize, but this would miss the possible
non-trivial topology of the fermionic vacuum. Therefore we will
proceed in a different way as follows.

We start with the full classical phase space of the quiver quantum
mechanics including all the fermions. Next we are going to impose
the constraints
\be \label{aux11}
\langle \Gamma_p,H(x_p)\rangle=0
\ee
which will restrict $x_p$ to take values in the bosonic solution
space. The constraint (\ref{aux11}), however, is not invariant under all
supersymmetries but only half of them. We can therefore impose
(\ref{aux11}) supersymmetrically as long as, at the same time, we
remove half the fermions. The resulting system still has two
supersymmetries left which one could, in principle, work out explicitly.
If we assume that the solutions space is K\"ahler (which may in
fact be a consequence of the two remaining supersymmetries), the
resulting supersymmetry transformations will necessarily look like
those of standard supersymmetric $\mathcal{N}=2$ quantum mechanics on a
K\"ahler manifold. Notice that so far we have not used the
symplectic structure for the fermions at all.

Though it would be interesting to work this out in more detail, we
finally expect that, after geometric quantization of the
supersymmetric quantum mechanical system, the supersymmetric
wave-functions will be ${\cal L}$-valued spinors which, at the same
time, are zero-modes of the corresponding Dirac operator.

Recall (see e.g. \cite{Lawson}) that on a
K\"ahler manifold ${\cal M}$ there is a canonical Spin${}^c$
structure where the spinors take values in $\Lambda^{0,\ast}({\cal
M})$. To define a spin structure we need to take a square root of
the canonical bundle $K=\Lambda^{N,0}({\cal M})$ and twist
$\Lambda^{0,\ast}({\cal M})$ by that. We also need to remember
that the coordinate part of the wave functions were sections of
a line bundle, ${\cal L}$. Thus altogether the spinors on the
solution space are given by sections of
\be {\cal L} \otimes \Lambda^{0,\ast}({\cal M}) \otimes K^{1/2}.
\ee
The Dirac operator is given by
\be D= \bar{\partial} + \bar{\partial}^{\ast}
\ee
and we have to look for zero modes of this Dirac operator. These
are precisely the harmonic spinors on ${\cal M}$ and therefore the
BPS states correspond to $H^{(0,\ast)}({\cal M}, {\cal L}\otimes
K^{1/2})$. By the Kodaira vanishing theorem, $H^{(0,n)}({\cal M},
{\cal L}\otimes K^{1/2})$ vanishes unless $n=0$. Thus, finally,
the BPS states are given by the global holomorphic sections of
${\cal L}\otimes K^{1/2}$. The only difference with the previous
purely bosonic analysis is that the line-bundle is twisted by
$K^{1/2}$.

To find the number of BPS states we can therefore follow exactly the same
analysis as in the bosonic case.  We just have to make sure that in the inner
product we use the norm appropriate for ${\cal L}\otimes K^{1/2}$. This can be
accomplished by inserting an extra factor of $(\det \partial_i
\partial_{\bar{j}} \calk)^{-1/2}$ in the inner product. For example, for the
two-sphere, this introduces an extra factor of $(1+\cos\theta)^{-1}$ in the
integral, reducing the number of states by one compared to the purely bosonic
analysis. This is in perfect agreement with \cite{Denef:2002ru}.

For the three-center case (and more generally for toric K\"ahler
manifolds) we find, after some manipulations, that
\be \label{fermion-measure}
(\det \partial_i
\partial_{\bar{j}} \calk)^{-1/2} \sim \exp\biggl(\sum_i \frac{\partial
g}{\partial x^i}\biggr) \sqrt{\det{} \biggl[ \frac{\partial^2
g}{\partial x_i
\partial x_j}\biggr]}
\ee
in terms of the function $g$ given for the three-center case in
(\ref{jan51}). To get this relation we used that $\calk$ is the
Legendre transform of $g$ (see Appendix \ref{app_toric}).
Evaluating this explicitly for the three-center case yields an
extra factor
\be
\frac{1}{j_+-j} \sqrt{\frac{1+\cos\theta}{1-\cos\theta}} \, a(j) \\
\ee
with
\be
a(j) = \sqrt{j (j_+ - j_-) + 2 (j_+ - j)(j - j_-)}
\ee
This result indicates that, in the presence of spinors, we should
take $n$ integral and $m$ half-integral with $-n\leq m+\frac{1}{2}
\leq n$ and $j_- \leq n \leq j_+ - 1$. Then, the
total number of normalizable wave-functions becomes
\be  \label{ndof2}
{\cal N} = (j_+ - j_-)(j_+ + j_-)
\ee
which does not have the unwanted shifts anymore!

For completeness let us provide the corrected form of the norm for
a wave function, including fermionic corrections,
\begin{align} \label{3norm_fixed}
|\psi_{m,n}|^2 \sim \int e^{-2j} \frac{(j_+-j)^{j_+
-1}}{(j-j_-)^{j_-}} &\left( j^2 \sin^2 \theta
\left( \frac{j-j_-}{j_+-j} \right) \right)^n \nonumber \\
&\left( \frac{1+\cos\theta}{1-\cos\theta} \right)^{m+\frac{1}{2}}
\, a(j) \, j\,dj\,d\cos\theta\,d\phi\,d\sigma.
\end{align}
Note that this  is only the norm {\em for non-scaling solutions} with $j_- > 0$.  The norm for
wave-functions on solution spaces with a scaling point ($j_- = 0$) is given in
section \ref{deepthroat}.

\subsubsection{Comparison to the Split Attractor Flow Picture}\label{comparison}

In the previous subsections we computed the number of states corresponding to
the position degrees of freedom of a given set of bound black hole centers. The
approach we developed amounts essentially to calculating the appropriate
symplectic volume of the solution space. To count the total number of BPS states
of a given total charge one needs to take into account the fact that the
different black hole centers may themselves carry internal degrees of
freedom and that there may be many multicenter realizations of the same
total charge. In the special case, however, when all the centers correspond to
zero entropy bits without internal degrees of freedom the position degrees of
freedom should account for all states. In this case it is interesting to compare
the number of states obtained in our approach, using geometric quantization,
with the number obtained by considering jumps at marginal stability as in
\cite{Denef:2007vg} (see also section \ref{afconjecture}).

To make this comparison we use the attractor flow conjecture which states that
to each component of solution space there corresponds a unique attractor flow
tree.  Given a component of solution space we can calculate its symplectic
volume and hence the number of states. Given the corresponding attractor flow
tree we can calculate the degeneracy using the wall crossing formula of
\cite{Denef:2007vg}.  To determine which attractor tree corresponds to which
solution space (as needed to compare the two state counts) we will have to
assume that part of the attractor flow conjecture holds.  Although this might
seem to weaken the comparison we should point out that the attractor flow tree
has no inherent meaning outside the context of the attractor flow conjecture
thus the need to assume the latter to relate the former to our solutions is not
surprising.  Moreover, the attractor flow conjecture (defined in
\cite{Denef:2000nb}\cite{Denef:2000ar}) is distinct from (and weaker than) the
wall crossing formula (defined in \cite{Denef:2007vg}) which relies on it.

As mentioned before (around eqn. (\ref{norm2}); see also the section about the
addition of fermions), in the two center case we get a perfect agreement between
the two calculations.  This is not so surprising because both approaches are, in
fact, counting the number of states in an angular momentum multiplet with $j=
\frac{1}{2} \langle \Gamma_1, \,\Gamma_2 \rangle - \frac{1}{2}$. Furthermore,
there is no ambiguity in specifying the split attractor tree.  Things become
more interesting in the three centers case where there are now naively three
attractor trees for a given set of centers.  According to the attractor flow
conjecture only one tree should correspond to any given solution space.  It is
possible to match solution spaces to attractor trees if we are willing to assume
part of the attractor flow conjecture.


Let us consider the three center attractor flow tree depicted in figure
\ref{fig_flow}.  For the given charges, $\Gamma_1$, $\Gamma_2$,
and $\Gamma_3$, there are, in fact, many different possible trees
but, in terms of determining the relevant number of states, the
only thing that matters is the branching order.  In figure
\ref{fig_flow} the first branching is into charges $\Gamma_3$ and
$\Gamma_4=\Gamma_1 + \Gamma_2$ so the degeneracy associated with
this split is $|\langle \Gamma_4, \Gamma_3\rangle|$ and the
degeneracy of the second split is $|\langle \Gamma_1, \Gamma_2
\rangle|$ giving a total number of states
\begin{equation}\label{tree_states}
    \mathcal{N}_{\textrm{tree}} = |\Gamma_{12}|\, |(\Gamma_{13} +
    \Gamma_{23})|
\end{equation}
where we have adopted an abbreviated notation, $\Gamma_{ij} = \langle \Gamma_i,
\Gamma_j \rangle$ and have dropped the factors of $\Omega(\Gamma_a)$ in
(\ref{afc_entropy}) (because we
are only interested in the spacetime contribution to the state count so we
consider centers with no internal states).

To compare this with the number of states arising from geometric quantization of
the solution space, (\ref{ndof2}), we need to determine $j_+$ and $j_-$. As
described in Appendix \ref{3center-moduli}, $j_+$ and $j_-$ correspond to two
different collinear arrangements of the centers and, in a connected solution
space, there can be only two such configurations.  To relate this to a given
attractor flow tree we will {\em assume part of the attractor flow conjecture};
namely, that we can tune the moduli to force the centers into two clusters as
dictated by the tree.  For the configuration in figure \ref{fig_flow}, for
instance, this implies we can move the moduli at infinity close to the first
wall of marginal stability (the horizon dark blue line) which will force
$\Gamma_3$ very far apart from $\Gamma_1$ and $\Gamma_2$. In this regime it is
clear that the only collinear configurations are
$\Gamma_1$-$\Gamma_2$-$\Gamma_3$ and $\Gamma_2$-$\Gamma_1$-$\Gamma_3$; it is not
possible to have $\Gamma_3$ in between the other two charges.  Since $j_+$ and
$j_-$ always correspond to collinear configurations they must, up to signs, each
be one of
\begin{align}
    j_1 &= \frac{1}{2}(\Gamma_{12} + \Gamma_{13} + \Gamma_{23}) \\
    j_2 &= \frac{1}{2}(-\Gamma_{12} + \Gamma_{13} + \Gamma_{23})
\end{align}
$j_+$ will correspond to the larger of $j_1$ and $j_2$ and $j_-$
to the smaller but, from the form of (\ref{ndof2}), we see
that this will only effect $\mathcal{N}$ by an overall sign (the state count
depends only on the absolute value of $\mathcal{N}$).  Thus
\begin{equation}
    \mathcal{N} = \pm (j_1 - j_2)(j_1 + j_2) = \pm \Gamma_{12}\, (\Gamma_{13} +
    \Gamma_{23})
\end{equation}
which nicely matches (\ref{tree_states}).

Of course to obtain this matching we have had to assume the attractor flow
conjecture itself (in part) so it does not serve as an entirely independent
verification.  Another drawback of this argument is that it is not applicable to
the decoupling limit described in \cite{deBoer:2008fk} where the asymptotic
moduli, $t^A$, are fixed to the {\em AdS-point} \cite{deBoer:2008fk}.  However,
it is possible to circumvent this limitation by gluing an asymptotic flat region
to the interior geometry. This can always be done by choosing the moduli in the
new added region to be equal to the asymptotic moduli, $t_{\text{AdS}}$, of the
original solution. The physicality of such gluing relies on two important
observations. The first one is that far away the centers behave like a one big
black hole with charges the sum of all charges carried by the centers. The
second important ingredient is that $t_{\text{AdS}}$ is equal to the attractor
value of the moduli associated to this big black hole. Doing so we are back to
the asymptotic flat geometry where we have the freedom to play the asymptotic
moduli once again. This is the same argument used in
\cite{deBoer:2008fk}\cite{Andriyash:2008it}  to generalize the existence
conjecture from asymptotic flat solutions to the decoupled limit.

Our result here provides another non-trivial consistency check for the
conjecture that {\it{there is a one-to-one map between split-attractor trees and
BPS solutions to $\mathcal{N}=2$ four dimensional supergravity}}
\cite{Denef:2007vg}.  By explicitly evaluating (\ref{ndof}), we find it is the
product of two contributions exactly as predicted by the split attractor flow
conjecture. Using geometric quantization it is clear that the three-center
entropy always factorizes into a product of splits along walls of marginal
stability matching an attractor flow tree (the only reason we need to assume
part of the attractor flow conjecture for the matching is to determine {\em
which} particular tree).  This is strong evidence in favor of the fundamental
underpinning of the attractor flow conjecture: namely, that by tuning moduli it
is always possible to disassemble multicenter configurations pairwise.

Let us make some further remarks on the results derived here.  The scaling solutions
corresponding to $\lambda \rightarrow 0$ have $j_- = 0$ even if the centers
don't align at this point. Therefore the connection to the wall crossing formula breaks down.
The procedure of geometric quantization itself, however, does not seem to suffer
any pathologies for these solutions.  The curvature scales always stay small
allowing us to trust the supergravity solutions.  Thus one can see the resulting
degeneracy as a good prediction. Although the symplectic form seems to
degenerate at $j = j_-$ this is, in fact, nothing but a coordinate artifact as
can be seen from studying the polytope associated with scaling solutions.

For fixed $j_{\pm}$ the Hilbert space is finite dimensional and it is not
possible localize the centers arbitrarily accurately.  Thus the supergravity
solutions can only be well approximated in the large $j_{\pm}$ limit.  In
Sections \ref{sec_infinity} and \ref{deepthroat} we will study the nature of ``classical'' states
defined in this limit.  We will be interested in particular in the
boundary of the solution spaces where classical pathologies such as infinitely
deep throats or barely bound centers (see \cite{deBoer:2008fk}) moving off to
infinity may appear.  We will show that quantum effects resolve these
pathologies since there is less then one unit of phase space volume in the
pathological regions (even for large finite charges) so the pathologies are
purely classical artifacts.

\section{Halo Degeneracies}\label{haldeg}
The first and simplest generalization beyond three centers consists of the case
of so-called halo configurations \cite{Denef:2002ru,Denef:2007vg}. These are
solutions consisting of a single center of charge $\G_1$ together with other
centers that all have a charge of the form $\G_i=a_i\G_\star$ with the $a_i$
positive integers\footnote{The overall sign of the $a_i$ can of course be
absorbed in $\G_\star$ and is thus a convention. They all have to have the same
sign because otherwise no valid attractor flow tree exists.}. These
configurations are characterized by a split tree of the form $\{\Gamma_1,N
\Gamma_\star\}$ with $N=\sum_a a n_a$, where $n_a$ is the number of centers of charge $a \G_\star$. As all $a$ and $n_a$ are positive integers every halo of total charge $N\G_\star$ thus corresponds to a specific partitioning of
$N$ and vice versa. It follows straightforwardly from the constraint equations
(\ref{consistency2}) that all the $\G_i$ centers orbit $\Gamma_1$ at the same
distance $r_{1i}=l$, given by
$$ l = \frac{\langle \Gamma_\star, \Gamma_1 \rangle}{\langle h, \Gamma_\star \rangle} $$
Note that this radius is independent of the different $a_i$, furthermore all the centers $\G_i$ can be placed arbitrarily on this sphere surrounding $\G_1$ as they don't interact among each other.

So for a halo configuration of $m$ orbiting centers the solution space is simply
the product of  $m$ identical S$^2$'s. When quantizing this system we have to
take into account that in case some $a_i=a_j$, the corresponding centers should
be treated as indistinguishable particles and we will have to quotient by the
appropriate permutation group. As all centers in the Halo are independent there
are two ways to proceed. Let us sketch the idea in the simplest case of a halo
consisting of $N$ equally charged centers, i.e all $a_i=1$. As we argued in
general in the previous sections, the Hilbert space $\calh$ for a given set of
charged centers\footnote{Note that this Hilbert space doesn't include the
degrees of freedom intrinsic to the centers, which is measured by their entropy.
In case none of the centers has entropy then $\calh$ is the full Hilbert space.}
is the space of sections of an appropriate line bundle on the solution space. So to
count the degeneracies of the simple halo of $N$ identical centers one could
simply apply this recipe and construct sections of that line bundle on the
solution space which, in this case, is $($S$^2)^N/S_N$. This is exactly what we will do
later in this section, but actually there is a more standard and slightly
simpler way to obtain the same result. Given that the $N$ particles are
independent and identical the Hilbert space for all the particles is nothing but
the (anti-)symmetrized product of the one particle Hilbert space, i.e
$\calh_N=\calh_1^N/S_N$. By the definition of $\calh_1$ this is the space of
sections of $\call_1^N/S_N$, where $\call_1$ is the one-particle line bundle. So
we can either first take the solution space corresponding to all the centers and
construct a wave function on it, or we can take the one particle (2-center)
solution space and construct a multiparticle wave function on it.

In the remainder of this section we will derive formulas for the degeneracies of
halos using both approaches and will see that the results agree.

\subsection{Degeneracy from (Anti-)Symmetrizing Wavefunctions on $\text{S}^2$}\label{antisym-wave}
In this first way of counting we will consider multiparticle wavefunctions on
the solution space of a single particle. This is very reminiscent of the
approach taken in \cite{Denef:2002ru}. To gain some intuition about a Halo
system consider first the simplest case of $N$ centers, each carrying a charge
$\Gamma_\star$. One can think of this system as $N$ (non-interacting)
``electrons'' surrounding a magnetic monopole of ``effective charge'' $\langle
\G_1,\G_\star\rangle$. As all the electrons are identical and noninteracting the
wavefunction of the full system will be a symmetrized product of a single
electron wave function. The single particle wave function is a section of a
line-bundle $\call$ on S$^2$, the position space of the electron.  The line
bundle $\call$ has Euler number  $|\langle \Gamma_1 \,, \Gamma_\star\rangle|$
(neglecting the fermionic nature of the wavefunction for a moment; see section
\ref{2case}). To generalize this to $N$ such particles we have to tensor the
bundle $N$ times and (anti-)symmetrize due to indistinguishability. The Euler
number of the resulting line bundle then gives the number of sections, i.e. the
number of states. It is more convenient to summarize these numbers for different
$N$ in terms of a generating function. Such a generating function for Euler
numbers of the symmetric product of a line bundle was given in e.g.
\cite{Dijkgraaf:1998zd}. The only difference with the discussion there is that,
in our case, the line bundles are fermionic in nature as described in section
\ref{fermion-dof}. Taking this point into account properly, following e.g.
\cite{Vafa:1994tf}, gives the following generating function
\begin{equation} \label{non-prim-1}
    \sum_N d_N q^N = (1 + q)^{|\langle \Gamma_1\,,\Gamma_\star \rangle|}
\end{equation}
where $d_N$ stands for the Euler number of the $N^{\mathrm{th}}$ symmetric
product $\mathcal{L}^N/S_N$ . Newton's binomial expansion yields
\be
d_N=\begin{pmatrix}|\langle
\Gamma_1\,,\Gamma_\star \rangle|\\N\end{pmatrix}\,.
\ee
The fact that the different centers in the
halo behave like fermions results in an upper bound of $|\langle
\Gamma_1\,,\Gamma_\star \rangle|$ for the
number of such centers; this is nothing but Pauli's exclusion principle at
work. This quantum effect is more subtle to understand from a supergravity
perspective \cite{Denef:2002ru}.

The generating function (\ref{non-prim-2}) can be generalized to include halos
given by an arbitrary partition $\{n_a\}$ of $N$, i.e. $N = \sum_a a\, n_a$, where $n_a$ is the number of centers carrying the same charge $\G_a = a \G_\star$.  It is not hard to see that
the generating function including such arbitrary partitions is given by
\begin{equation} \label{non-prim-2}
    \sum_N D_N q^N = \prod_{k>0} (1 + q^k)^{k|\langle \Gamma_1\,,\Gamma_\star
	\rangle|}\,,
\end{equation}
where $D_N$ is the degeneracy of all halos with total charge $N\G_\star$. The
degeneracies can by found by expanding the product:
\be
D_N=\sum_{P}\prod_a\begin{pmatrix}a\,|\langle
\Gamma_1\,,\Gamma_\star \rangle|\\n_a\end{pmatrix}\,,
\ee
where the sum is over all possible partitions $P=\{n_a\}$ of $N$.  This agrees with the
fact that the total degeneracy of a given partition is the product of the
degeneracies of each group of identical terms in the partition. The
degeneracies of such a group of identical halo charges was exactly what we
calculated in (\ref{non-prim-1}) to be a binomial coefficient.

The formula (\ref{non-prim-2}) is the similar to (5.6) in \cite{Denef:2007vg}
but there are also some obvious differences. The first one comes from the fact
that \cite{Denef:2007vg} is calculating an index while we are counting states without relative
signs. The second difference is that we are neglecting the degeneracies
associated to internal degrees of freedom of each individual center.

\subsection{Degeneracy from Wavefunctions on $(\text{S}^2)^n/S_n$}

We can now reproduce the above results in a more straightforward way using
techniques we developed in earlier sections. Here we will directly quantize the
solution space corresponding to a given halo. As discussed before, the solution
space for halos will be the product of various two center spaces, i.e. spheres,
quotiented out by the appropriate permutation group, due to the identification
of identical centers.  An elegant way to describe this space, that can also be
generalized to more involved configurations of the next section, is by using the
technology of toric orbifolds that was developed in \cite{lerman-1995,
abreu-2001} (see Appendix \ref{app_toric}).

Let us start again with the simplest halo configuration, that of $N$ centers of
charge $\Gamma_\star$ orbiting a charge $\G_1$. As discussed in the beginning of
this section the classical solution space for this configuration is the product
of $N$ S$^2$'s. The polytope describing each of those is simply an interval, see
e.g. \cite{abreu-2000},
$$ -j \leq x_i \leq j $$
where $j = \frac{1}{2} |\langle \Gamma_1, \Gamma_\star \rangle|$ is the size of
the two-center angular momentum and $x_i = j \cos \theta_i$. The polytope
associated to the product (S$^2$)$^N$ is then an $N$-dimensional hypercube.

Because we are dealing with $N$ identical centers we need to quotient the total
space by the permutation group $S_N$. The resulting space is a toric orbifold
\cite{lerman-1995}. Such toric orbifolds are uniquely described by labelled
polytopes. These are polytopes where each facet, i.e. ($d-1$)-dimensional face,
comes with a positive integer $m$, where the points on the facet are
$\mathbb{Z}_{m}$ singularities. In our case of S$^N/S_N$ the polytope is given
by the convex hull of the following facets with their corresponding labels:
  \begin{itemize}
     \item two facets with label 1 given by
        \begin{itemize}
          \item[i)] the facet  $-j = x_1 \leq x_2 \leq .... \leq x_N \leq j$.
          \item[ii)] the facet $-j \leq x_1 \leq x_2 \leq .... \leq x_N = j$.
        \end{itemize}
     \item ($N-1$) facets with label 2 given by
        \begin{itemize}
          \item[iii)] the facets  $-j \leq x_1\leq .... \leq x_i = x_{i+1}\leq .... \leq x_N \leq j$ (one for each $i = 1,...,N-1$)
        \end{itemize}
  \end{itemize}
This polytope is the quotient of the original hypercube by the orbifold group
$S_N$ and the labels can be easily understood by considering the fixed points of
the group action.

Given a labeled polytope there are canonical complex coordinates on the
corresponding toric orbifold \cite{abreu-2001}. As in the non-orbifold case they
are constructed from a single function $g$ which, in our specific case, is given by
\begin{equation}
      g(x) = \frac{1}{2} (x_1 + j) \ln (x_1 + j) + \frac{1}{2} (j-x_n) \ln (j - x_n) + \sum_{i=1}^{N-1} (x_{i+1} - x_i) \ln 2 (x_{i+1} - x_i)\,.
\end{equation}

The complex coordinates are then given in terms of this function as $\exp
(\partial_i g(x) + i \phi_i)$ leading, in the above case, to
  \begin{eqnarray}\label{orb-complex}
    z_1^2 &\sim& \frac{1}{j} \; \frac{1 + \cos \theta_1}{(\cos \theta_2 - \cos \theta_1)^2} \; e^{2 i \phi_1} \\
    z_i^2 &\sim& \left( \frac{\cos \theta_i - \cos \theta_{i-1}}{\cos \theta_{i+1} - \cos \theta_i} \right)^2 \; e^{2 i \phi_i}\,, \qquad i = 2, ..., N-1 \\
    z_n^2 &\sim& \frac{1}{j} \; \frac{(\cos \theta_n - \cos \theta_{N-1})^2}{1-\cos \theta_N} \; e^{2 i \phi_N}
  \end{eqnarray}
These are the complex coordinates for which the symplectic form (\ref{con3}) is a K\"ahler form.

A basis for the Hilbert space of holomorphic sections of the line bundle with
$c_1(\call)=\tilde\Omega/2\pi$, is then formed by the wave functions $\psi_{m} =
\prod_{i=1}^{N} z_i^{m_i}$.  The range of $m=(m_i)$ is constrained by requiring
that the $\psi_m$ are normalizable. Due to the fermionic nature of the
wavefunctions the appropriate integration measure to calculate the norm is
$e^{-\calk} (\det \partial_i \partial_{\bar{j}} \calk)^{-1/2} $ (as in section
\ref{fermion-dof}). The K\"ahler potential can be calculated from $g$
(see \cite{abreu-2001})  and is, in our case,
  \begin{equation}
     \calk \sim - j \ln [(1+ \cos \theta_1) (1- \cos \theta_N)] + j (\cos \theta_N - \cos \theta_1)
  \end{equation}
With some manipulation we determine that
   $$(\det \partial_i
\partial_{\bar{j}} \calk)^{-1/2} \sim \exp\biggl(\sum_i \frac{\partial
g}{\partial x^i}\biggr) \det{}^{1/2}\biggl[ \frac{\partial^2 g}{\partial x_i
\partial x_j} \biggr]$$
and furthermore use that
\begin{eqnarray}
    \exp\biggl[{\sum_{i=1}^N \frac{\partial g}{\partial x_i} } \biggr] &=& \sqrt{\frac{1+ \cos \theta_1}{1 - \cos \theta_N}} \\
    \det \biggl[ \frac{\partial^2 g}{\partial x_i \; \partial x_j} \biggr] &=& [1 + \frac{1}{2} (\cos \theta_N - \cos \theta_1)] \frac{1}{1-\cos \theta_1} \; \frac{1}{1-\cos \theta_N} \; \prod_{i=1}^{n-1} \frac{1}{\cos \theta_{i+1} - \cos \theta_i}\nonumber
 \end{eqnarray}
Putting everything together we find the following norm for the wavefunctions

  \begin{align}
     |\psi_m|^2 \sim \int e^{-j (y_N - y_1)}&[1 + \frac{1}{2} (\cos \theta_N - \cos \theta_1)]^{1/2} (1 + y_1)^{j+m_1} (1-y_N)^{j - m_N -1}\nonumber\\
      &(y_2 - y_1)^{2 (m_2 - m_1) -1/2} ..... (y_n - y_{N-1})^{2(m_n - m_{N-1})-1/2}\,.
  \end{align}
Normalizability then implies the following domain for $m$:

  \begin{equation} \label{domain}
    -j \leq m_1 < m_2<....<m_N <j
  \end{equation}

This can be interpreted as picking a representative of a totally antisymmetric
$N$-tensor, which is the expected behavior of $N$ identical fermions. The
degeneracy is thus nothing other than the binomial coefficient, which means that
the generating function is given by
  \begin{equation} \label{non-prim-3}
     \sum_N d_N q^N = (1 + q)^{|\langle \Gamma_1, \Gamma_\star \rangle|}\,,
  \end{equation}
where we have used the fact that  the angular momentum $j$ is proportional to
the intersection product of the charges. So we find the same result as in the
previous approach (see (\ref{non-prim-1})).

We can now also extend this construction to more general halos. One can reduce
the more general case given by an arbitrary partition of $N$, to
products of our previous simpler example by grouping all equal terms in the
partition. Writing partitions of $N$ as $N = \sum_a a\,n_a$, the solution space is then of the form
$\prod_a$(S$^2)^{n_a}/S_{n_a}$.  The derivation of the complex coordinates and
the normalizability constraint then proceeds analogously to the previous
example. Using the indices $m^a=(m^a_i)$, $i=1\ldots u_a$ to label the
wavefunctions $\psi_{m^a}=\prod_i (z_i)^{m_i^a}$ one finds that they are
normalizable iff
\begin{equation} \label{non-prim-non-scaling}
    \qquad -a \,j \leq m^a_1 < m^a_2<....<m^a_N < a\,j
\end{equation}
The degeneracies of all the halos with halo charge $N\G_\star$ can then be
captured using the following generating function:
\begin{equation} \label{non-prim-4}
  \sum_N D_N q^N = \prod_{k>0} (1 + q^k)^{k |\langle \Gamma_1, \Gamma_\star \rangle|}
\end{equation}
which coincides with eqn. (\ref{non-prim-2}) again showing the equivalence of
the two derivations.

Using this generating function we can derive the large $N$ behaviour of the
degeneracies $D_N$. Let us define
\be
F(\beta)=\log\sum_ND_Ne^{-\beta N}=|\langle \Gamma_1, \Gamma_\star \rangle|\,\sum_{n=1}^{\infty}\frac{(-1)^{n-1}}{n}\frac{e^{-\b n}}{(1-e^{-\beta n})^2}\,,
\ee
where in the last equality we used expression (\ref{non-prim-4}) and some rather
straightforward manipulations of the resulting series. The physical
interpretation of $F$ and $\b$ is that of the free energy and inverse
temperature. As is familiar from thermodynamic systems we can find the large
energy, or in this case large $N$, degeneracies by approximating the free energy
at large temperatures and then Legendre transform to find the corresponding
entropy. In the large temperature limit $\beta \ll 1$ the free energy is
approximately
\be
F(\beta)\approx \frac{3}{4}\,|\langle \Gamma_1, \Gamma_\star \rangle|\,\zeta(3)\,\frac{1}{\b^2}\,,
\ee
with $\zeta(3)\approx 1.202$ the Riemann zeta function evaluated in 3. From the
free energy we can calculate the average ``energy`` $N$ in terms of $\beta$ by the
standard formula
\be
N=-\frac{\partial F}{\partial\b}=\frac{3}{2}\,|\langle \Gamma_1, \Gamma_\star \rangle|\,\zeta(3)\,\frac{1}{\b^3}\,.\label{nb}
\ee
Note that indeed the large temperature regime corresponds to that of large $N$.
Using the relation between free energy and entropy we can then find an
expression for the large $N$ degeneracies:
\be
S(N)=\log D_N=\beta N - F=\left(\frac{3}{16}\,|\langle \Gamma_1, \Gamma_\star \rangle|\,\zeta(3)\, N^2\right)^{\frac{1}{3}}\,,
\ee
where we used the relation (\ref{nb}) between $N$ and $\b$. The conclusion is
thus that for a halo of total charge $N\G_\star$ there are of the order
$N^{2/3}$ states when $N\gg 1$. Note that this is slightly larger than the
growth of $N^{1/2}$ that comes from the partitioning of $N$, due to the presence
of additional angular momentum  degrees of freedom that also grow with $N$. As
the angular momentum is due to the electromagnetic coupling of the halo
particles this ``entropy enhancement'' is in some ways similar to that observed in
e.g. \cite{Gaiotto:2004ij} and \cite{Bena:2008nh}.

\section{Dipole Halos}\label{deconstruction}

So far we have restricted our analysis to generic solutions with two or three
centers and a first generalization to an arbitrary number of centers in the
specific case of a single charge surrounded by a halo of identical centers. The
next further natural generalization is to bind the halo to an additional center.
More precisely we can embed the halo split tree $\{\G_1,N\G_\star\}$ in a larger
tree of the form\footnote{In this notation figure \ref{fig_flow} would be
$\{\G_3, \{\G_1, \G_2\}\}$.} $\{\G_2,\{\G_1,N\G_\star\}\}$. If the moduli are
very near the wall of marginal stability the additional center $\G_2$ will be
very far apart from the halo and the degeneracies will simply factorize
\cite{Denef:2007vg}, i.e.
$D_{N}=(\langle\G_2,\G_1\rangle+N\langle\G_2,\G_\star\rangle)D_N^{\text{halo}}$.
Note that to derive this degeneracy we use the attractor flow conjecture which
only works for non-scaling solutions so for scaling solutions we will have to
resort to other methods.  For more arbitrary values of the moduli the center
$\G_2$ will be closer to the halo and deform it. In certain cases it can even
deform so much that the topology of the split changes to
$\{\G_1,\{\G_2,N\G_\star\}\}$. Such a change can happen when crossing a wall of
threshold stability \cite{deBoer:2008fk}\cite{Denef:2007vg} and in that case the
number of states does not change (even though the topology of the tree changes).

In this section we will reproduce the degeneracy as inferred from the simple
flow tree arguments above by our quantization methods. We will do this in the
specific example of a D6-$\antiDp{6}$-D0 system, which seems closely related to
the microstates of the D4-D0 black hole \cite{Denef:2007yt}.  There exist such
configurations with a purely fluxed D6-$\antiDp{6}$ pair and an arbitrary number of
anti\footnote{In our conventions it is anti-D0's that bind to D4 branes.  We
will however often just refer to them as D0's.}-$D0$'s. Depending on the sign
of the B-field these D0's bind to the D6 or anti-D6 respectively. When we take
the B-field to be zero at infinity the system is at threshold and the D0's are
free to move in the equidistant plane between the D6 and anti-D6. This system
and its behaviour under variations of the asymptotic moduli was studied in
detail in appendix B of \cite{deBoer:2008fk}.

As the number of states will be independent of the asymptotic moduli (as long as
we don't cross a wall of marginal stability) we are free to choose them such
that the solution space has its simplest form. The symplectic form on solution
space is most easily calculated at threshold which, in our example, corresponds
to $B|_\infty=0$ \cite{deBoer:2008fk}. For a set of D0's with charges $\GmA =
\{0, 0, 0, -q_a \}$ with all the $q_a$ positive and $\sum_a q_a = N$, bound to a
D6, $\G_6=(1,\frac{p}{2},\frac{p^2}{8},\frac{p^3}{48})$, and $\antiDp{6}$, $\G_{\bar 6}=(-1,\frac{p}{2},-\frac{p^2}{8},\frac{p^3}{48})$,  the constraint equations take the following form at
threshold:
\begin{align}
	-\frac{q_a}{x_{6a}} + \frac{q_a}{x_{\overline{6}a}} = 0 \label{d6d6d0_c1}\\
	-\frac{I}{r_{6\bar6}} + \sum_a \frac{q_a}{x_{6a}} = -\beta \label{d6d6d0_c2}
\end{align}
Here $I = - \langle \GmSix, \GmSixb \rangle=\frac{p^3}{6}$ is given in terms of
the total D4-charge $p$ of the system and $\beta =
\langle \GmSix, h \rangle$ with $I, \beta > 0$.  From the first line we
indeed see the D0's lie in the plane equidistant from the $D6$ and
$\overline{D6}$, as we are at threshold, and so we can simply write $x_a :=
x_{6a} = x_{\overline{6}a}$.

An explicit expression for the symplectic form (\ref{con3}) can be obtained
using the following coordinate system.  We define an orthonormal frame
$(\hat{u}, \hat{v}, \hat{w})$ fixed to the D6-$\antiDp{6}$ pair, such that the
D6-$\antiDp{6}$ lie along the $w$ axis and with the D0's lying in the $u$-$v$
plane.  Rotations of the system can then be interpreted as rotations of the
$(\hat{u}, \hat{v}, \hat{w})$ frame with respect to a fixed $(\hat x,\hat y,\hat
z)$ frame. We will parameterize these overall rotations in the standard fashion
by two angles, $(\theta,\phi)$.  We can furthermore specify the location of the
$a$'th D0 with respect to D6-$\antiDp{6}$ pair by two additional angles,
$(\theta_a,\phi_a)$. The first angle, $\theta_a$, is the one between
$\vec{x}_{6\overline{6}}$ and $\vec{x}_{6a}$, while $\phi_a$ is a polar angle in
the $u$-$v$ plane (see figure \ref{fig_coords}).  Our $2N+2$ independent coordinates
on solution space are thus $\{\theta, \phi, \theta_1, \phi_1, \dots, \theta_N,
\phi_N\}$.

\FIGURE{
\includegraphics[page=4,scale=1.2]{pictures_quantizing.pdf}
\caption{The coordinate system used to derive the D6-$\antiDp{6}$-N D0
symplectic form.  The coordinates ($\theta$, $\phi$) define the orientation of
the ($\hat{u}, \hat{v}, \hat{w}$) axis with respect to the fixed, reference,
($\hat{x}, \hat{y}, \hat{z}$) axis.  The D6-$\antiDp{6}$ lie along the $\hat{w}$
axis (with the origin between them) and the D0's lie on the $\hat{u}$-$\hat{v}$
plane at an angle $\phi_a$ from the $\hat{u}$-axis.  The radial position of
each D0 in the $\hat{u}$-$\hat{v}$ plane is encoded in the angle $\theta_a$
(between $\vec{x}_{6\overline{6}}$ and $\vec{x}_{6a}$).  }\label{fig_coords}
}

The standard Euclidean coordinates of the centers are then given in terms of the angular coordinates by
\begin{align}
	\vec{x}_6 &= \frac{j}{\beta} \hat{w} \qquad &\hat{u} &= \cos \phi\, \hat{x} -
	\sin \phi \,\hat{y} \\
	\vec{x}_{\overline{6}} &= - \frac{j}{\beta} \hat{w} \qquad &\hat{v} &= \sin
	\phi \cos \theta \,\hat{x} + \cos \phi \cos \theta \,\hat{y} - \sin\theta \,
	\hat{z}\\
	\vec{x}_a &= \frac{j \, \tan \theta_a}{\beta} (\cos \sigma_a \hat{u} + \sin
	\sigma_a \hat{v}) \qquad &\hat{w} &= \sin\phi \sin\theta \, \hat{x} + \cos\phi
	\sin\theta \, \hat{y} + \cos \theta \, \hat{z}
\end{align}
The angular momentum, $j(\theta_a, \phi_a)$, is a function of the other
coordinates rather than an independent coordinate (when $N=1$ it can be traded for
$\theta_1$ as demonstrated in the general three center
case), and is given by
\be \label{deconst-ang-moment}
j = \frac{I}{2} - \sum_a q_a \cos\theta_a\,.
\ee
Using this explicit coordinatization it is straight-forward, though tedious, to evaluate the symplectic
form (\ref{con3}). The result is relatively simple:
\be \label{deconst-symp-form}
\Omega = -\frac{1}{4} d \biggl[ 2j \cos \theta \, d\phi + 2 \sum_a  q_a
\cos\theta_a \, d\sigma_a\biggr]
\ee
with $d$ denoting the exterior derivative.  Note that, as is manifest from our
angular coordinatization, we are still in the toric setting with each additional
center introducing an additional $U(1)$ coordinate.  Note also that for a single D0
this reduces to (\ref{fsym}) when $\theta_1$ is traded for $j$.

In case $N > I/2$ it is possible to combine a sufficient number of centers and
form a scaling throat.  When there is more than one D0 center this spoils the
simplicity of the polytope and the toric manifold becomes singular. This is one
of the reasons for which we will restrict our analysis in this paper to the
non-scaling case $I/2>N$, or, as in section \ref{deepthroat}, restrict to three
center scaling solutions. We hope, however, to return to more general scaling
solutions in the near future.

\subsection{Degeneracy using Toric Techniques}
In the following we are going to use the same toric techniques as in the
previous sections to calculate the degeneracy associated to the non-scaling
D6-$\antiDp{6}$-D0 states, i.e. those for which $N < I/2$. From the associated
symplectic form, eqn. (\ref{deconst-symp-form}), one easily reads the toric
coordinates to be
\begin{equation} \label{deconst-polytope-coord}
    y = j \; \cos \theta \,, \qquad \qquad x_a = q_a \, \cos\theta_a \geq 0
\end{equation}
Because some of the centers can be identical we need to orbifold our polytope
by the appropriate symmetric group. Before treating the configuration given by a
generic partitioning $\{q_a\}$ of $N$ i.e. $N = \sum_a n_a q_a$, let us focus on the simple case of $n$
centers carrying the same D0 charge, $-q$, so that $N=q\,n$. In this case the
polytope without identifications is given by
\begin{eqnarray} \label{deconst-poly}
    - \left(\frac{I}{2} - \sum_a x_a\right) \leq y \leq \left( \frac{I}{2} - \sum_a x_a \right) \nonumber\\
       0 \leq x_1 \leq x_2 \leq .... \leq x_n \leq q\, , \qquad \sum_a q \leq \frac{I}{2}
\end{eqnarray}
where we have used the explicit form of the angular momentum $j$ from eqn.
(\ref{deconst-ang-moment}) in the first inequality. We can then orbifold this
polytope and introduce labels to the facets of the quotiented polytope. In the
current example this amounts to
 \begin{itemize}
     \item four facets with label 1 given by
        \begin{itemize}
          \item[i)] the facet  $x_1=0$,
          \item[ii)] the facet $x_a=q$,
          \item[iii)] the facet $y = \frac{I}{2} - \sum_a x_a$,
           \item[iv)] the facet $y = -\frac{I}{2} +\sum_a x_a$,
        \end{itemize}
     \item ($n-1$) facets with label 2 given by
        \begin{itemize}
          \item[v)] the facets  $x_{a+1} - x_a = 0$, for $a=1, ..., n-1$.
        \end{itemize}
  \end{itemize}

Given this labelled polytope we can then again construct the corresponding
complex coordinates through the function
\begin{eqnarray}
    g = \frac{1}{2} \biggl(\frac{I}{2} + y - \sum_a x_a\biggr) \ln \biggl(\frac{I}{2}
	+ y - \sum_a x_a\biggr) + \frac{1}{2} \biggl(\frac{I}{2} - y - \sum_a
	x_a\biggr) \ln
	\biggl(\frac{I}{2} - y - \sum_a x_a\biggr) \nonumber \\
         + \frac{1}{2} x_1 \ln x_1 + \frac{1}{2} (q - x_n) \ln (q - x_n) + \sum_{a=1}^{n-1} (x_{a+1} - x_a) \ln 2 (x_{a+1} - x_a)\nonumber
\end{eqnarray}
Explicitly they can be calculated by $z_i = \exp{(\partial_i g + i \phi_i)}$, leading to
\begin{align}
    z_0^2 &\sim \frac{I/2 + y - \sum_a x_a}{I/2 - y - \sum_a x_a} \; e^{2 i \sigma_0}\nonumber \\
    z_1^2 &\sim \frac{1}{(I/2 + y -\sum_a x_a) (I/2 - y -\sum_a x_a)} \; \frac{x_1}{(x_2 - x_1)^2} \; e^{2 i \sigma_1}\nonumber \\
    z_n^2 &\sim \frac{1}{(I/2 + y -\sum_a x_a) (I/2 - y -\sum_a x_a)} \; \frac{(x_n - x_{n-1})^2}{(q - x_n)} \; e^{2 i \sigma_n} \\
    z_i^2 &\sim \frac{1}{(I/2 + y -\sum_a x_a) (I/2 - y -\sum_a x_a)} \; \left(\frac{x_i - x_{i-1}}{x_{i+1} - x_ i}\right)^2 \; e^{2 i \sigma_i} \, , \qquad i = 2, ... , n-1\nonumber
\end{align}
The next step is to construct the K\"ahler potential, which turns out to be equal to
\begin{eqnarray}
    \calk =& - \frac{I}{2} \ln \biggl(\frac{I}{2} + y - \sum_a x_a\biggr)
	\biggl(\frac{I}{2} - y - \sum_a x_a\biggr) \nonumber \\
	&- 2 \sum_a  x_a + x_n - x_1 - q \ln (q - x_n)
\end{eqnarray}
A basis for the Hilbert space is then given by normalizable functions $\psi_{m}
= \prod_{i=0}^n z_i^{m_i}$ where the norm is given by
$$|\psi_{m}|^2 \sim \int d\,vol\, e^{-\calk} \sqrt{\det\partial_i \partial_{\bar{j}} \calk}\,|\psi_{m}(z)|^2 \,.\label{normcond}$$
As in our previous examples the determinant can be computed by the identity
$$ \sqrt{\det\partial_i \partial_{\bar{j}} \calk} \sim \exp[\partial_i g] \; \sqrt{\det \partial_i \partial_j g}\,. $$
The first term is simply $\prod_i |z_i|$.  The second term is more involved but
can be shown to be
$$ \det \partial_i \partial_j g \sim  \biggl(\frac{I}{2} + y - \sum_a
x_a\biggr)^{-1} \biggl(\frac{I}{2} - y - \sum_a x_a\biggr)^{-1} (x_1)^{-1} (q - x_n)^{-1} \prod_{a=1}^{n-1} (x_{a+1} - x_a)^{-1}  $$
Using these expressions to analyze the normalizability constraint
(\ref{normcond}) then leads to the following constraint on the possible
exponents $m=(m_i)$:
\begin{eqnarray} \label{cond-1}
    &0 \leq m_1 < m_2 < .... < m_n < q \, ,& \nonumber \\
           &- \left( \frac{I -1}{2} - \sum_a \left(m_a + \frac{1}{2}\right) \right) \leq m_0 + \frac{1}{2} \leq \left( \frac{I -1}{2} - \sum_a \left(m_a + \frac{1}{2}\right) \right)&
\end{eqnarray}
where $n$ is the number of D0-centers carrying charge $q$. The total
number of normalizable states is thus
\begin{equation}
     d_{n,q} = \sum_{m = n (n-1)/2}^{(I - n - 1)/2} b_m^n (q) [(I-n) - 2 m]
\end{equation}
where the coefficient $b_m^n (q)$ is the number of ways to write $m$ as a sum of $n$ strictly ordered positive integers all smaller than $q$.

Let us now generalize the simple example of $n$ equally charged D0-centers to an
arbitrary partition of $N$. Label the different groups of equally charged
centers by an index $a$ and the charge of individual centers in this group by
$q_a$ (i.e. $q_a\neq q_b$ iff $a\neq b$). With $n_a$ we then denote the number of
centers with charge $q_a$ so that the total D0-charge $N$ carried by the D0 centers
is given by
$$N = \sum_a n_a q_a$$
Labeling the centers in a given group $a$ by an additional index $i=1,\cdots,n_a$ we can
simply generalize the conditions (\ref{cond-1}) by applying them to each group
of equally charged centers separately. The conditions on the powers
$m^a=(m^a_i)$ then become
\begin{eqnarray}
& 0 \leq m^a_1 < m^a_2 < .... < m^a_{n_a} < q_a\,, \nonumber& \\
        &   - \left[ \frac{I - 1}{2} - \sum_{a,i} \left( m^a_i + \frac{1}{2}\right) \right] \leq m_0 + \frac{1}{2} \leq \left[ \frac{I - 1}{2} - \sum_{a,i} \left( m^a_i +\frac{1}{2} \right) \right]\,. &\label{cond-2}
\end{eqnarray}
A first step towards counting all possible states with total charge $N$ in
D0-centers is to count the degeneracy for a fixed partitioning of $N$. The
number of solutions to the constraints (\ref{cond-2}) is given by
\begin{eqnarray} \label{deconstruction-deg-1}
   d_{\{n_a, q_a\}} &=& \sum_{{\text{all allowed }} m_i^a} \left[ I -  2 \sum_{a,i} \left(m_i^a + \frac{1}{2} \right) \right] \nonumber\\
                &=& I \, \prod_a \left( \begin{array}{c}
                            q_a \\ n_a \end{array} \right)  - 2 \sum_{{\text{all allowed }} m_i^a} \sum_{i,a} \left(m_i^a + \frac{1}{2} \right)
\end{eqnarray}
We can calculate the sum of the last terms by introducing the quantities
$$ l_i^a = q_a - 1 - m_{n_a - i}^a $$
and noting that then
\begin{eqnarray}
   &0 \leq l^a_1 < l^a_2 < .... < l^a_{n_a} < q_a\, ; &\\
            &\sum_{{\text{all allowed }} m_i^a} \sum_{a,i} \left(m^a_i + \frac{1}{2}\right) = N \, \prod_a \left( \begin{array}{c}
                            q_a \\ n_a \end{array} \right) - \sum_{{\text{all allowed }} l_i^a} \sum_{i,a} \left( l^a_i + \frac{1}{2}\right)&\label{eqn}
\end{eqnarray}
where we used that $\sum_a n_a q_a = N$.  As $l_i^a$ and $m_i^a$ satisfy the
same conditions, eqn. (\ref{eqn}) simply implies that
$$ \sum_{{\text{all possible }} m_i^a} \sum_i \left(m^a_i + \frac{1}{2}\right) = \sum_{{\text{all possible }} l_i^a} \sum_i \left(l^a_i + \frac{1}{2}\right) = \frac{N}{2} \left( \begin{array}{c}
                            q_a \\ n_a \end{array} \right)\,.  $$
Using this last equality we see that degeneracy (\ref{deconstruction-deg-1}) is given by
\be\label{deconstruction-deg-2}
d_{\{n_a,q_a\}} = (I - N)\,\prod_{a} \left( \begin{array}{c}
q_a \\ n_a \end{array} \right)  = (I - \sum_a n_a q_a)\,\prod_{a} \left( \begin{array}{c}
q_a \\ n_a \end{array} \right)
\ee
So we indeed find back the result we derived using attractor tree arguments: the
degeneracy is that of the corresponding halo, multiplied by
$\langle\G_1,\G_2\rangle+N\langle\G_1,\G_\star\rangle=I-N$.

Counting all the different degeneracies for all possible partitions of a
given total halo charge $N\G_\star$, gives rise to the following generating
function:
\begin{eqnarray} \label{non-scaling-gen-fct}
    \mathcal{Z}(q) = \sum_{N}  D_N q^N &=& \sum_{\{n_a\}\,\{q_a\}} (I - \sum_a n_a q_a)\, \prod_{a} \left( \begin{array}{c}
                            q_a \\ n_a \end{array} \right) q^{n_a \, q_a} \nonumber\\
                    &=& (I - q \partial_q) \prod_{q_a} \left[\sum_{n_a} \left( \begin{array}{c}
                                                                            q_a \\ n_a \end{array} \right) q^{n_a q_a} \right] \nonumber\\
                    &=& (I- q\partial_q) \prod_k (1+q^k)^k\,.
\end{eqnarray}

Using this generating function we can again estimate the large $N$ growth of
$D_N$, but, as the only difference from the (non-dipole) halo case is a
pre-factor of order $(I-N)$, we see that the growth of $\log D_N\sim N^{2/3}$
also holds in this case (modulo logarithmic corrections). Note that, although we
find an exponential number of states, the $N^{2/3}$ scaling of the entropy
is far less than the expected $N$ (at $N \sim I$) for a black hole with
charges $I/4 < N < I/2$ (for $N < I/4$ the state is polar and no
single center black hole exists with this total charge \cite{Denef:2007vg}). It
would, however, be extremely interesting to do a similar counting for scaling
solutions ($N \ge I/2$) which do admit single center black hole realizations and compare this
to the black hole entropy.  We hope to come back to this in the near future.

\section{Quantum Structure of Solutions {\bf{I}}: Centers Near Infinity}\label{sec_infinity}

Having determined the structure of the quantum states associated with the two
and three center solutions we can now investigate some potential problems with
the classical solutions that we expect to be resolved by quantum effects.  For
instance, although we have already argued that the apparent non-compactness of
the three center phase space for some choices of charges and moduli is not a problem since
the symplectic volume of the phase space is finite we would still like to
see that centers cannot move off to infinity once quantum effects are taken into
account.  Likewise, as has already been observed in \cite{Bena:2006kb} and
\cite{Denef:2007vg}, scaling solutions can develop an infinitely deep throat
classically but we expect the quantization of phase space to cap this throat off
at some finite value.

To investigate these issues we would like to consider the expectation value of
the harmonic functions (\ref{harmonics}) defining the solutions.  Of course this
will depend on the particular state we are considering and there are, in
general, many possible states one can construct so making any general statement
is quite difficult.  We do not, however, need detailed properties of $\langle
H(r)\rangle$, only its behaviour in various asymptotic limits.  In this section
we are going to discuss the non-compact case leaving the scaling solutions
to the next section. Our main concern here will be the $r$ dependence at large
$r$, which we will be able to extract for a general pure state.

Let us consider the two independent constraint equations
(\ref{3cntr_constraint}) for the three center case once more
\begin{align}
	\frac{a}{u} - \frac{b}{v} &=	\frac{\Gamma_{12}}{r_{12}} -
	\frac{\Gamma_{31}}{r_{31}} = \langle h, \Gamma_1 \rangle
	=: \alpha \\
	\frac{b}{v} - \frac{c}{w} &= \frac{\Gamma_{31}}{r_{31}} -
	\frac{\Gamma_{23}}{r_{23}} = - \langle h,
	\Gamma_3 \rangle =: \beta
\end{align}
where we have (re)introduced a hopefully obvious short-hand notation.
We would like to see when one center can move off to infinity.  We will assume
that $a$, $b$ and $c$ are non-zero.  In order to satisfy the
triangle inequalities while taking at least one center to infinity we must have either two or three of the distances $u$,
$v$, and $w$ become infinite.  Let us, for definiteness, try to set $u$ and $v$
to infinity which corresponds to centers 2 and 3 staying a finite distance apart
while center 1 moves off to infinity.  From the constraint equation its clear
that this can only be done if $\alpha = 0$.  We can also consider the
case when all three centers move infinitely far apart, which can only occur if
$\beta = 0$ as well, but this is a somewhat trivial case as the angular momentum will
then, by (\ref{angularmomentum2}), vanish, as will the symplectic form, so the
space cannot be quantized without adding additional degrees of freedom (momenta).

\FIGURE{
\includegraphics[page=1,scale=0.65]{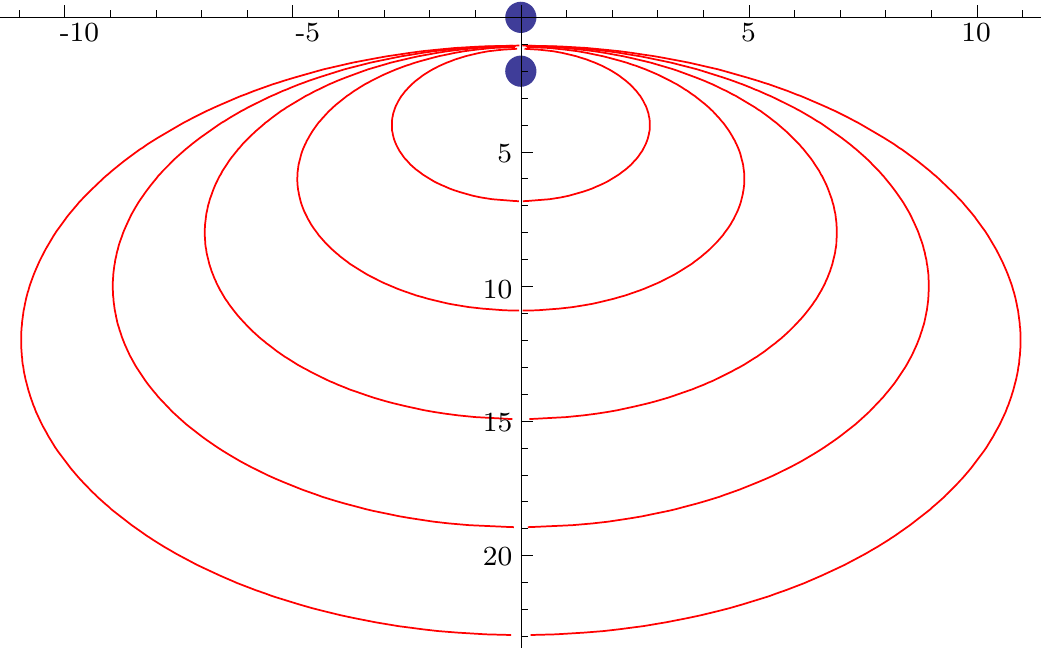} \qquad
\includegraphics[page=2,scale=0.65]{pictures_quantizing.pdf}
\caption{ {\bf (Left)} Possible locations for $\vec{x}_1$ are given above.  The
centers $\vec{x}_3$ and $\vec{x}_2$ sit at $(0,0)$ and $(0, -2)$ in the $(x,
z)$-plane respectively
and the different orbits represent different values of $a/b$.  In the figure
above $a/b < 1$ and as $a/b \rightarrow 1$ the orbit increases until it becomes
the line $z = -1$ when $a = b$ (not depicted). {\bf (Right)} Normalized
probability densities $|\psi_{m, n}(j)|$ plotted as a function of $j$ alone
(neglecting angular dependence) for some values of $n$ with $J_- \leq n \leq J_+$.  In
this example $J_+ = 18$ and $J_- = 6$.}\label{fig_orbits}
}

Thus we restrict to the case $\alpha = 0$, $\beta \neq 0$ which gives

\begin{equation}
	\frac{r_{12}}{r_{31}} = \frac{u}{v} = \frac{a}{b}
\end{equation}

\noindent When translated into coordinates this equation defines an ellipse for
the possible locations of center 1 but it is easy to see from this that $u$ and
$v$ must be bounded unless $a=b$, in which case the orbit is not an ellipse but
rather center 1 must lie on the plane between centers 2 and 3.   Hence all
unbound three center solutions will be of this form (where center 1 is fixed on
the plane normal to the axis defined by the other two centers).  Orbits for
various values of $a/b$ are depicted in figure \ref{fig_orbits}. Note that this proves the claim of \cite{deBoer:2008fk} that for three centers 'non-compactness' can only appear when the asymptotic moduli are at threshold stability, as $\alpha=0, \ a=b$ is exactly the definition of threshold stability \cite{deBoer:2008fk} in different notation\footnote{More precisely: $a=b\Leftrightarrow \langle\Gamma_1,\Gamma_2+\Gamma_3\rangle=0$ and $\alpha=0\Leftrightarrow \mathrm{Im}(Z_1\bar Z_{2+3})=0$.}.

Recall that $\sum_p \langle h, \Gamma_p \rangle = 0$ so if we define $\gamma =
\langle h, \Gamma_2 \rangle$ then $\alpha = 0$ implies $\beta = \gamma$.  Using
(\ref{angularmomentum2}) we find that

\begin{equation}
	\vec{j} = \frac{\beta}{2} \, r_{23} \, \hat{z} = \frac{\beta w}{2} \, \hat{z}
\end{equation}

\noindent with the positive $z$ axis defined by $\vec{x}_3 - \vec{x}_2$.

Let us consider the other equation

\begin{equation}
	\frac{b}{v} - \frac{c}{w} = \beta = 1
\end{equation}

\noindent where we rescale the coords so $\beta = 1$.\footnote{We can permute the
centers to force a positive sign since $\beta$ appears in one of the three
contraint equations and $-\beta$ on the other and we are free to use any two of
the three.}  Note that this forces $c < 0$ if we want to allow $v \rightarrow
\infty$.  $b$ can be positive or negative but we will assume $b < 0$ for
concreteness (this will not alter the analysis).

Using the fact that $r_{23}/2 \leq r_{13} \leq \infty$ we find

\begin{equation}
	j_+ = \frac{|c|}{2} \qquad j_- = \textrm{Min}\biggl(\frac{|c|}{2} - |b|,\,
	0\biggr)
\end{equation}

\noindent with $j_+$ reached at $r_{13} \rightarrow \infty$ and $j_-$ at
$r_{13}=|c|/2 - |b|$ (unless this is less than zero\footnote{Since $a=b$,
$b=c/2$ corresponds exactly to the beginning of the scaling regime and is, in
fact, nothing more than $N = I/2$ in the specific example of D6$\antiDp{6}$D0.
Although we work here with $j_- > 0$ it should not matter much since that large
$n$ states we consider in this section have little support in the small $j$
regime.}).

We can now consider the expectation value of
\begin{equation}
	H(r) = \frac{1}{|\vec{r} - \vec{r}_1|}
\end{equation}
The asymptotic behaviour of this function is particularly important in the
decoupling limit considered in \cite{deBoer:2008fk} since if $r_1 \rightarrow
\infty$ the classical solutions will no longer be asymptotically AdS${}_3$.  Thus we
would like to check if there are wave-functions for which $\langle H(r)
\rangle$ does not decay as $r^{-1}$.  If such states exist they would spoil they
asymptotics of our solutions, particularly in a decoupled AdS$_3$ limit as in
\cite{deBoer:2008fk}, and it would be hard to interpret them physically.

We are interested in studying wavefunctions localized near infinity so
we could restrict our attention to the states with the largest angular momentum,
$n=j_+-1$, but it turns out to be tractable to study more general pure states
$\phi = \sum_{n, m} c_{n, m} \psi_{n, m}$ (though we will find contributions
from $n < j_+-1$ are more strongly suppressed near infinity as suggested by figure
\ref{fig_orbits}, right).  For this state the expectation value is given by

\begin{align}
	\langle \phi| H(r) | \phi \rangle &= C \sum_{n', m', n, m}
	\int_{r_1^-}^{\infty}\, h(r, r_1)\, f_{n', m', n, m}(r_1)
	\, dr_1 \, d\cos\theta \, d\phi \, d\sigma \label{harm_exp}\\
	h(r, r_1) &= \frac{1}{ (r^2 + r_1^2 - 2 r r_1 \cos \alpha(\theta, \phi,
	\sigma) )^{1/2}} \\
	&= \left\{ \begin{array}{ll}
	\sum_{l=0}^\infty C_{(l)}(\theta, \phi, \sigma) \frac{r_1^l}{r^{l+1}}
	\qquad r > r_1 \\
	\sum_{l=0}^\infty C'_{(l)}(\theta, \phi, \sigma) \frac{r^l}{r_1^{l+1}}
	\qquad r < r_1 \end{array} \right. \label{harm_ang} \\
	f_{n', m', n, m}(r_1) &= e^{-2 \lambda j_+} (j_+ - \lambda j_+)^{j_+ - N/2
	-1}
	(\lambda j_+ - j_-)^{N/2 - j_-} g_{n', m', n, m}(\theta, \phi, \sigma)
	\nonumber \\
	&\sqrt{ (\lambda j_+) (j_+ - j_-) + 2 j_+ (1 -
	\lambda)(\lambda j_+ - j_-)  }
	\frac{\lambda(r_1)^{2 N + 1}}{(r_1 - b)^2} \label{rad_meas}\\
	\lambda(r_1) &= \frac{r_1}{r_1 - b} \label{r_func}
\end{align}

\noindent Here $N = n + n'$ so $j_- \leq N/2 \leq j_+ - 1$ and $\vec{r}_1 =
\vec{r}_{13}$ because $\vec{x}_3$ is at the origin.  The function $f_{n', m', n,
m} = c^*_{n', m'} \, c_{n, m}  \,\psi_{n', m'}^* \, \psi_{n, m} \, dj/dr$ which
we re-write in terms of $r_1$ using $j = \lambda(r_1) J_+$.\footnote{Note that
we have absorbed a factor of $(r_1 - b)^{-2}$ from the jacobian $dj/dr_1$ into
our definition of $f(r_1)$.}  Since we are only interested in the large $r$
behaviour of this function we can integrate out the angular dependence and also
neglect constant factors (both of which have been absorbed into the function
$g$).

The integral splits into two parts given by $r > r_1$ and $r < r_1$.  Since
$\lambda(r_1) \sim 1 + b/r_1 + \dots$ for large $r_1$, in the second
region  we see that $f(r_1) \sim r_1^{-1 - (J_+ - N/2)}$ so, after performing the
angular integrals, we are left with
\begin{equation}
	\sum_{l,k=0}^\infty C_{(l,k)} r^{l} \int_r^\infty r_1^{-l - k - 2 - (J_+ -
	N/2)}
\end{equation}
The expansion in $k$ comes from expanding $\lambda(r_1)$ in powers of
$r_1^{-1}$.  Clearly the $r_1 > r$ region only contributes negative powers of
$r$ to $\langle H(r) \rangle$.

In the region $r_1 < r$ we cannot expand $\lambda(r_1)$ since $r_1$ may not be
larger than $|b|$ but we can split this integral once more into two regions:
$r_- \leq r_1 \leq \tilde{r}$ and $\tilde{r} < r_1 \leq r$ for some $\tilde{r}
\gg |b|$.  In the first region the integration domain is $r$-independent so the $r$
dependence is simply $r^{-l-1}$.  In the second region we can repeat the
analysis for the $r_1 > r$ integral, expanding $\lambda$ in $r_1^{-1}$, and we
find a similar greater than $r_1^{-1}$ fall-off.

As was mentioned above, such configurations (with $\alpha = 0$ and $a/b = 1$)
lie on walls of {\em Threshold Stability} as discussed in \cite[Appendix
B]{deBoer:2008fk}.  In fact, our computation nicely agrees with the fact that,
as follows from the wall crossing formula \cite{Denef:2007vg}, no states
actually decay when crossing such a wall of threshold stability.  More loosely
speaking, our calculation roughly excludes the possibility of ``states running
off to infinity''.  This is important e.g. in the consistency of the decoupling
limit discussed in \cite{deBoer:2008fk} where the limit itself forces the moduli to a
wall of threshold stability for many charge configurations and we do not want
this to spoil the asymptotics of the solution.  It is also more important in a
more general context as unbounded centers do not admit an easy physical
interpretation.

\section{Quantum Structure of Solutions {\bf{II}}: Scaling Solutions} \label{deepthroat}

We have seen in the previous section that the naive run away behaviour of some
solutions is, in fact, a classical artifact which vanishes once quantum effects
are correctly accounted for.  The other potentially paradoxical region in
solution space is near a scaling point, where an infinitely deep throat develops
in spacetime \cite{Bena:2006kb, Denef:2007vg,Bena:2007qc} (see also section
\ref{scaling_sol}). This throat is particularly enigmatic in the context of
AdS/CFT. An infinite throat suggests that the bulk excitations localized deep in
the throat will give rise to a continuum of states in the dual CFT.

To make this connection with the dual $\caln=(0,4)$ CFT we uplift the
4-dimensional solutions to 5-dimensions and take the decoupling limit described
in \cite{deBoer:2008fk}.  The quantization procedure described so far will carry
on {\em mutatis mutandis} to the uplifted solutions because they have exactly
the same solution space. We will start with a general discussion but then
specialize to a working example given by a three center D6$\antiDp{6}$D0 scaling
solution where the charge of the D0 center, $N$, satisfies
$N>\langle\G_6,\G_{\bar6}\rangle/2$. We begin with some details on the structure
of scaling three-center solution spaces since, as alluded to in section
\ref{3-centers}, there are some subtle differences in the geometry of scaling
and non-scaling solution spaces. After constructing the appropriate wave
functions from the K\"ahler geometry we will use them to estimate the depth of
the throat.  We will argue that these throats get capped at some scale
$\epsilon$ and that this also sets the mass gap for the CFT.  We will perform an
estimate indicating that $\epsilon\sim N/\langle\G_6,\G_{\bar6}\rangle$ and
argue that the corresponding mass gap in the CFT has the familiar $1/c$
behaviour when $\epsilon$ is of order 1.

\subsection{Quantizing the Three Center Scaling Solutions}

As was done in the non-scaling case we must first construct the appropriate polytope for these
solutions (see figure \ref{fig_polytope}). The only property that differentiates
these solution spaces is that $j_-=0$ (this is the {\em scaling point}). As a
result, the associated polytope differs slightly from the non-scaling one; for
instance, the first inequality in (\ref{polytope}) is redundant. This may seem to
be a small modification but it actually changes the topology of the solution space, taking the limit from non-scaling to scaling corresponds to a blow down and a non contractible $S^2$ vanishes. Furthermore as we will discuss later, the probability densities
at the boundary of solution space, $j=j_-$, will have a very different
behavior in the $j_- =0$ case (figure \ref{fig_lim}) than in the $j_-\neq 0$  case (see figure
\ref{fig_orbits}).

Using the coordinates $x=j$ and  $y = j \cos \theta$ as in section
\ref{3-centers}, the scaling solution's polytope is defined by
\be \label{scaling-polytope}
    j_+ - x \geq 0\,, \qquad x+y\geq 0\,, \qquad x-y \geq 0
\ee
The construction of the complex coordinates is achieved through the
function $g$ (see appendix \ref{app_toric}). We will only need an expression for their norm squared, which is
given by
\begin{equation}
|z_1|^2=\frac{j^2\sin^2\theta}{j_+-j}\,,\qquad|z_2|^2=\frac{1+\cos\theta}{1-\cos\theta}\,.
\end{equation}
To get the wave functions one requires that they have a finite norm using the
measure $e^{ - \calk}$, modified by fermionic corrections as discussed in
section \ref{fermion-dof}.  $\calk$, as usual, stands for the K\"ahler
potential. One can calculate that
\begin{eqnarray}
{\cal K}&=&j-j_+\log(j_+-j)\,,\\
\sqrt{\det\left(\partial_i\partial_{\bar j}{\cal K}\right)}&=&\sqrt{\frac{1+\cos\theta}{1 - \cos \theta}}\,\frac{\sqrt{2j_+-j}}{j_+-j}\,.
\end{eqnarray}
Putting everything together, we end up with the following norm
\begin{align}
   |\psi_{n,m}|^2 \sim \int e^{-j} \, &\sqrt{2 j_+ - j}\, (j_+-x)^{j_+ -1 -n}\,
   j^{2n + 1} \nonumber \\
   &(1+ \cos \theta)^{n+(m +1/2)} \, (1 - \cos \theta)^{n-(m+1/2)}\, dj\, d\cos\theta\label{scalpsi}
\end{align}
Requiring that the norm is finite imposes the following restrictions
\be\label{scaling-norm}
    0\leq n\leq j_+ -1\,, \qquad -n \leq m+1/2\leq n
\ee
So the number of states is given by
$$ \caln = j_+^2 $$
Unfortunately, we cannot compare this prediction to wall-crossing because it is
not clear how to treat scaling solutions within the framework of the attractor
flow conjecture \cite{Denef:2007vg}.  On the other hand this proves the
usefulness of the tools we developed in this paper as they provide the only
known way to compute the number of BPS states for scaling
solutions\footnote{\label{foot_deconstruct}While the degeneracies computed for
scaling solutions cannot be checked against the attractor flow conjecture there
is, nonetheless, a tantalizing connection with a particular open string
computation.  If we consider a configuration with a pure fluxed D6 and
$\antiDp{6}$ center with charges $\{\pm1, p/2, \pm p^2/8, p^3/48\}$  and a
single D0-center as in \cite{Denef:2007yt} (see also section
\ref{deconstruction} where the non-scaling version of these configurations were
considered) then the number of states, $\caln$, is proportional to $p^6$.
Additional D0 centers will increase the state count by a factor of $p^3$ each.
This is very reminiscent of \cite{Gaiotto:2004ij,Denef:2007yt} where orbiting
D0's puff up into D2's which can each occupy one of $\sim p^3$ Landua-levels in
the Calabi-Yau. We hope to make this connection more precise in the near future
and see if it can lead to a derivation of black hole entropy in terms of zero
entropy supergravity solutions.}.

Another important property that is worth mentioning is that the probability
density, given by the integrand of (\ref{scalpsi}), vanishes at $j=0$.  This
suggests that, although classically the coordinate locations of the centers can
be arbitrarily close together, quantum mechanically this is not true any more.
The probability that centers sit on top of each other is zero which implies that
there is a minimum non-vanishing expected inter-center distance.  Since the depth of the
throat is related to the coordinate distance between the centers it follows that
the throat will be capped off once quantum effects are taken into account.  In
the following section we will study this phenomena quantitatively and make some
predictions for the depth of the throat and the corresponding mass gap in the
CFT.

\subsection{Macroscopic Quantum Effects}

Before we analyse a specific type of scaling solution in some more quantitative
detail let us point out some general features of any three center
solution space with a scaling point. As long as we are interested in spherically
symmetric quantities, e.g. sizes and distances, we can neglect the angular
part of (\ref{scalpsi}) as it will drop out after normalization. For such
questions we can effectively use wave functions on $j$-space, which depend
only on the quantum number $n$ and the value of $j_+$.  The corresponding
probability densities are
\begin{equation}
\langle n,j_+;j|n,j_+;j\rangle=\frac{e^{-j}j^{2n+1}(j_+-j)^{j_+-n-1}\sqrt{2j_+-j}}{\int_{0}^{j_+}\, e^{-j}j^{2n+1}(j_+-j)^{j_+-n-1}\sqrt{2j_+-j}\,dj}\,,\label{scalprobs}
\end{equation}
These are plotted for $n=0,\,5,\,10,\,15,\,20$ and $j_+=21$ in figure \ref{fig_lim}
(left).

\FIGURE{
\includegraphics[page=6,scale=0.6]{pictures_quantizing.pdf} \,
\includegraphics[page=7,scale=0.6]{pictures_quantizing.pdf}
\caption{ {\bf (Left)} Plot of the probability densities (\ref{scalprobs}) for $j_+=21$. The blue, orange, red, purple and green curves correspond to respectively $n=0,\,5,\,10,\,15,\,20$.
{\bf(Right)} Plot of the probability density corresponding to the lowest state
$n=0$ for the values $j_+=5,\,50$ (blue) and $j_+=\infty$ (green) where the
curve at $j_+=50$ is already barely distinguishable from the limiting curve
$j_+=\infty$ (see eqn. (\ref{limdim})). Note that the probability distribution
vanishes on $j=0$.}\label{fig_lim}
}

The state that interests us the most is the $n=0$ state as this has the greatest
support near the classical scaling point, $j=0$. Note that when there is no
scaling point, i.e. $j_-\neq0$, the lowest state, with $n=j_-$, peaks on $j_-$
(see figure \ref{fig_orbits}).  In the scaling case the lowest state actually
has zero support on $j_-=0$ (see figure \ref{fig_lim}, right).  This seems to
indicate that the scaling solutions, supergravity solutions where all centers
coincide in coordinate space and $j=0$, are not well defined classical solutions
as they correspond to a point in the phase space where no wave function has finite
support.  To get an idea of how well this classical point can be
approximated by a quantum expectation value we calculate $\langle j\rangle$ in
the lowest state, $|0,j_+;j\rangle$.

For a scaling solution space characterized by $j_+$ this expectation value is given by
\begin{equation}
\langle j\rangle_{j_+}=\int_0^{j_+}\langle 0,j_+;j|j|0,j_+;j\rangle dj=\frac{\int_0^{j_+}e^{-j}j^{2}(j_+-j)^{j_+-1}\sqrt{2j_+-j}\,dj}{\int_{0}^{j_+}\, e^{-j}j(j_+-j)^{j_+-1}\sqrt{2j_+-j}\,dj}
\end{equation}
In general this expression is not analytically tractable.  We are, however,
particularly interested in the supergravity regime which coincides with
$j_+\rightarrow\infty$. In this limit the expression can be simplified
considerably by using the well know expression for the exponential,
$\lim_{j_+\rightarrow\infty}(1-\frac{j}{j_+})^{j_+}=e^{-j_+}$, giving
\begin{equation}
\lim_{j_+\rightarrow\infty}\langle 0,j_+;j|0,j_+;j\rangle=4 e^{-2j}j\,,\label{limdim}
\end{equation}
this is the green curve plotted in figure \ref{fig_lim} (right).
Using this limiting behaviour it is straightforward to calculate that
\begin{equation}
\langle j\rangle_\infty=1\,.
\end{equation}
In other words, even in the lowest state the expected value of $j$ is one
quantum, i.e $\hbar$.  Moreover, because the depth of the throat grows very
rapidly in the region $j \sim 0$ many macroscopically different configurations
sit within the range $0 < j < 1$ so, even though $j\sim 1$ is only one plank
unit away from the scaling point the corresponding geometry (expectation value
of the metric) is very different.

As further evidence for the formation of a cap we will also compute the
behaviour of the harmonic functions appearing in the metric in this same state.
To make this computation simpler we will work with the system introduced in
section \ref{sec_infinity} ($\alpha =0$ and $a=b$ but now we take $|b| > |c|/2$)
and we will also use the notation used in there.  We will first be interested in
determining the $r$-dependence of the (expectation values of the) functions
\begin{equation}
	h_1(r) = \frac{1}{|\vec{r} - \vec{r}_1|} \qquad h_2(r) = \frac{1}{|\vec{r} - \vec{r}_2|}
\end{equation}
where $r_1$ and $r_2$ have the same meaning as in section \ref{sec_infinity}
($\vec{r}_3 = -\vec{r}_2$ so we need only work with one of them).  Note that the
functions $h_i(r)$ appear directly in the harmonics $\{H^0(r), H^A(r), H_A(r),
H_0(r)\}$ but which harmonics they appear in depends on the specific form of
$\Gamma_i$ (which we do not fix at this point).

To compute the $r$-dependence of $h_i(r)$ we proceed very much as in section
\ref{sec_infinity}, and indeed the computation is mostly analogous,
\begin{align}
	h_i(r, r_i) &= \left\{ \begin{array}{ll}
	\sum_{l=0}^\infty C_{(l)}(\theta, \phi, \sigma) \frac{r_i^l}{r^{l+1}}
	\qquad r > r_i \\
	\sum_{l=0}^\infty C'_{(l)}(\theta, \phi, \sigma) \frac{r^l}{r_i^{l+1}}
	\qquad r < r_i \end{array} \right. \label{harm_ang_sc} \\
	\int_0^{j_+}\langle 0, j_+;j | h_i(r) | 0, j_+; j \rangle dj&= \frac{1}{L} \sum_{l}
	\biggl[
	r^{-l-1} \int_{0}^{r}\, C_{(l)} \, r_i^l\, f(r_i) \, dr_i + \nonumber \\
	&r^{l} \int_{r}^{\tilde{r}}\, C'_{(l)}\,  r_i^{-l-1}\, f(r_i) \, dr_i +
	r^{l} \int_{\tilde{r}}^{r_+}\, C'_{(l)}\,  r_i^{-l-1}\, f(r_i) \, dr_i
	\biggl]
	\label{harm_exp_sc}\\
	f(r_i) &= j(r_i)\,e^{-2 j(r_i)} \, \sqrt{2 J_+ - j(r_i)}\,
	\frac{dj(r_i)}{dr_i} \label{rad_meas_sc}\\
	j(r_1) &\sim \frac{r_1}{r_1 - b} j_+ \label{r1_func} \\
	j(r_2) &\sim  r_2 \label{r2_func}
\end{align}
Here we are only interested in the regime $r \ll |b|$ and we define $\tilde{r}$
such that $r < \tilde{r} \ll |b|$.  $r_+$ is defined by $j(r_+) = J_+$. We have
also approximated $(1-\frac{j}{J_+})^{J_+} = e^{-j}$.  The factor of $1/L$ above
is the normalization of the wavefunction which we will not
need in this particular computation.  The ``$\sim$'' in the last
two equations reflects an ambiguity by a constant prefactor that depends on the
moduli at infinity (which we set to one in section \ref{sec_infinity} by
rescaling the coordinates).

The integrals in (\ref{harm_exp_sc}) can be solved (in terms of
$\Gamma$-functions) by expanding in  $r_1/|b|$ or $r_2/J_+$ yielding an answer
in terms of a power-series in $r$.  The lowest order term in the series is a
constant so we find $\langle h_i(r) \rangle \sim a_i + b_i r^\alpha$ with
$\alpha > 0$.  This implies that the harmonics in the metric have the same small
$r$ behaviour so for $r \sim 0$, near the scaling point, we can evaluate the
behaviour of the five dimensional metric  in the decoupling limit (see e.g.
\cite[eqn.  (2.15)]{deBoer:2008fk}) and find the metric does not develop a
throat.

\subsubsection*{Cutting the throat}

Finally we would like to translate our insight above, that quantum mechanically
one cannot reach the scaling point, and hence no infinitely deep throat
develops, into a rough quantitative estimate of a mass gap in the dual CFT.

We will do this in the particular example of the D6$\antiDp{6}$D0 three center
solution introduced in section \ref{deconstruction}, which is scaling when
$N>I/2$. In section \ref{deconstruction} we considered the non-scaling version
of these solutions where $N < I/2$ and we considered multiple D0 centers; here
we work with a single D0 center and scaling charges, but we will use the
notation of that section.  Note that, from eqns
(\ref{d6d6d0_c1})-(\ref{d6d6d0_c2}), this system is actually a particular
instance of the general construction of section \ref{sec_infinity} since
$\alpha=0$ and $a=b$ (in the notation of section \ref{sec_infinity}).  Thus our
computation of the harmonic functions in the previous subsection applies to
this system.

The calculation will proceed in two steps. First, we want to estimate at which
scale the quantum smearing cuts off the naive infinite throat. Second, we will
translate this scale into a mass gap (in the dual CFT) by analysing a scalar field on
a toy model geometry with a throat cut off at this scale. The scale at which  we
expect a deviation from the naive infinite throat is of the order of the
minimum expected inter-center coordinate distance.  In the case of the D6$\antiDp{6}$D0
scaling solution $j_+=\frac{I}{2}$ and, furthermore, the angular momentum is
related to the inter-center distances by (see the constraint eqns. (\ref{d6d6d0_c1}) with $\beta=\frac{1}{4}$ for asymptotic AdS space \cite{deBoer:2008fk})
\begin{equation}
r_{6\bar6}=8 j\,,\qquad r_{0}=\frac{8jN}{I-2j}\simeq8j\,\frac{N}{I}\,,
\end{equation}
where, in the second expression, we have made a large $I$ approximation. As
these expressions are linear in $j$ we see that their expectation value in the
$n=0$ state is directly given in terms of $\langle j\rangle_\infty=1$. So we
find that quantum mechanically one expects the throat to be cut of at a scale of
order $\epsilon\sim\frac{N}{I}>1/2$.

A nice check on this estimate is to determine the charge dependence of the
constant term in $\langle h(r)\rangle$ in the $r \sim 0$ limit since, by a small
$r$ expansion,
\be
h(r) = \frac{1}{|\vec{r} - \vec{r}_i|} = \frac{1}{r_i} + \mathcal{O}(r)
\ee
Note that this leading, $r$-independent, term feeds directly into the metric at small $r$ so is a
very relevant physical quantity to compute.  A careful computation of this
leading term, taking into account the normalization $L$ in eqn.
(\ref{harm_exp_sc}), yields
\be
\langle h_1(r)\rangle \sim \frac{\gamma\, j_+}{|b|} + \mathcal{O}(r)
\ee
with $\gamma$ a small number of order one.  For the D6$\antiDp{6}$D0 system this
gives $\langle \epsilon^{-1}\rangle  := \langle r_1^{-1} \rangle \sim
\frac{I}{N}$ which confirms our previous computation.  This shows that the
estimate of $\epsilon$ is relatively robust and does not depend very
strongly upon which particular quantum expectation value we use to compute it.
Moreover, as the metric and solutions are defined via the harmonics this
expectation value very directly relates to the quantum expectation value of the
metric.

Now that we have understood the charge dependence of $\epsilon$ we wish to
translate this into a mass gap in the dual CFT.  The computation of the mass gap
in terms of a scalar wave equation on a capped-throat geometry is somewhat
technical and has thus been relegated to appendix \ref{appendix3}.  Here we will
quote the final result for the mass gap $\Delta(L_0+\bar L_0)$ in terms of $\epsilon$:
\begin{equation}
\Delta (L_0+\bar L_0)\sim\frac{\epsilon}{c}>\frac{1}{2c}.\label{massgap}
\end{equation}
with $c=p^3=6I$, the central charge of the dual CFT.  In the regime
where $N \approx I/2$ (so the inequality above is saturated) this matches the
expectation from the long string picture that the lowest energy excitation in
the CFT is of order $1/c$ (see e.g.  \cite{Mathur:2005zp}).  Whether the
different scaling in the Cardy regime, $N \gg I$, reflects new physics of these
solutions or is an artifact of our toy model geometry (see appendix
\ref{appendix3}) would be interesting to explore.

It would also be interesting to find an interpretation of the dependence of the
mass gap on the parameter $\epsilon\sim N/I\sim\frac{L_0}{c}$ which seems
closely related to $h=(L_0-c/24)/c$.  The latter played an interesting role as
an order parameter in phase transitions in the dual CFT, as discussed in
\cite{deBoer:2008fk}. It was suggested there that this phase transition is due
to a large number of winding modes being turned on. Although this is
speculative, it might, itself, hint at a ``long string'' picture for the CFT at
small $h$. It is interesting that the result (\ref{massgap}) might also hint at
such a picture. In any case these results and speculations only further
highlight the importance of better understanding the dual $\caln=(0,4)$ CFT.

\section{Conclusions}\label{conclusions}

In this paper we have studied spaces of solutions of the field equations of
${\cal N}=2$ (${\cal N}=1$) supergravity in four (five) dimensions that describe
multicentered black hole configurations. For suitable choices of the charges of
the individual centers the resultant geometry is completely smooth so our setup
also applies to spaces of smooth solutions.  We have determined the appropriate
symplectic form on these solution spaces and used that to quantize them and
determine both the number of states and the explicit expressions for the
wave-functions in several examples.

The solutions we consider are all stationary, but in general not static, and, as
such, they carry non-trivial angular momentum. Because of this momentum we did
not need to include small velocities for the centers in order to obtain a
non-degenerate phase space (as in e.g. \cite{Michelson:1999dx}). In other words,
the space of stationary solution really is a phase space and not a configuration
space. However, for some choices of the charges, we do find that the phase space
is degenerate. This happens, for instance, when all centers carry D0 and D4
brane charge so that all inner products $\langle \Gamma_p,\Gamma_q\rangle $ are
identically zero (note that in this case there is also no angular momentum). In
situations like these one could try to include small velocities for the centers
in order to arrive at a well-defined phase space. It is clearly an interesting
question whether this modified system will give rise to BPS states upon
quantization. Superficially, the momenta increase the energy while leaving the
charges invariant, and they therefore violate the BPS condition. However, if the
Hilbert space has a continuous spectrum of momenta it is possible that there is
a BPS bound state at zero momentum in the spectrum. This is difficult to analyze
in general, but in our case we don't expect this to happen, at least not in
asymptotically AdS spaces since AdS effectively provides a box and will
therefore put an IR cutoff on the admissible momenta. It is crucial, for this
argument, to be in the RR sector of the CFT. As was argued in
\cite{deBoer:2008fk}, time independent solutions in the RR sector which include
a purely fluxed D6 and $\antiDp{6}$ center can be re-interpreted as possibly
time-dependent solutions in the NS sector which no longer include the purely
fluxed D6 and $\antiDp{6}$ centers. Therefore it is quite likely that there are
interesting time-dependent BPS solutions in the NS-sector. We believe that the
index computation of e.g.  \cite{BrittoPacumio:2000sv}, where multicentered
black hole configurations including velocities were considered, can also be
reinterpreted along these lines and it would be interesting to work this out in
more detail. Thus our proposal is that solution spaces with a degenerate
symplectic form should not be thought of as describing proper BPS bound states.

This immediately leads to another issue; namely, we know that, for example, $N$
D0-branes can form a marginal bound state \cite{Banks:1996vh,Sethi:2000ba}, but
the symplectic form for such a configuration (for example in the presence of a
D4-brane) would vanish identically. This clearly conflicts with the statements
of the preceding paragraph.

We would like to argue that the resolution of this inconsistency lies in the
fact that the marginal bound state of D0-branes cannot be understood purely from
a low-velocity expansion in the Coulomb branch.  Rather, the presence of the
non-abelian degrees of freedom is essential for the bound state to exist. In
other words, the marginal bound state has significant support in the Higgs
branch. This is supported both by the analysis of \cite{Sethi:2000ba} as well as
by the size of the bound state (see e.g. \cite{Polchinski:1999br}). Again, it
would be interesting to explore this further. In this paper, we have taken the
point of view that a solution containing a marginal bound state of e.g. $N$
D0-branes should be counted separately from a similar solution where the
marginal bound state has been replaced by $N$ individual D0-branes, and we
obtained satisfactory results.

One of the main results of our paper is that we have explicitly shown how
scaling solutions, which classically give rise to arbitrarily deep throats in
space-time, are modified in the quantum theory. The deep throats are capped off
in such a way that the spectrum of the dual CFT remains gapped, and we have
estimated the size of the gap to be of order $1/c$. This resonates well with the
``fractionation'' idea described in \cite{Mathur:2005zp}. Here, the results
suggest the possible existence of a long string picture, whose length is $p^3$
times as long as that of the original string. In any case, it is striking that
our results show that geometries that are perfectly smooth and can have
arbitrarily low curvature can nevertheless be ruled out as acceptable classical
geometries. The reason, in our case, is that although the solutions develop a
very deep throat in space-time, they are only responsible for a tiny fraction of
the total phase space, somewhat analogously to what was observed in
\cite{Mathur:2007sc}.  There is a nice qualitative picture one can use to
understand this.  The scaling solutions require localization of the centers
within an arbitrarily small coordinate distance.  If the centers did not warp
the geometry, standard quantum mechanics (the uncertainty principle) would tell
us this is not possible.  The back-reaction of the centers, however, stretches
this infinitesimal coordinate distance to macroscopic sizes.  The novel physics
emerges from the realization that the symplectic form does not seem to feel this
``stretching'' so the phase space volume stays small.

To our knowledge, this is one of the first explicit examples where we can see
that quantum gravitational effects are much stronger than what one would expect
based on semiclassical reasoning. It is tempting to speculate that a similar
phenomenon may take place at the horizon of a black hole and this is one of the
key ingredients in resolving the black hole information paradox.

The simplest scaling solutions are described by quiver quantum mechanical
systems where the quiver has a closed loop. What we have studied is the
quantization of the Coulomb branch of the theory. One could imagine that down
the throat, where the branes approach each other very closely, a Higgs branch
description is more appropriate, since the open string stretching between the
branes become very light. However, in examples one finds that the number of
states in the Higgs branch is much larger and can have exponential growth
\cite{Denef:2007vg}. In these cases the Coulomb and Higgs branches are therefore
either completely decoupled, or alternatively, only part of the Higgs branch can
be seen on the Coulomb branch. This point clearly also deserves a better
understanding.

For a multicentered black hole configuration the quantization we are doing
effectively counts the number of configurational degrees of freedom, but it
clearly ignores any entropy the individual black holes might have. Therefore the
resulting states do not have a straightforward interpretation as states in the
dual CFT. The quantization of spaces of smooth solutions, on the other hand,
does lead to BPS states which should have a dual description as BPS states in
the CFT as well. At the same time, multicentered black hole configurations
should also be describable by suitable ensembles of states in the CFT. One may
therefore wonder to which multicentered black hole configuration a state
obtained from quantizing a given space of smooth solution belongs. This is in
particular of relevance for the ``fuzzball'' proposal \cite{Lunin:2001jy} (see
also review \cite{Mathur:2005zp,Bena:2007kg,Skenderis:2008qn}) which, in its
most optimistic form, would imply that all states that make up multicentered
black holes can be obtained by quantizing suitable spaces of smooth supergravity
solutions.

In \cite{Denef:2007vg} it was proposed that the Hilbert space of states of a given total charge can be partitioned in sub-spaces each corresponding to a given attractor flow tree. Furthermore we expect the states on a solution space of smooth centers to be elements of this Hilbert space. This seems to imply that such states should be related to a unique attractor tree. This is especially non-trivial in the case of smooth scaling solutions.
Most naively one could propose the following relationship between multicentered
black hole attractor trees  and {\em smooth} ``multi-scaling'' solutions.  Let us assume that for a given set of charges there is only one way (in a given connected component of solution space) to cluster the charges such that the individual clusters form ``local scaling solutions''.  By local scaling
solutions we mean that in some regime in solution space all the charges in a
given cluster can be brought arbitrarily close together to form an infinitely
deep throat.  Since there are multiple clusters there is also a regime in
solution space where the charges in each cluster are brought together to form
several separated throats.  So our assumption is essentially that within a given connected component of
solution space it is not possible to re-arrange the charges into a different set
of clusters that still enjoy this property.  Let us explore this possibility briefly as
it may have interesting consequences.  For instance, if this is true then it
would be natural that the states arising from the quantization of such a
connected component correspond to microstates of
the multicentered black hole composed of centers each carrying a charge
corresponding to one of the clusters that can form a scaling throat.

The split attractor flow picture might then be useful in classifying this
correspondence (between multi-black hole configurations and smooth multi-scaling
solutions).  Recall that all non-scaling multicenter solutions, smooth or not,
are described by a suitable attractor flow tree of which the existence is
necessary and (conjectured to be) sufficient for the existence of the
multicenter solution. For the corresponding multi-scaling solutions the
existence of the same attractor flow tree is necessary as well (although known not to be sufficient). In particular, when some
centers locally form a scaling solution, which can have an arbitrarily deep
throat, they cannot be distinguished individually in the attractor flow tree
(because no amount of moduli tuning will force the centers apart and this is the
basis of the attractor tree construction). The tree is thus conjectured
\cite{Denef:2007vg} to end on the attractor point corresponding to the sum of
the charges that locally form a scaling solution.  The existence of the tree is
necessary because when all the clusters have turned into infinitely deep throats
the spacetime geometry away form an individual cluster is very well approximated
by the geometry of a multi-black-hole solution with the same charges so if the
latter has pathologies then so will the former.

A tentative proposal is therefore that a given smooth solution belongs to the
multicentered black hole configuration which has the same attractor flow tree.
The attractor tree either describes a multicentered black hole configuration
with one black hole for each endpoint of the tree or it can be ``resolved'' to a
similar configuration with each black hole replaced by a ``local'' smooth
scaling solution.  In particular, the smooth solutions that belong to a single
black hole are those where all the centers can simultaneously disappear down a
single, arbitrarily deep, throat; smooth solutions that belong to a two-centered
black hole are those that can disappear down two throats; etc.  So roughly each
group of smooth centers that locally form a scaling solution belongs to a single
black hole constituent.  We should emphasize that while this is an appealing
picture, very little is yet known about the structure of scaling solutions with
more than three centers and this would be a prerequisite to making any such
proposal precise.

To test this idea and to see whether one can possibly account for the entropy of
a single black hole in this way, we need to find scaling solutions with many
centers, quantize the corresponding space of solutions, and count the resulting
number of states. The most promising example to study is the one consisting of
D6$\antiDp{6}$ and many D0's studied in section~6. In section~6 we counted the
number of solutions in the non-scaling regime and found perfect agreement with
the wall-crossing formula. These solutions can also be arranged in such a way
that they are scaling, but we have not yet completed a computation of the number
of states one can obtain in this way (see also footnote \ref{foot_deconstruct}).
We hope to report on this in the near future.

It is also interesting to note that our solution space quantization is somewhat reminiscent
of the ``near horizon probe'' proposal of \cite{Gaiotto:2004ij,Kim:2005yb,Denef:2007yt} and hence might lead to some insights in the exploration of the link between this proposal and the fuzzball picture \cite{Raeymaekers:2007ga,Raeymaekers:2008gk}.

Finally, there are several other loose ends and open problems to explore. These
include a proper understanding of the dual CFT and the right way to describe the
CFT duals of the multicentered black hole configurations. We hope that the
results in this paper, such as the computation of the gap, will allow us to make
some progress in this direction. We would also like to have a better
understanding of the role of the fermions in the quantization, and like to know
whether the solution space always comes out to be K\"ahler or even toric (which
is supported by the agreement with and the derivation of the wall-crossing
formula in \cite{Denef:2007vg}). It would also, perhaps, be interesting to compute
the symplectic form directly in supergravity and understand why terms that do
not come from the gauge field cancel (see discussion in the beginning of section
\ref{symplecticandquantization}).  We leave these, and several other issues,
to a future study.

\section*{Acknowledgements}
The authors would like to thank Frederik Denef for initial collaboration and
discussion during later stages of the project.  We also gratefully acknowledge
useful discussions with I.~Bena, M.~Berkooz, T.~Levi, S.~Mathur, G.~Moore,
K.~Papadodimas, B.~ Pioline, and S.~Raju.

DVdB is an Aspirant of the FWO Vlaanderen and is partially supported by
the European Community's Human Potential Programme under contract
MRTN-CT-2004-005104 `Constituents, fundamental forces and symmetries of the
universe', by the FWO - Vlaanderen, project G.0235.05 and by the Federal Office
for Scientific, Technical and Cultural Affairs through the `Interuniversity
Attraction Poles Programme - Belgian Science Policy' P6/11-P.  The work of JdB,
SES and IM is supported financially by the Foundation of Fundamental Research on
Matter (FOM).

\appendix

\section{The Three Center Solution Space} \label{3center-moduli}

In this appendix we will analyze some properties of the moduli
space of three-center solutions. Our starting point will be the
set of equations (\ref{consistency2}) which we will rewrite as
follows
\bea
\frac{a}{x} - \frac{b}{y} & = & c_2 - c_1 \nonumber \\
\frac{b}{y} - \frac{c}{z} & = & c_3 - c_2 \nonumber \\
\frac{c}{z} - \frac{a}{x} & = & c_1 - c_3. \label{jjj1}
\eea
Here $a,b,c$ represent the inner products $\langle
\Gamma_a,\Gamma_b\rangle$, $x,y,z$ are the lengths of the three
sides of the triangle spanned by $\mathsf{x}_a$, and
$c_2-c_1=\langle h,\Gamma_1\rangle$ etc. The constants $c_a$ are
not uniquely fixed, as we shift them by a fixed amount without
modifying the above equations. Still, expressing things in terms
of $c_a$ allows for a somewhat more symmetric treatment.

The first important remark is that up to an $SO(3)$ rotation,
$x,y,z$ uniquely determine the solution. In other words, the
quotient of the solution space by $SO(3)$ is precisely the set of
solutions $x,y,z$ of (\ref{jjj1}).

Second, we should keep in mind that $x,y,z$ are the sides of a
triangle, i.e. they should be nonnegative numbers that satisfy the
triangle inequality $x+y\geq z$ and its cyclic permutations.

In our discussion of the solution space quantization, the size of
angular momentum will play an important role, as we will use it as a
coordinate on the solution space. In terms of the variables used in
(\ref{jjj1}), the angular momentum is given by (see
(\ref{angularnorm}))
\be \label{jjj2}
J^2 = - \frac{1}{4} \left( x^2 (c_2-c_1)(c_1-c_3) + y^2
(c_3-c_2)(c_2-c_1) + z^2 (c_1-c_3)(c_3-c_2) \right).
\ee
We would in particular like to know whether $|J|$ is a good
single-valued coordinate on the solution space and what range of
values it takes.

It is easy to write down the general solution to (\ref{jjj1}) in
terms of a single free parameter $\lambda$:
\be
x=\frac{a}{\lambda-c_1}, \qquad y= \frac{b}{\lambda-c_2}, \qquad z
= \frac{c}{\lambda-c_3}. \label{jjj3}
\ee
This is the general solution if $a,b,c$ are not equal to zero. If
all three are zero, or two out of three are zero, there are either
no solutions to (\ref{jjj1}) or the space of solutions is at
least two-dimensional. In either case the symplectic form becomes
degenerate and most likely these solution spaces do not give rise to
BPS states\footnote{In order to obtain a non-degenerate symplectic
form we would have to add additional degrees of freedom to the
theory, which would act like momenta for some of the coordinates.
A nonzero momentum will typically increase the energy while the
charges remain fixed, violating the BPS condition. Therefore we
expect no BPS states in this case.}. Finally if one of $a,b,c$ is
zero, say $a=0$, then either there are no solutions or one finds a
fixed value for $y,z$ from (\ref{jjj1}), while $x$ is not
constrained by (\ref{jjj1}). However, $x$ is constrained by the
triangle inequalities so that the solution space becomes
\be
a=0, \,\,\, b\neq 0, \,\,\, c\neq 0, \,\,\, \Longrightarrow
y,z\,\,\, {\rm fixed}, \,\,\, |y-z|\leq x \leq y+z.
\ee

We now continue with the case where $a,b,c$ are not equal to zero so that the
solutions are of the form (\ref{jjj3}). We again need to distinguish a few
cases. The most degenerate case is when $c_1=c_2=c_3$. Then either the moduli
space is empty or one-dimensional, but in the latter case the angular momentum
vanishes identically everywhere on the solution space and thus the symplectic form
is trivially degenerate.

The next case is $a,b,c$ nonzero and two of the $c_i$ equal to
each other. Using the permutation symmetry of (\ref{jjj1}) and the
possibility to simultaneously change the signs of
$a,b,c,c_i,\lambda$, we can distinguish three different cases: (i)
$c_3>c_1=c_2$ and $a,b,c>0$, (ii) $c_1=c_2>c_3$ and $a,b,c>0$, and
(iii) $c_3>c_1=c_2$, $a,b>0$ and $c<0$. Positivity of $x,y,z$
requires that $\lambda\in I_1=(c_3,\infty)$ for cases (i),(ii) and
$\lambda \in I_1=(c_1,c_3)$ in case (iii). Next we denote by $I_2$
the set of solutions of the triangle inequalities
\be
\frac{a+b}{\lambda-c_1} > \frac{c}{\lambda-c_3} >
\frac{|a-b|}{\lambda - c_1}.
\ee
It is easy (though somewhat tedious) to see that $I_1\cap I_2$ is
either empty, an interval of the form $[\lambda_-,\lambda_+]$, an
interval of the form $[\lambda_-,\infty)$, an interval of the form
$(c_1,\lambda_+]$, or an interval $(c_1,\infty)$. The endpoints
$\lambda_+$ and $\lambda_-$ always correspond to a point where a
triangle inequality is saturated. The interval extends all the way
to infinity only if\footnote{Actually, this can possibly also
happen when $a+b=c$ or $c=|a-b|$.} $a+b>c>|a-b|$, i.e. when
$a,b,c$ satisfy triangle inequalities, which can only happen in
case (i) and (ii). In these cases there is a scaling throat with
$x,y,z \rightarrow 0$. The interval starts at $c_1$ only if we are
in case (ii) or (iii) and $a=b$, and in this case the solution space
includes configurations where a center can move off to infinity.

From the point of view of angular momentum, the case where one of
the centers moves away to infinity (e.g. $x,y\rightarrow\infty$)
can be viewed as a case where the triangle inequalities $x+z\geq
y$ and $y+z\geq x$ are both saturated. Therefore, in all cases we
have analyzed so far, the solution space contained just a single
connected component described by single interval of possible
values of $\lambda$, and at the endpoints of the interval either
one has a scaling solution with vanishing angular momentum, or a
solution that saturates at least one triangle inequality. Whenever
this happens, we always find that
\be
|J|^2 = \frac{1}{4} (\pm a + \pm b + \pm c)^2
\ee
for suitable choices of the signs, as can be seen easily e.g. from
(\ref{angularmomentum}).

It remains the analyze the generic case with all $c_i$ different from each
other. Up to an overall sign flip and a permutation, there are two cases,
which are (iv) $c_1<c_2<c_3$ and $a,b,c>0$ and (v) $c_1<c_2<c_3$ and $a,b>0$,
$c<0$. Positivity of $x,y,z$ in case (iv) implies $\lambda>c_3$ and implies $c_2<\lambda<c_3$ in case (v) . The
main problem is to analyze the triangle inequalities. They can be analyzed
qualitatively by sketching $x+y-z$, $x-y+z$ and $-x+y+z$ as a function of
$\lambda$. We know that each of these functions can have most two zeroes as a
function of $\lambda$, and we know its behavior near the three poles at
$\lambda=c_1,c_2,c_3$. We will skip the details, but one finds that the moduli
space consists of at most two components, each of which corresponds to a certain
interval of possible values of $\lambda$. At the boundaries of each interval a
triangle inequality is saturated. Notice that in case (iv) one of the components
can be of the form $[\lambda_-,\infty)$. This is possible whenever $a,b,c$
themselves satisfy triangle inequalities. If this happens, at $\lambda=\infty$
there is a scaling solution.

To summarize, the solution space in all cases consists of at most
two components, corresponding to two intervals of possible values
of $\lambda$. At the endpoints of the interval some triangle
inequality is saturated. This can include configurations where one
of the centers moves off to infinity (cases (ii) and (iii) above,
with $a=b$), and scaling solutions where $\lambda\rightarrow
\infty$ (cases (i), (ii) and (iv) with $a,b,c$ obeying triangle
inequalities).

Finally, we would like to show that $J^2$ is a good coordinate on
each component of the moduli space of solutions to (\ref{jjj1}).
In order to do so, we compute $dJ^2/d\lambda$. According to
(\ref{jjj3}), $dx/d\lambda=-x^2/a$ and similarly for $y,z$. If we
differentiate (\ref{jjj2}), use these relations, and finally
replace $c_i$ by the left hand side of the original equations
(\ref{jjj1}), we obtain
\be
2\frac{dJ^2}{d\lambda} =
\frac{x^3}{a}\left(\frac{a}{x}-\frac{b}{y}\right)\left(\frac{c}{z}-\frac{a}{x}\right)
+\frac{y^3}{b}\left(\frac{b}{y}-\frac{c}{z}\right)\left(\frac{a}{x}-\frac{b}{y}\right)+
\frac{z^3}{c}\left(\frac{c}{z}-\frac{a}{x}\right)\left(\frac{b}{y}-\frac{c}{z}\right).
\ee
We rewrite this as
\be \label{jjj4}
-2 a b c x y z \frac{dJ^2}{d\lambda} = n_0 a^2 + n_1 a + n_2 =
n_0\left(a+\frac{n_1}{2n_0}\right)^2 + \frac{ 4 n_2 n_0-n_1^2}{4
n_0}
\ee
with $n_0,n_1,n_2$ certain $a$-independent polynomials. The right
hand side of (\ref{jjj4}) is positive if $n_0$  and $4 n_2 n_0 -
n_1^2$ are positive. By explicit computation we find
\bea
n_0 & = & \left(z^2 b + \frac{c y}{2 z}(x^2 -y^2 - z^2) \right)^2
+ \frac{c^2 y^2}{4 z^2}\theta \nonumber \\
4 n_2 n_0 - n_1^2 & = & b^2 c^2 x^2 (bz-cy)^2 \theta
\eea
where
\be
\theta=(x+y+z)(x+y-z)(x-y+z)(-x+y+z).
\ee
Since $\theta>0$ if all triangle inequalities are satisfied, we
have indeed shown that $J^2$ is a monotonous function of $\lambda$
and that $J^2$ is a good coordinate on each component of the
solution space.

\section{Aspects of Toric Geometry} \label{app_toric}

In this appendix we review some techniques in toric geometry that we use
throughout the paper. By construction we start in our description of solution
space in the main text from a symplectic point of view. It is however more
convenient for geometrical quantization to have a K\"ahler description, which
can always be made in the case of a symplectic toric manifold. The main formulas
in this appendix are thus the expressions (\ref{cc},\ref{km}) for the complex
coordinates and K\"ahler potential in terms of the symplectic coordinates on a
symplectic toric manifold. Before giving these formulas we review some of the
basics of symplectic toric manifolds and symplectic toric orbifolds.

\subsubsection*{Polytopes}
As is customary we will refer to the convex hull of a finite number of points in
$\mathbb{R}^n$ as a {\it polytope}. The boundary of such a polytope is itself
the union of various lower dimensional polytopes that are called {\it faces}. In
particular a zero-dimensional face is called a ${\it vertex}$, a one-dimensional
face an {\it edge} and a $n-1$-dimensional face a {\it facet}. Note that we can
view any polytope as the intersection of a number of affine half spaces in
$\mathbb{R}^n$. A polytope $P$ can thus be uniquely characterized by a set of
inequalities, namely $\vec x\in P$ iff $\forall i=1,\ldots, \#(\text{facets})$
\be
      \langle \vec c_{i}, \vec x\rangle \geq \lambda_i\ \ \Leftrightarrow\ \ \sum_j c_{ij} x_j \geq \lambda_i\,.\label{ineqs}
\ee
Given a polytope we will call the set $\vec c_i\in\mathbb{Z}^n$, given by the inward pointing normals to the various facets, the {\it normal fan}.

An $n$-dimensional polytope is called a {\it Delzant polytope} if it satisfies the following three conditions
\begin{itemize}
\item {\bf simplicity:} in each vertex exactly $n$ edges meet,
\item {\bf rationality:} each of the $n$ edges that meet at the vertex $p$ is of the form $p+t u_i$ with $t\in\mathbb{R^+}$ and $u_i\in\mathbb{Z}^n$,
\item {\bf smoothness:} for each vertex the $u_i$ form a $\mathbb{Z}$-basis of $\mathbb{Z}^n$.
\end{itemize}
The polytope is called {\it rational} instead of Delzant if we replace in the third condition the requirement of a $\mathbb{Z}$-basis by that of a $\mathbb{Q}$-basis.

\subsubsection*{Symplectic Toric Manifolds}
Before giving the precise technical definition of a symplectic toric manifold, let us first sketch the idea. Roughly speaking a toric manifold is a $\mathbb{T}^n$ fibration over a given $n$-dimensional polytope, such that at each facet a single $U(1)$ inside the $\mathbb{T}^n$ shrinks to zero size. On the intersections of the different facets multiple $U(1)$'s collapse, e.g. at the vertices all circles have shrunk. On the interior of the polytope the toric manifold is simply of the form $P^\circ\times\mathbb{T}^n$ and the full toric manifold is the compactification of this space. On the interior there is thus a standard set of coordinates of the form $(x_i,\theta_i)$ with $x_i\in P^\circ$ and $\theta_i\in\mathbb{T}$ and the manifold comes with a standard symplectic form $\Omega=\sum_idx_i\wedge d\theta_i$. It is of course rather non-trivial that this manifold can be smoothly compactified, but when the polytope is Delzant this is the case. Let us now state the above ideas more precisely.

A {\it symplectic toric manifold} is a compact connected $2n$-dimensional symplectic manifold $(M,\Omega)$ that allows an effective Hamiltonian action of an $n$-dimensional torus $\mathbb{T}^n$. Remember that the action of a Lie group on a symplectic manifold is called Hamiltonian if there exists a {\it moment map} $\mu$ from the manifold to the dual Lie algebra that satisfies
\be
d\langle\mu(p),X\rangle=\Omega(\cdot,\tilde X)\,,
\ee
with $p\in M$, $X$ a generator of the Lie algebra and $\tilde X$ the corresponding vectorfield. Furthermore the moment map should be equivariant with respect to the group action, i.e. $\mu(g(p))=\mathrm{Ad}^*_g\circ\mu(p)$, with $\mathrm{Ad}^*$ the coadjoint representation.

By a theorem of Delzant \cite{delzant} every symplectic toric manifold is uniquely characterized by a Delzant polytope. Given a symplectic toric manifold the corresponding polytope is given by the image of the moment map. To conversely reconstruct the manifold from the polytope is slightly more involved and relies on the technique of symplectic reduction, we refer readers interested in further details to e.g. \cite{acds}. Note that the normal fan to the polytope can be interpreted as a {\it fan}, which is used to characterize toric varieties in algebraic geometry, see e.g. \cite{Hori:2003ic} for a nice introduction. This can be useful to identify a symplectic manifold given by a polytope and furthermore provides an embedding in projective space.

\subsubsection*{K\"ahler toric manifolds}
What will be of use in this paper is that Delzant's construction also associates a set of canonical complex coordinates to every symplectic toric manifold, effectively implying that every closed symplectic toric manifold is actually a K\"ahler manifold. As throughout the paper we will make use of the explicit construction of these complex coordinates in terms of the symplectic coordinates $(x_i,\theta_i)$, we will detail the general procedure here, be it without proofs or motivation. Those can be found in references \cite{Guillemin,abreu-2000}.

As mentioned above (\ref{ineqs}) any polytope $P$ is characterized by a set of inequalities. Given this combinatorial data of the polytope one can define associated functions
\be
    l_i(x) = \sum_j c_{ij} x_j - \lambda_i\,, \qquad\qquad l_{\infty} = \sum_{i,j} c_{ij} x_j\,,\label{fun}
\ee
which are everywhere positive on $P$. Using these functions one can define a 'potential' as follows
\be
   g(x) = \frac{1}{2} \sum_i l_i (x) \log l_i(x)\,.
\ee
In case the polytope is Delzant, it is shown in \cite{Guillemin} that this potential defines good complex coordinates on the toric manifold as follows
\be
    z_i = \exp\left(\frac{\partial}{\partial x_i} g(x) + i \theta_i \right)\,. \label{cc}
\ee
Furthermore a K\"ahler potential for the corresponding K\"ahler metric $\Omega(\cdot,J\cdot)$ is given by
\be
    \calk = \sum_i \lambda_i \log l_i(x) + l_{\infty}\,.\label{km}
\ee
It follows from the construction \cite{Guillemin,abreu-2000} that $\calk$ is the Legendre transform of $g$, i.e. $\calk(z)=\frac{\partial g}{\partial x}x-g(x)$. This can be used to derive that
\be
   \det \partial_i \partial_{\bar{j}} \calk = \exp(\sum_i \frac{\partial g}{\partial x_i}) \, \det \frac{\partial^2 g}{\partial x_i \partial x_j}\,,
\ee
which will be a useful formula in the bulk of the paper.

\subsubsection*{Toric orbifolds}
As we also consider quotients of symplectic toric manifolds by a permutation group in this paper, it will be necessary to introduce the generalization of the above construction of complex coordinates to that of symplectic toric orbifolds. As in the manifold case, a {\it symplectic toric orbifold} is a $2n$-dimensional symplectic orbifold that allows a Hamiltonian $\mathbb{T}^n$ action. As was shown in \cite{lerman-1995} such symplectic toric orbifolds are in one to one correspondence to labeled rational polytopes. Such a labeled rational polytope is nothing but a rational polytope with a natural number attached to each facet. The label $m_i$ denotes that the $i$'th facet is a $\mathbb{Z}_{m_i}$ singularity. Again the explicit construction of the toric orbifold from the labeled polytope is rather involved and we refer those who are interested to \cite{lerman-1995}. The labeled polytope corresponding to the quotient of a symplectic toric manifold by a group respecting the torus action, is however easy to find. It is given by the quotient of the original polytope and attaching a label $m$ to each facet that is a $\mathbb{Z}_m$ fixed point under the group action.

Given a labeled rational polytope one can construct complex coordinates on the toric orbifold in a way similar to the manifold case. The functions $l_i$ from (\ref{fun}) are generalized to \cite{lerman-1995,abreu-2001}
\be
   l_i(x) = m_i \left(\sum_j c_{ij} x_j - \lambda_i\right)\,, \qquad\qquad l_{\infty} = \sum_{i,j} m_i c_{ij} x_j\,,
\ee
where $m_i$ is the label attached to the facet orthogonal to the vector $\vec c_i$.
The construction of the complex coordinates and the k\"ahler potential from these functions then carries on analogously to (\ref{cc},\ref{km}).

\section{Gravitational Throats and CFT Mass Gaps}\label{appendix3}
Having an infinitely deep throat in AdS spaces seems paradoxal as it suggests a continuous spectrum on the CFT side, as one can make arbitrarily small energy excitations by localizing them deep enough in the throat. In this appendix we would like to calculate the correspondence between the size of the mass gap and a smoothly capped off throat in the bulk. We are going to approximate the finite throat by a toy model metric with a throat that is crudely cut off at a scale $\epsilon$ and try to solve the scalar wave equation in this background. Due to the technical difficulty of the full problem we resort to matching the far ($r\gg\epsilon$) and near ($r\ll\epsilon$) region solutions. By doing so we are able to relate the inverse of the depth of the throat to the mass gap on the CFT. The result we find agrees with what one expects for CFT's with a long string picture. We use this result in section \ref{deepthroat} to relate the cutoff scale $\epsilon$ we calculate there to the size of a mass gap.

To get a reasonable guess to what the capped off geometry would be, we use that far away from the tip of the throat the geometry looks like the one of a D4D2D0 BTZ black hole. So our starting point is the following metric (see (4.1) in \cite{deBoer:2008fk})
\be ds^2=-\frac{r}{U} dtd\psi + \frac{1}{4} \frac{r+C}{U} d\psi^2
+ U^2 \frac{dr^2}{r^2}
\ee
with $C=\frac{S^2}{\pi^2 C^3}$ some constant determined by the charges. Let us assume this metric to be a good approximation for $r\geq \epsilon$, whereas we
take
\be
ds^2=-\frac{\epsilon}{U} dtd\psi +  \frac{r^2 U^2}{\epsilon^2} d\psi^2
+ \frac{U^2}{\epsilon^2} dr^2
\ee
valid for $r\leq \epsilon$. We find the following radial equations for a free scalar field in these two 'sub-geometries', in the outer region
\be
\left(-\partial_r \frac{r^2}{U^2} \partial_r + m^2 -
\frac{\omega^2 (C+r)U}{r^2} - \frac{4 U \omega k}{r} \right)\phi=0
\ee
and in the inner region
\be
\left(-\partial_r \frac{\epsilon^2}{U^2} \partial_r + m^2 -
\frac{\omega^2 r^2 U^4}{\epsilon^4} - \frac{4 U \omega
k}{\epsilon} \right)\phi=0.
\ee
Let us solve these equations. The first one has two solutions but
only one is normalizable at infinity. Define $\lambda$ via
\be
\frac{1}{4} - \lambda^2 = -m^2 U^2
\ee
and also
\be
\xi\equiv -i \frac{\omega}{2} \sqrt{\frac{U}{C}} (\omega + 4 k)
\ee
and the variable
\be
z = 2\omega i \sqrt{UC} /r
\ee
then the field equation for $\phi$ becomes
\be
(\partial_z^2 + (-\frac{1}{4} + \frac{\xi}{z} +
\frac{\frac{1}{4}-\lambda^2}{z^2} ))\phi=0
\ee
and the solution in the outer region in terms of Whittaker
functions is (see e.g. eq (84) in \cite{KeskiVakkuri:1998nw})
\be
\phi(z) = M_{\xi,\lambda}(z) +M_{-\xi,\lambda}(-z).\label{wsol1}
\ee
These two terms are proportional to each other but this answer is
the linear combination which yields a real answer. The large $z$
behavior is
\be
\phi(z) \sim \frac{e^{z/2}
z^{-\xi}}{\Gamma(\frac{1}{2}+\lambda-\xi)} + c.c.
\ee

Next we turn to the cap region. Here, the field equation is
usually solved in terms of parabolic cylinder functions. We need
the right linear combination which vanishes as $r\rightarrow 0$ to
be smooth there. Define
\be
a=\frac{\epsilon m^2}{2 U} - 2 k,\qquad r = x
\sqrt{\frac{\epsilon^3}{2\omega U^3}}
\ee
then the field equation is $(\partial_y^2 + \frac{y^2}{4}
-a)\phi=0$ and then we find the following large $x$ behavior of the
solution (notice that $r=\epsilon$ is at large $x$)
\be
\phi \sim \exp(-i x^2/4) x^{ia-1/2} \frac{(-1)^{3/8} 2^{-1/4-ia/2}
e^{a\pi/4} \sqrt{\pi}}{\Gamma(\frac{3}{4} + \frac{ia}{2})} + c.c.\label{wsol2}
\ee
Notice that the coefficients are important, only for this
particular coefficient (and including the c.c.) this function has the
asymptotics of the regular solution.

In order to match the solutions (\ref{wsol1}) and (\ref{wsol2}) at $r=\epsilon$, both the fields
and their first derivatives should match. A priori this need not
be possible since we already have unique solutions up to overall
normalization. Since the overall normalization is not fixed, the
condition that we need to impose is
\be \label{matching}
\frac{\partial z}{\partial r}
\frac{\phi'(z)}{\phi(z)}|_{r=\epsilon} = \frac{\partial
x}{\partial r} \frac{\phi'(x)}{\phi(x)}|_{r=\epsilon} .
\ee
In the small $\epsilon$-limit, the derivatives of the exponentials
give the largest contributions to $\phi'$, thus we only need to
differentiate those.

To proceed, we define $e^{i\rho}$ to be the phase of
$1/\Gamma(1/2+\lambda-\xi)$, and $e^{i\sigma}$ to be the phase of
\be
\frac{(-1)^{3/8} 2^{-1/4-ia/2}
e^{a\pi/4} \sqrt{\pi}}{\Gamma(\frac{3}{4} + \frac{ia}{2})} .
\ee
Then we find
\be
\phi_{\rm outer} \sim \cos(\frac{z}{2i} - (\xi/i) \log(z/i) +\rho)
,\qquad \phi_{\rm inner} \sim \frac{1}{\sqrt{x}}
\cos(-\frac{x^2}{4} + a \log(x) +\sigma)
\ee
up to irrelevant overall normalizations. We now evaluate the
matching condition keeping only the leading terms for small
$\epsilon$, i.e. we only differentiate $z/2i$ and $-x^2/4$. After
doing this, various factors of $\epsilon$ and $\omega$ happily
cancel and we are left with the matching condition (evaluated at
$r=\epsilon$)
\be
-\frac{U^{5/2}}{C^{1/2}} \tan(-\frac{x^2}{4} + a \log(x) +\sigma)
= \tan(\frac{z}{2i} - (\xi/i) \log(z/i) +\rho) .
\ee
As we increase $\omega$, but keeping $\omega$ small enough
so the matching strategy remains sensible, the
first tangent seems to oscillate most rapidly (as long as
$U^3>2\sqrt{UC}$, otherwise the other tangent seems to win).
Because of this oscillatory nature we get a gapped spectrum. A
crude estimate for the gap can therefore be made by looking at the values
of $\omega$ for which the first tangent makes a $\pi$ period.
Since $x^2/4$, evaluated at $r=\epsilon$, is $\omega
U^3/(2\epsilon)$, we find the following rough estimate:
\be
\Delta \omega \sim \frac{2\epsilon \pi}{U^3}.
\ee
Since the eigenvalue of $\partial_t$ is like that of
$L_0+\bar{L}_0$, without any further factors, we finally conclude
that
\be
\Delta (L_0+\bar{L}_0) \sim \frac{2\epsilon \pi}{U^3}.
\ee
In case $\epsilon$ is of order 1 the gap in the conformal weights scales
like $1/U^3$ which is precisely $1/c$, with $c$ the central charge of the dual CFT, as one would get from long string fractionation. This suggests that long strings might play an important role in the physics of four dimensional black holes and the dual $N=(0,4)$ SCFT.

\bibliographystyle{jhep}

\begin{thebibliography}{10}

\bibitem{Strominger:1996sh}
A.~Strominger and C.~Vafa, {\it Microscopic origin of the bekenstein-hawking
  entropy},  {\em Phys. Lett.} {\bf B379} (1996) 99--104,
  [\href{http://xxx.lanl.gov/abs/hep-th/9601029}{{\tt hep-th/9601029}}].

\bibitem{Lin:2004nb}
H.~Lin, O.~Lunin, and J.~M. Maldacena, {\it Bubbling ads space and 1/2 bps
  geometries},  {\em JHEP} {\bf 10} (2004) 025,
  [\href{http://xxx.lanl.gov/abs/hep-th/0409174}{{\tt hep-th/0409174}}].

\bibitem{Mandal:2005wv}
G.~Mandal, {\it Fermions from half-bps supergravity},  {\em JHEP} {\bf 08}
  (2005) 052, [\href{http://xxx.lanl.gov/abs/hep-th/0502104}{{\tt
  hep-th/0502104}}].

\bibitem{Lunin:2001fv}
O.~Lunin and S.~D. Mathur, {\it Metric of the multiply wound rotating string},
  {\em Nucl. Phys.} {\bf B610} (2001) 49--76,
  [\href{http://xxx.lanl.gov/abs/hep-th/0105136}{{\tt hep-th/0105136}}].

\bibitem{Grant:2005qc}
L.~Grant, L.~Maoz, J.~Marsano, K.~Papadodimas, and V.~S. Rychkov, {\it
  Minisuperspace quantization of 'bubbling ads' and free fermion droplets},
  {\em JHEP} {\bf 08} (2005) 025,
  [\href{http://xxx.lanl.gov/abs/hep-th/0505079}{{\tt hep-th/0505079}}].

\bibitem{Maoz:2005nk}
L.~Maoz and V.~S. Rychkov, {\it Geometry quantization from supergravity: The
  case of 'bubbling ads'},  {\em JHEP} {\bf 08} (2005) 096,
  [\href{http://xxx.lanl.gov/abs/hep-th/0508059}{{\tt hep-th/0508059}}].

\bibitem{Donos:2005vs}
A.~Donos and A.~Jevicki, {\it Dynamics of chiral primaries in ads(3) x s**3 x
  t**4},  {\em Phys. Rev.} {\bf D73} (2006) 085010,
  [\href{http://xxx.lanl.gov/abs/hep-th/0512017}{{\tt hep-th/0512017}}].

\bibitem{Rychkov:2005ji}
V.~S. Rychkov, {\it D1-d5 black hole microstate counting from supergravity},
  {\em JHEP} {\bf 01} (2006) 063,
  [\href{http://xxx.lanl.gov/abs/hep-th/0512053}{{\tt hep-th/0512053}}].

\bibitem{Balasubramanian:2005mg}
V.~Balasubramanian, J.~de~Boer, V.~Jejjala, and J.~Simon, {\it The library of
  babel: On the origin of gravitational thermodynamics},  {\em JHEP} {\bf 12}
  (2005) 006, [\href{http://xxx.lanl.gov/abs/hep-th/0508023}{{\tt
  hep-th/0508023}}].

\bibitem{Lunin:2001jy}
O.~Lunin and S.~D. Mathur, {\it Ads/cft duality and the black hole information
  paradox},  {\em Nucl. Phys.} {\bf B623} (2002) 342--394,
  [\href{http://xxx.lanl.gov/abs/hep-th/0109154}{{\tt hep-th/0109154}}].

\bibitem{Lunin:2002qf}
O.~Lunin and S.~D. Mathur, {\it Statistical interpretation of bekenstein
  entropy for systems with a stretched horizon},  {\em Phys. Rev. Lett.} {\bf
  88} (2002) 211303, [\href{http://xxx.lanl.gov/abs/hep-th/0202072}{{\tt
  hep-th/0202072}}].

\bibitem{Mathur:2002ie}
S.~D. Mathur, {\it {A proposal to resolve the black hole information paradox}},
   {\em Int. J. Mod. Phys.} {\bf D11} (2002) 1537--1540,
  [\href{http://xxx.lanl.gov/abs/hep-th/0205192}{{\tt hep-th/0205192}}].

\bibitem{Lunin:2002bj}
O.~Lunin, S.~D. Mathur, and A.~Saxena, {\it {What is the gravity dual of a
  chiral primary?}},  {\em Nucl. Phys.} {\bf B655} (2003) 185--217,
  [\href{http://xxx.lanl.gov/abs/hep-th/0211292}{{\tt hep-th/0211292}}].

\bibitem{Denef:2002ru}
F.~Denef, {\it Quantum quivers and hall/hole halos},  {\em JHEP} {\bf 10}
  (2002) 023, [\href{http://xxx.lanl.gov/abs/hep-th/0206072}{{\tt
  hep-th/0206072}}].

\bibitem{Denef:2007vg}
F.~Denef and G.~W. Moore, {\it Split states, entropy enigmas, holes and halos},
   \href{http://xxx.lanl.gov/abs/hep-th/0702146}{{\tt hep-th/0702146}}.

\bibitem{Bena:2004tk}
I.~Bena and P.~Kraus, {\it Microscopic description of black rings in ads/cft},
  {\em JHEP} {\bf 12} (2004) 070,
  [\href{http://xxx.lanl.gov/abs/hep-th/0408186}{{\tt hep-th/0408186}}].

\bibitem{Bena:2007qc}
I.~Bena, C.-W. Wang, and N.~P. Warner, {\it Plumbing the abyss: Black ring
  microstates},  \href{http://xxx.lanl.gov/abs/0706.3786}{{\tt 0706.3786}}.

\bibitem{Mathur:2007sc}
S.~D. Mathur, {\it Black hole size and phase space volumes},
  \href{http://xxx.lanl.gov/abs/0706.3884}{{\tt 0706.3884}}.

\bibitem{Mathur:2005zp}
S.~D. Mathur, {\it The fuzzball proposal for black holes: An elementary
  review},  {\em Fortsch. Phys.} {\bf 53} (2005) 793--827,
  [\href{http://xxx.lanl.gov/abs/hep-th/0502050}{{\tt hep-th/0502050}}].

\bibitem{deBoer:2008fk}
J.~de~Boer, F.~Denef, S.~El-Showk, I.~Messamah, and D.~Van~den Bleeken, {\it
  {Black hole bound states in AdS$_3$ x S$^2$}},
  \href{http://xxx.lanl.gov/abs/0802.2257}{{\tt 0802.2257}}.

\bibitem{Maldacena:1997de}
J.~M. Maldacena, A.~Strominger, and E.~Witten, {\it Black hole entropy in
  m-theory},  {\em JHEP} {\bf 12} (1997) 002,
  [\href{http://xxx.lanl.gov/abs/hep-th/9711053}{{\tt hep-th/9711053}}].

\bibitem{Minasian:1999qn}
R.~Minasian, G.~W. Moore, and D.~Tsimpis, {\it {Calabi-Yau black holes and
  (0,4) sigma models}},  {\em Commun. Math. Phys.} {\bf 209} (2000) 325--352,
  [\href{http://xxx.lanl.gov/abs/hep-th/9904217}{{\tt hep-th/9904217}}].

\bibitem{Denef:2000nb}
F.~Denef, {\it Supergravity flows and d-brane stability},  {\em JHEP} {\bf 08}
  (2000) 050, [\href{http://xxx.lanl.gov/abs/hep-th/0005049}{{\tt
  hep-th/0005049}}].

\bibitem{Bates:2003vx}
B.~Bates and F.~Denef, {\it Exact solutions for supersymmetric stationary black
  hole composites},  \href{http://xxx.lanl.gov/abs/hep-th/0304094}{{\tt
  hep-th/0304094}}.

\bibitem{Gaiotto:2005gf}
D.~Gaiotto, A.~Strominger, and X.~Yin, {\it New connections between 4d and 5d
  black holes},  {\em JHEP} {\bf 02} (2006) 024,
  [\href{http://xxx.lanl.gov/abs/hep-th/0503217}{{\tt hep-th/0503217}}].

\bibitem{Bena:2005va}
I.~Bena and N.~P. Warner, {\it Bubbling supertubes and foaming black holes},
  {\em Phys. Rev.} {\bf D74} (2006) 066001,
  [\href{http://xxx.lanl.gov/abs/hep-th/0505166}{{\tt hep-th/0505166}}].

\bibitem{Berglund:2005vb}
P.~Berglund, E.~G. Gimon, and T.~S. Levi, {\it Supergravity microstates for bps
  black holes and black rings},  {\em JHEP} {\bf 06} (2006) 007,
  [\href{http://xxx.lanl.gov/abs/hep-th/0505167}{{\tt hep-th/0505167}}].

\bibitem{Behrndt:1997ny}
K.~Behrndt, D.~Lust, and W.~A. Sabra, {\it {Stationary solutions of N = 2
  supergravity}},  {\em Nucl. Phys.} {\bf B510} (1998) 264--288,
  [\href{http://xxx.lanl.gov/abs/hep-th/9705169}{{\tt hep-th/9705169}}].

\bibitem{Shmakova:1996nz}
M.~Shmakova, {\it Calabi-yau black holes},  {\em Phys. Rev.} {\bf D56} (1997)
  540--544, [\href{http://xxx.lanl.gov/abs/hep-th/9612076}{{\tt
  hep-th/9612076}}].

\bibitem{Bena:2007kg}
I.~Bena and N.~P. Warner, {\it Black holes, black rings and their microstates},
   \href{http://xxx.lanl.gov/abs/hep-th/0701216}{{\tt hep-th/0701216}}.

\bibitem{Cheng:2006yq}
M.~C.~N. Cheng, {\it More bubbling solutions},  {\em JHEP} {\bf 03} (2007) 070,
  [\href{http://xxx.lanl.gov/abs/hep-th/0611156}{{\tt hep-th/0611156}}].

\bibitem{Balasubramanian:2006gi}
V.~Balasubramanian, E.~G. Gimon, and T.~S. Levi, {\it {Four Dimensional Black
  Hole Microstates: From D-branes to Spacetime Foam}},  {\em JHEP} {\bf 01}
  (2008) 056, [\href{http://xxx.lanl.gov/abs/hep-th/0606118}{{\tt
  hep-th/0606118}}].

\bibitem{Denef:2000ar}
F.~Denef, {\it {On the correspondence between D-branes and stationary
  supergravity solutions of type II Calabi-Yau compactifications}},
  \href{http://xxx.lanl.gov/abs/hep-th/0010222}{{\tt hep-th/0010222}}.

\bibitem{Michelson:1999dx}
J.~Michelson and A.~Strominger, {\it Superconformal multi-black hole quantum
  mechanics},  {\em JHEP} {\bf 09} (1999) 005,
  [\href{http://xxx.lanl.gov/abs/hep-th/9908044}{{\tt hep-th/9908044}}].

\bibitem{Bena:2006kb}
I.~Bena, C.-W. Wang, and N.~P. Warner, {\it {Mergers and typical black hole
  microstates}},  {\em JHEP} {\bf 11} (2006) 042,
  [\href{http://xxx.lanl.gov/abs/hep-th/0608217}{{\tt hep-th/0608217}}].

\bibitem{Lee:1990nz}
J.~Lee and R.~M. Wald, {\it Local symmetries and constraints},  {\em J. Math.
  Phys.} {\bf 31} (1990) 725--743.

\bibitem{Raju:2007uj}
S.~Raju, {\it Counting giant gravitons in ads(3)},
  \href{http://xxx.lanl.gov/abs/0709.1171}{{\tt 0709.1171}}.

\bibitem{Crnkovic:1986ex}
C.~Crnkovic and E.~Witten, {\it Covariant description of canonical formalism in
  geometrical theories}, . Print-86-1309 (PRINCETON).

\bibitem{Zuckerman:1989cx}
G.~J. Zuckerman, {\it Action principles and global geometry}, . Print-89-0321
  (YALE).

\bibitem{Ivanov:1990jn}
E.~A. Ivanov and A.~V. Smilga, {\it {Supersymmetric gauge quantum mechanics:
  Superfield description}},  {\em Phys. Lett.} {\bf B257} (1991) 79--82.

\bibitem{Diaconescu:1997ut}
D.-E. Diaconescu and R.~Entin, {\it {A non-renormalization theorem for the d =
  1, N = 8 vector multiplet}},  {\em Phys. Rev.} {\bf D56} (1997) 8045--8052,
  [\href{http://xxx.lanl.gov/abs/hep-th/9706059}{{\tt hep-th/9706059}}].

\bibitem{Ritter:2002zg}
W.~G. Ritter, {\it {Geometric quantization}},
  \href{http://xxx.lanl.gov/abs/math-ph/0208008}{{\tt math-ph/0208008}}.

\bibitem{EcheverriaEnriquez:1999jr}
A.~Echeverria-Enriquez, M.~C. Munoz-Lecanda, N.~Roman-Roy, and
  C.~Victoria-Monge, {\it {Mathematical foundations of geometric
  quantization}},  {\em Extracta Math.} {\bf 13} (1998) 135--238,
  [\href{http://xxx.lanl.gov/abs/math-ph/9904008}{{\tt math-ph/9904008}}].

\bibitem{Alday:2006nd}
L.~F. Alday, J.~de~Boer, and I.~Messamah, {\it The gravitational description of
  coarse grained microstates},  {\em JHEP} {\bf 12} (2006) 063,
  [\href{http://xxx.lanl.gov/abs/hep-th/0607222}{{\tt hep-th/0607222}}].

\bibitem{Guillemin}
V.~Guillemin, {\it Kaehler structures on toric varieties},  {\em J.
  Differential Geometry} {\bf 40} (1994) 285--309.

\bibitem{abreu-2000}
M.~Abreu, {\it Kahler geometry of toric manifolds in symplectic coordinates},
  2000.

\bibitem{Aichelburg:1987hy}
P.~C. Aichelburg and F.~Embacher, {\it {SUPERGRAVITY SOLITONS. 1. GENERAL
  FRAMEWORK}},  {\em Phys. Rev.} {\bf D37} (1988) 338.

\bibitem{Aichelburg:1987ia}
P.~C. Aichelburg and F.~Embacher, {\it {SUPERGRAVITY SOLITONS. 4. EFFECTIVE
  SOLITON INTERACTION}},  {\em Phys. Rev.} {\bf D37} (1988) 2132.

\bibitem{vaisman:1994}
I.~Vaisman, {\it {Super-Geometric Quantization}},  {\em Phys. Rev.} {\bf D37}
  (1988) 338.

\bibitem{Lawson}
H.~B. Lawson, Jr. and M.-L. Michelsohn, {\em Spin geometry}, vol.~38 of {\em
  Princeton Mathematical Series}.
\newblock Princeton University Press, Princeton, NJ, 1989.

\bibitem{Andriyash:2008it}
E.~Andriyash and G.~W. Moore, {\it {Ample D4-D2-D0 Decay}},
  \href{http://xxx.lanl.gov/abs/0806.4960}{{\tt arXiv:0806.4960}}.

\bibitem{Dijkgraaf:1998zd}
R.~Dijkgraaf, {\it {Fields, strings, matrices and symmetric products}},
  \href{http://xxx.lanl.gov/abs/hep-th/9912104}{{\tt hep-th/9912104}}.

\bibitem{Vafa:1994tf}
C.~Vafa and E.~Witten, {\it {A Strong coupling test of S duality}},  {\em Nucl.
  Phys.} {\bf B431} (1994) 3--77,
  [\href{http://xxx.lanl.gov/abs/hep-th/9408074}{{\tt hep-th/9408074}}].

\bibitem{lerman-1995}
E.~Lerman and S.~Tolman, {\it Hamiltonian torus actions on symplectic orbifolds
  and toric varieties},  1995.

\bibitem{abreu-2001}
M.~Abreu, {\it Kaehler metrics on toric orbifolds},  {\em J. Differential
  Geometry} {\bf 58} (2001) 151--187.

\bibitem{Gaiotto:2004ij}
D.~Gaiotto, A.~Strominger, and X.~Yin, {\it {Superconformal black hole quantum
  mechanics}},  {\em JHEP} {\bf 11} (2005) 017,
  [\href{http://xxx.lanl.gov/abs/hep-th/0412322}{{\tt hep-th/0412322}}].

\bibitem{Bena:2008nh}
I.~Bena, N.~Bobev, C.~Ruef, and N.~P. Warner, {\it {Entropy Enhancement and
  Black Hole Microstates}},  \href{http://xxx.lanl.gov/abs/0804.4487}{{\tt
  arXiv:0804.4487}}.

\bibitem{Denef:2007yt}
F.~Denef, D.~Gaiotto, A.~Strominger, D.~Van~den Bleeken, and X.~Yin, {\it Black
  hole deconstruction},  \href{http://xxx.lanl.gov/abs/hep-th/0703252}{{\tt
  hep-th/0703252}}.

\bibitem{BrittoPacumio:2000sv}
R.~Britto-Pacumio, A.~Strominger, and A.~Volovich, {\it {Two-black-hole bound
  states}},  {\em JHEP} {\bf 03} (2001) 050,
  [\href{http://xxx.lanl.gov/abs/hep-th/0004017}{{\tt hep-th/0004017}}].

\bibitem{Banks:1996vh}
T.~Banks, W.~Fischler, S.~H. Shenker, and L.~Susskind, {\it {M theory as a
  matrix model: A conjecture}},  {\em Phys. Rev.} {\bf D55} (1997) 5112--5128,
  [\href{http://xxx.lanl.gov/abs/hep-th/9610043}{{\tt hep-th/9610043}}].

\bibitem{Sethi:2000ba}
S.~Sethi and M.~Stern, {\it {The structure of the D0-D4 bound state}},  {\em
  Nucl. Phys.} {\bf B578} (2000) 163--198,
  [\href{http://xxx.lanl.gov/abs/hep-th/0002131}{{\tt hep-th/0002131}}].

\bibitem{Polchinski:1999br}
J.~Polchinski, {\it {M-theory and the light cone}},  {\em Prog. Theor. Phys.
  Suppl.} {\bf 134} (1999) 158--170,
  [\href{http://xxx.lanl.gov/abs/hep-th/9903165}{{\tt hep-th/9903165}}].

\bibitem{Skenderis:2008qn}
K.~Skenderis and M.~Taylor, {\it {The fuzzball proposal for black holes}},
  \href{http://xxx.lanl.gov/abs/0804.0552}{{\tt arXiv:0804.0552}}.

\bibitem{Kim:2005yb}
S.~Kim and J.~Raeymaekers, {\it {Superconformal quantum mechanics of small
  black holes}},  {\em JHEP} {\bf 08} (2005) 082,
  [\href{http://xxx.lanl.gov/abs/hep-th/0505176}{{\tt hep-th/0505176}}].

\bibitem{Raeymaekers:2007ga}
J.~Raeymaekers, {\it {Near-horizon microstates of the D1-D5-P black hole}},
  {\em JHEP} {\bf 02} (2008) 006,
  [\href{http://xxx.lanl.gov/abs/0710.4912}{{\tt arXiv:0710.4912}}].

\bibitem{Raeymaekers:2008gk}
J.~Raeymaekers, W.~Van~Herck, B.~Vercnocke, and T.~Wyder, {\it {5D fuzzball
  geometries and 4D polar states}},
  \href{http://xxx.lanl.gov/abs/0805.3506}{{\tt arXiv:0805.3506}}.

\bibitem{delzant}
T.~Delzant, {\it {Hamiltoniens p\'eriodiques et images convexes de
  l'application moment}},  {\em Bull. Soc. Math. France} {\bf 116} (1988) no.
  3, 315--339.

\bibitem{acds}
A.~Cannas~da Silva, {\it {Symplectic Toric Manifolds}},
  \href{http://xxx.lanl.gov/abs/http://www.math.ist.utl.pt/~acannas/Books/tori%
c.pdf}{{\tt http://www.math.ist.utl.pt/~acannas/Books/toric.pdf}}.

\bibitem{Hori:2003ic}
K.~Hori {\em et.~al.}, {\it {Mirror symmetry}}, . Providence, USA: AMS (2003)
  929 p.

\bibitem{KeskiVakkuri:1998nw}
E.~Keski-Vakkuri, {\it {Bulk and boundary dynamics in BTZ black holes}},  {\em
  Phys. Rev.} {\bf D59} (1999) 104001,
  [\href{http://xxx.lanl.gov/abs/hep-th/9808037}{{\tt hep-th/9808037}}].

\end{thebibliography}
\providecommand{\href}[2]{#2}\begingroup\raggedright\endgroup

\end{document}